\documentclass[a4paper,11pt]{article}

\usepackage{jcappub}
\usepackage{lineno}

\usepackage{amsmath,amssymb,amsfonts}
\usepackage{mathtools}
\usepackage{bm}

\usepackage{graphicx}
\usepackage{subcaption}
\usepackage{booktabs}
\usepackage{array}
\usepackage{fontawesome5}
\usepackage{multirow}

\usepackage[numbers,sort&compress]{natbib}
\usepackage[
    colorlinks=true,
    linkcolor=blue,
    citecolor=blue,
    urlcolor=blue
]{hyperref}

\usepackage{xcolor}
\usepackage{enumitem}

\newcommand{\dd}{\mathrm{d}}

\title{
Fourier-Preconditioned Path Deformations for Multi-Field Vacuum Tunnelling
}
\author[1,2]{Suriyah R. Kannagi,}
\author[1]{Aadarsh Singh,}
\author[1]{Sudhir K. Vempati\note{Corresponding author.}}

\affiliation[1]{
Centre for High Energy Physics,
Indian Institute of Science,
C. V. Raman Avenue,
Bangalore 560012,
India
}

\affiliation[2]{
Perimeter Institute for Theoretical Physics,
31 Caroline Street North,
Waterloo, ON N2L 2Y5,
Canada
}

\emailAdd{srajalingamkannagi@perimeterinstitute.ca,
 aadarshsingh@iisc.ac.in, vempati@iisc.ac.in}

\date{\today}

\abstract{We present an endpoint-safe Fourier method for multi-field vacuum tunnelling.
The field-space tunnelling path is written as a straight-line interpolation
between the false and true vacua, plus sine-mode deformations that vanish at the
endpoints. This gives a finite-dimensional path optimisation problem, which we
implement using automatic differentiation in the JAX numerical framework.
The method is studied both as a standalone variational ansatz for curved
tunnelling paths and as a preconditioner for existing bounce solvers. On the
OptiBounce benchmark potential for $N_\phi=3,\ldots,20$ and on a nested
random-coefficient potential family up to $N_\phi=50$, the Fourier result
agrees with FindBounce, OptiBounce, and CosmoTransitions at the sub-percent
level in the regular benchmark cases, while requiring only a modest number of
modes. We also compare several endpoint-safe basis families and find that
Fourier sine modes provide a robust default for smooth tunnelling paths.
When used as an initialiser, the Fourier path supplies useful geometric
information to existing solvers before the final bounce calculation is carried
out. In the CosmoTransitions tests, this reduces the number of steps in subsequent path
deformation, while in the FindBounce point-injection tests, it gives large
runtime improvements in the high-dimensional cases up to 90 $\%$. These results suggest that
endpoint-safe Fourier paths provide a useful bridge between simple analytic path
ans\"atze and fully numerical multi-field bounce algorithms.}

\begin{document}

\maketitle

\flushbottom

\section{Introduction}
\label{sec:introduction}

False-vacuum decay is a basic non-perturbative process in quantum field
theory, with applications ranging from electroweak vacuum stability to
first-order phase transitions in the early Universe. In the semiclassical
description, the decay rate per unit volume is controlled by the Euclidean
action of a bounce configuration,
\begin{equation}
    \frac{\Gamma}{V}
    \sim
    A\,\exp\!\left[-S_E[\bm{\phi}_b]\right],
    \label{eq:decay_rate_intro}
\end{equation}
where $\bm{\phi}_b$ is the bounce solution and $A$ is the fluctuation
prefactor~\cite{Coleman:1977py,Callan:1977pt}. At zero temperature, the relevant
solution is $O(4)$ symmetric, while at finite temperature, the corresponding
thermal bounce is usually $O(3)$ symmetric~\cite{Linde:1981zj,Linde:1983}. An
accurate calculation of the bounce action is therefore important for vacuum
stability studies, bubble nucleation, and gravitational-wave predictions from
cosmological phase transitions~\cite{Isidori:2001bm,Degrassi:2012ry,
Buttazzo:2013uya,Caprini:2015zlo,Caprini:2019egz}.

For a single scalar field, the bounce problem reduces to a one-dimensional
radial boundary-value problem. Even there, shooting methods can become delicate
in thin-wall regimes, nearly flat potentials, or potentials with multiple
scales. In theories with several scalar fields, the problem is harder: the
bounce traces a curve in field space, and the tunnelling trajectory is not known
in advance. A straight-line interpolation between the false and true vacua is a
simple starting point, but it need not follow the true tunnelling valley and can
give a poor representation of the action.
This difficulty has motivated a number of numerical approaches, and several
publicly available codes now address the multi-field bounce from complementary
angles. The Coleman--Glaser--Martin (CGM) construction reformulates the bounce
as a reduced minimisation problem by removing the scale instability associated
with the negative mode~\cite{Coleman:1977th}. Building on related ideas,
CosmoTransitions implements an iterative path-deformation method for multi-field
phase transitions and bubble profiles~\cite{Wainwright:2011kj}, while FindBounce
evaluates multi-field bounce actions through a polygonal bounce
construction~\cite{Guada:2020xnz}. Working directly from the reduced problem,
SimpleBounce solves the CGM equations through a flow
equation~\cite{Sato:2019axv}, and BubbleProfiler uses a perturbative
gradient-flow method for multi-field profiles~\cite{Athron:2019nbd}. OptiBounce
likewise targets the reduced problem, combining a pseudo-spectral
Gauss--Legendre collocation scheme with nonlinear
optimisation~\cite{BARDSLEY}.
These tools have made multi-field tunnelling calculations practical, but the
choice of path representation and initialisation remains important.
A complementary viewpoint is provided by the tunnelling-potential formalism
introduced by Espinosa and developed together with Konstandin for the
multi-field case~\cite{Espinosa:2018hue,Espinosa:2018szu}. In this
approach, the tunnelling action is written directly as a field-space functional
in terms of an auxiliary tunnelling potential $V_t$. The full formulation
reproduces the Euclidean bounce action at its minimum, and approximate
tunnelling potentials can already give accurate action estimates for typical
potentials. This makes the formalism especially useful for constructing
reduced field-space objectives and for understanding multi-field tunnelling as a
path problem.

In this work, we develop an endpoint-safe Fourier path-deformation method for
multi-field tunnelling. The path is written as a straight interpolation between
the false and true vacua plus a finite set of sine-mode deformations that vanish
at both endpoints,
\begin{equation}
    \bm{\phi}(t)
    =
    (1-t)\bm{\phi}_F
    +
    t\bm{\phi}_T
    +
    \sum_{k=1}^{N_m}
    \bm a_k \sin(k\pi t),
    \qquad
    t\in[0,1].
\end{equation}
The endpoint conditions are therefore satisfied analytically for any values of
the coefficients. We minimise a fixed-$V_t$ reduced functional over these
Fourier coefficients using automatic differentiation in JAX---a Python framework
for automatic differentiation and just-in-time compilation of array
programs~\cite{Bradbury:2018jax}---together with the limited-memory
Broyden--Fletcher--Goldfarb--Shanno algorithm with box constraints (L-BFGS-B), a
quasi-Newton optimiser well suited to smooth, moderately sized
problems~\cite{Byrd:1995lbfgs}. The number of Fourier modes is increased
adaptively, mode by mode, until the relative improvement in the reduced action
falls below a prescribed tolerance.
The method has two uses. First, it can be treated as a standalone
finite-dimensional variational ansatz for the tunnelling path. In this role, it
provides a simple way to study mode convergence and to identify curved
field-space trajectories. Second, the optimised Fourier path can be used as a
preconditioner for existing solvers. In this more conservative use, the Fourier
calculation is not the final bounce calculation; it supplies an improved
initial path or intermediate path information to codes such as CosmoTransitions \cite{Wainwright:2011kj}
and FindBounce \cite{Guada:2020xnz}.

The main results of this paper are as follows. First, in a two-field example,
we show that the Fourier-deformed path follows the same low-potential valley as
the path obtained from CosmoTransitions. We then validate the method on the
published OptiBounce benchmark potential for $N_\phi=3,\ldots,20$ fields,
finding sub-percent agreement with FindBounce, OptiBounce, and
CosmoTransitions in the regular cases. We also test a nested random-coefficient
potential family up to $N_\phi=50$, where the JAX--Fourier result agrees with
FindBounce wherever the independent comparison completes, while using only a
modest number of modes in the high-dimensional cases. A comparison of
endpoint-safe basis families show that Fourier sine modes are a robust default
for smooth tunnelling paths, with local B-spline and hybrid Fourier-local bases
providing useful cross-checks.
We also use the Fourier path as an initialiser for existing solvers. For
CosmoTransitions, a one-mode Fourier initialisation reduces the subsequent
path-deformation work in the curved-valley benchmarks while leaving the final
action stable. For FindBounce, a conservative one-point Fourier-informed
initialisation substantially reduces the runtime in the displayed high-dimensional
examples, with actions remaining comparable to the straight-path initialisation.
These tests support the use of Fourier-deformed paths as a practical
preconditioning layer for established bounce algorithms.

The paper is organised as follows. Section~\ref{sec:tunnelling_setup} reviews
the Euclidean bounce, the Coleman--Glaser--Martin reduced problem, and the
tunnelling-potential formulation. Section~\ref{sec:fourier_method} introduces
the endpoint-safe Fourier path ansatz and the fixed-$V_t$ reduced functional. The JAX implementation is mentioned in detail in Appendix~\ref{app:jax_implementation}. Section~\ref{sec:benchmarks} presents the standalone
benchmarks, including the two-field visualisation, OptiBounce validation,
high-dimensional random benchmark, and basis comparison. Section~\ref{sec:solver_initialization}
studies Fourier initialisation of CosmoTransitions and FindBounce. We conclude
in Section~\ref{sec:conclusions}.
\section{Recap of Vacuum tunnelling and path-based formulation}
\label{sec:tunnelling_setup}

\subsection{Euclidean bounce action}

We consider a theory of $N_\phi$ real scalar fields $\bm{\phi}$ with canonical
kinetic terms and scalar potential $V(\bm{\phi})$. The Euclidean action is
\begin{equation}
    S_E[\bm{\phi}]
    =
    \int \dd^4 x
    \left[
        \frac{1}{2}
        \partial_\mu \bm{\phi}\cdot\partial_\mu \bm{\phi}
        +
        V(\bm{\phi})
    \right].
    \label{eq:euclidean_action_full}
\end{equation}
If the potential has a metastable false vacuum $\bm{\phi}_F$, the decay rate per
unit volume is controlled semiclassically by the bounce action,
\begin{equation}
    \frac{\Gamma}{V}
    \sim
    A\,\exp\!\left[-S_E[\bm{\phi}_b]\right],
    \label{eq:decay_rate}
\end{equation}
where $\bm{\phi}_b$ is the bounce configuration and $A$ is the fluctuation
prefactor~\cite{Coleman:1977py,Callan:1977pt}.
At zero temperature, the bounce is $O(4)$ symmetric. Writing
$\bm{\phi}=\bm{\phi}(r)$ with $r=\sqrt{x_\mu x_\mu}$, the Euclidean action
reduces to
\begin{equation}
    S_E[\bm{\phi}]
    =
    2\pi^2
    \int_0^\infty
    \dd r\, r^3
    \left[
        \frac{1}{2}
        \left|
        \frac{\dd\bm{\phi}}{\dd r}
        \right|^2
        +
        V(\bm{\phi})
    \right].
    \label{eq:euclidean_action_o4}
\end{equation}
The corresponding bounce equations are
\begin{equation}
    \frac{\dd^2\bm{\phi}}{\dd r^2}
    +
    \frac{3}{r}
    \frac{\dd\bm{\phi}}{\dd r}
    =
    \bm{\nabla}V(\bm{\phi}),
    \label{eq:bounce_eom}
\end{equation}
with boundary conditions
\begin{equation}
    \left.
    \frac{\dd\bm{\phi}}{\dd r}
    \right|_{r=0}
    =
    \bm{0},
    \qquad
    \lim_{r\to\infty}\bm{\phi}(r)=\bm{\phi}_F .
    \label{eq:bounce_bc}
\end{equation}
The first condition follows from regularity at the origin, while the second
condition ensures that the configuration approaches the false vacuum at large
Euclidean radius.

\subsection{Variational viewpoint and the CGM reduced problem}
The bounce is a stationary point of the Euclidean action, but not an ordinary
minimum. It has one negative mode, associated physically with the instability of
the false vacuum and mathematically with dilations of the bounce radius
\cite{Coleman:1977py,Callan:1977pt,Coleman:1977th}. This feature is one reason
why direct minimisation of $S_E$ can be numerically delicate
\cite{Coleman:1977th,Espinosa:2018hue,Espinosa:2018szu}.

A useful way to see the role of this negative mode is to consider the
one-parameter family of dilated configurations
\begin{equation}
    \bm{\phi}_\lambda(r)
    =
    \bm{\phi}(r/\lambda).
\end{equation}
with $\lambda$ being the scaling factor. Writing the radial Euclidean action as
\begin{equation}
    S_E[\bm{\phi}]
    =
    T[\bm{\phi}]
    +
    \mathcal{V}[\bm{\phi}],
\end{equation}
where
\begin{equation}
    T[\bm{\phi}]
    =
    2\pi^2
    \int_0^\infty
    \dd r\, r^3
    \frac{1}{2}
    \left|
    \frac{\dd\bm{\phi}}{\dd r}
    \right|^2,
    \qquad
    \mathcal{V}[\bm{\phi}]
    =
    2\pi^2
    \int_0^\infty
    \dd r\, r^3
    V(\bm{\phi}),
\end{equation}
one obtains, in four dimensions,
\begin{equation}
    T[\bm{\phi}_\lambda]
    =
    \lambda^2 T[\bm{\phi}],
    \qquad
    \mathcal{V}[\bm{\phi}_\lambda]
    =
    \lambda^4 \mathcal{V}[\bm{\phi}].
    \label{eq:dilation_scaling}
\end{equation}
Stationarity with respect to $\lambda$ gives the virial relation
\begin{equation}
    T[\bm{\phi}_b]
    =
    -2\mathcal{V}[\bm{\phi}_b].
    \label{eq:virial_relation}
\end{equation}

Coleman, Glaser, and Martin showed that the scale instability can be removed
by replacing the original saddle-point problem with a reduced minimisation
problem~\cite{Coleman:1977th}. For $d=4$, the corresponding reduced functional is
\begin{equation}
    f[\bm{\phi}]
    =
    \frac{
        T[\bm{\phi}]^2
    }{
        -\mathcal{V}[\bm{\phi}]
    },
    \label{eq:cgm_objective}
\end{equation}
defined for configurations with $\mathcal{V}<0$. At the bounce,
\begin{equation}
    S_E[\bm{\phi}_b]
    =
    \frac{1}{4}f[\bm{\phi}_b].
    \label{eq:cgm_action_relation}
\end{equation}
This result is conceptually important for the present work because it shows
that, after treating the scale instability appropriately, the bounce problem
can be reformulated as a genuine minimisation problem. Modern algorithms use
this lesson in different ways: OptiBounce directly minimises the CGM reduced
functional \cite{BARDSLEY}, while path-deformation, used by cosmoTransition \cite{Wainwright:2011kj}, and tunnelling-potential methods construct \cite{Espinosa:2018hue}
related but distinct reduced descriptions of the multi-field tunnelling problem.

\subsection{Path formulation and tunnelling-potential method}

In a multi-field theory, the bounce traces a curve in field space. Let $s$ be
the arc-length coordinate along this curve,
\begin{equation}
    \dd s^2
    =
    \sum_{i=1}^{N_\phi}
    \dd\phi_i^2 .
    \label{eq:arc_length_definition}
\end{equation}
Along a fixed path $\bm{\phi}(s)$, the potential becomes an effective
one-dimensional potential
\begin{equation}
    V(s)
    \equiv
    V(\bm{\phi}(s)).
\end{equation}
The kinetic term along the bounce can then be written as
\begin{equation}
    \sum_i
    \left(
        \frac{\dd\phi_i}{\dd r}
    \right)^2
    =
    \left(
        \frac{\dd s}{\dd r}
    \right)^2 .
    \label{eq:kinetic_arc_length}
\end{equation}

The tunnelling-potential formulation provides an alternative field-space
description of the bounce problem~\cite{Espinosa:2018hue,Espinosa:2018szu}.
One defines the tunnelling potential along the path by
\begin{equation}
    V_t(s)
    =
    V(s)
    -
    \frac{1}{2}
    \left(
        \frac{\dd s}{\dd r}
    \right)^2 .
    \label{eq:tunnelling_potential_definition}
\end{equation}
Equivalently,
\begin{equation}
    V(s)-V_t(s)
    =
    \frac{1}{2}
    \left(
        \frac{\dd s}{\dd r}
    \right)^2 .
    \label{eq:kinetic_from_vt}
\end{equation}
Following the tunnelling-potential formulation of
Refs.~\cite{Espinosa:2018hue,Espinosa:2018szu}, $V_t$ measures the difference
between the potential energy along the path and the radial kinetic energy of
the bounce.
For an $O(d)$-symmetric bounce, the tangential equation along the path is
\begin{equation}
    \frac{\dd^2 s}{\dd r^2}
    +
    \frac{d-1}{r}
    \frac{\dd s}{\dd r}
    =
    \frac{\dd V}{\dd s}.
    \label{eq:tangential_bounce}
\end{equation}
Differentiating Eq.~\eqref{eq:tunnelling_potential_definition} with respect to
$s$ and using Eq.~\eqref{eq:tangential_bounce}, one can eliminate the radial
coordinate in favour of field-space quantities. This gives
\begin{equation}
    r(s)
    =
    (d-1)
    \frac{
        \sqrt{2\,[V(s)-V_t(s)]}
    }{
        -\dd V_t/\dd s
    },
    \label{eq:r_from_vt}
\end{equation}
where the sign convention is chosen such that $-\dd V_t/\dd s>0$ along the
tunnelling direction.
The Euclidean action can then be written as the field-space functional
\cite{Espinosa:2018hue,Espinosa:2018szu}
\begin{equation}
    S_d[V_t,\bm{\phi}]
    =
    \frac{(d-1)^{d-1}}{d}\,
    \Omega_d
    \int
    \dd s\,
    \frac{
        \left[
            2\left(V(s)-V_t(s)\right)
        \right]^{d/2}
    }{
        \left(
            -\dd V_t/\dd s
        \right)^{d-1}
    } .
    \label{eq:reduced_tunnelling_action_d}
\end{equation}
For the zero-temperature case $d=4$, this reduces to
\begin{equation}
    S_4[V_t,\bm{\phi}]
    =
    54\pi^2
    \int
    \dd s\,
    \frac{
        \left[
            V(s)-V_t(s)
        \right]^2
    }{
        \left(
            -\dd V_t/\dd s
        \right)^3
    } .
    \label{eq:reduced_action_exact}
\end{equation}

Equation~\eqref{eq:reduced_action_exact} is the starting point for our
construction. In the full tunnelling-potential formalism, the physical bounce is
obtained by minimising this action with respect to both the path and the
tunnelling potential, and the result reproduces the standard Euclidean bounce
action. The practical advantage of this formulation is that it turns the bounce
problem into a field-space minimisation problem and also permits accurate
approximations based on simple trial forms for $V_t$. In the full formulation, $V_t$ is varied
together with the path and the exact bounce action is recovered at the minimum
\cite{Espinosa:2018hue,Espinosa:2018szu}. Espinosa and Konstandin also showed
that simple approximate tunnelling potentials can already provide accurate
action estimates in typical potentials. In the present implementation, we do not
solve the full variational problem for $V_t$. Instead, as described in
Sec.~\ref{sec:reduced_action_ansatz}, we use a fixed smooth representative
profile for $V_t$ and minimise the corresponding reduced functional over
boundary-condition-preserving Fourier deformations of the path. Thus, the tunnelling-potential formalism supplies the reduced-action framework,
while the new ingredient here is the endpoint-safe Fourier parametrisation and
its use both as a standalone path ansatz and as an initialiser for existing
multi-field tunnelling solvers.
\section{Fourier-deformed tunnelling paths}
\label{sec:fourier_method}

\subsection{Boundary-condition-preserving Fourier ansatz}
\label{subsec:fourier_ansatz}

We parametrise the tunnelling path by a dimensionless variable $t\in[0,1]$,
where $t=0$ corresponds to the false vacuum and $t=1$ corresponds to the true
vacuum. The simplest path connecting the two vacua is the straight line
interpolation
\begin{equation}
    \bm{\phi}_{\rm str}(t)
    =
    (1-t)\bm{\phi}_F
    +
    t\bm{\phi}_T .
    \label{eq:straight_path_again}
\end{equation}
with $\phi_T$ and $\phi_F$ denoting the true and false vacuum, respectively. In a multi-field potential, however, the dominant tunnelling trajectory need not
follow this straight line. We therefore deform the straight path using a finite
set of Fourier sine modes,
\begin{equation}
    \bm{\phi}(t)
    =
    (1-t)\bm{\phi}_F
    +
    t\bm{\phi}_T
    +
    \sum_{k=1}^{N_m}
    \bm a_k \sin(k\pi t),
    \label{eq:fourier_path_ansatz}
\end{equation}
where $\bm a_k$ is an $N_\phi$-component vector of variational coefficients and
$N_m$ is the number of retained modes. Equivalently, for the $i$th field,
\begin{equation}
    \phi_i(t)
    =
    (1-t)(\phi_F)_i
    +
    t(\phi_T)_i
    +
    \sum_{k=1}^{N_m}
    a_{ik}\sin(k\pi t),
    \qquad
    i=1,\ldots,N_\phi .
    \label{eq:fourier_path_component}
\end{equation}
The sine basis is useful because each deformation mode vanishes at both
endpoints,
\begin{equation}
    \sin(k\pi t)=0
    \qquad
    \text{for}
    \qquad
    t=0,1 .
    \label{eq:sine_endpoint_condition}
\end{equation}
As a result, the endpoint conditions
\begin{equation}
    \bm{\phi}(0)=\bm{\phi}_F,
    \qquad
    \bm{\phi}(1)=\bm{\phi}_T
\end{equation}
are satisfied exactly for any values of the coefficients $\bm a_k$. The
minimisation over Fourier coefficients, therefore, never moves the path endpoints
away from the chosen vacua. This removes the need to impose endpoint constraints
during the numerical optimisation.
The number of Fourier modes controls the resolution of the path deformation.
The lowest modes describe broad, smooth bending of the trajectory in field
space, while higher modes add progressively shorter-scale structure. Truncating
the expansion at finite $N_m$ therefore gives a controlled finite-dimensional
variational problem: increasing $N_m$ systematically enlarges the space of
allowed paths. In practice, the low modes are especially important because they
capture the large-scale deviation from the straight-line path, which is the
dominant correction in many smooth multi-field potentials.
The total number of variational parameters is
\begin{equation}
    N_{\rm var}=N_\phi \times N_m ,
\end{equation}
corresponding to one Fourier coefficient for each field and each retained mode.
The straight-line path is recovered by setting all coefficients to zero,
\begin{equation}
    a_{ik}=0
    \qquad
    \text{for all}
    \qquad
    i,k .
\end{equation}
Thus, the Fourier ansatz contains the straight path as a special point in
coefficient space, and the minimisation searches for action-lowering
deformations around it.

\subsection{Tunnelling-potential reduced functional}
\label{sec:reduced_action_ansatz}

We use the tunnelling-potential formulation as a field-space variational
framework for assigning an action to a candidate path
\cite{Espinosa:2018hue,Espinosa:2018szu}. From eq.\eqref{eq:reduced_action_exact} we have
\begin{equation}
    S[V_t]
    =
    54\pi^2
    \int
    \dd s\,
    \frac{
        \left[V(\bm{\phi}(s))-V_t(s)\right]^2
    }{
        \left[-\dd V_t/\dd s\right]^3
    } .
\end{equation}
This is the central field-space functional of the tunnelling-potential method
\cite{Espinosa:2018hue,Espinosa:2018szu}. In the present work, we use this
known reduced-action structure as an objective for optimising
Fourier-deformed paths, rather than as a new independent
tunnelling-potential solver.
In the full tunnelling-potential method, $V_t$ is determined self-consistently by
minimising the tunnelling-potential action. However, one of the practical
advantages of the formulation is that simple approximations to $V_t$ can already
give accurate estimates of the tunnelling action for representative potentials.
Following this logic, we use a fixed smooth interpolation for $V_t(t)$ and
minimise the resulting functional over Fourier-deformed paths.

We shift the potential by the false-vacuum energy,
\begin{equation}
    \widetilde V(\bm{\phi})
    =
    V(\bm{\phi})-V(\bm{\phi}_F),
\end{equation}
so that $\widetilde V(\bm{\phi}_F)=0$ and
$\widetilde V(\bm{\phi}_T)<0$. Motivated by the endpoint structure of the
tunnelling-potential formulation \cite{Espinosa:2018hue,Espinosa:2018szu}, we use the following fixed smooth profile in
the numerical implementation:
\begin{equation}
    \widetilde V_t(t)
    =
    t^2(3-2t)\,\widetilde V(\bm{\phi}_T),
    \qquad
    t\in[0,1].
    \label{eq:vt_cubic_ansatz}
\end{equation}
This smoothstep form satisfies
\begin{equation}
    \widetilde V_t(0)=0,
    \qquad
    \widetilde V_t(1)=\widetilde V(\bm{\phi}_T),
\end{equation}
and has a vanishing first derivative at both endpoints. It therefore provides a
simple monotonic tunnelling-potential profile with the correct endpoint
behaviour, giving a well-defined reduced functional for optimising the path.

With this choice, the only variational parameters are the Fourier coefficients
in the path ansatz. The reduced functional becomes
\begin{equation}
    \widehat S(\{a_{ik}\})
    =
    S\!\left[
        \widetilde V_t,\,
        \bm{\phi}(t;\{a_{ik}\})
    \right]
    =
    54\pi^2
    \int
    \dd s\,
    \frac{
        \left[
            \widetilde V(\bm{\phi}(s;\{a_{ik}\}))
            -
            \widetilde V_t(s)
        \right]^2
    }{
        \left[
            -\dd \widetilde V_t/\dd s
        \right]^3
    } .
    \label{eq:reduced_action_coefficients}
\end{equation}
with \(s\) denoting the arc-length coordinate along the path parametrized by \(t\). The optimised Fourier path is obtained from
\begin{equation}
    \{a_{ik}^{\star}\}
    =
    \operatorname*{arg\,min}_{\{a_{ik}\}}
    \widehat S(\{a_{ik}\}) .
    \label{eq:fourier_coefficient_minimization}
\end{equation}
Here $\widetilde V$ denotes the potential shifted so that the false vacuum has
zero energy, while $\widetilde V_t$ denotes the fixed tunnelling-potential
profile used in the numerical objective.
For the numerical evaluation, we discretise the interval $t\in[0,1]$ using a
uniform grid with $N_q=260$ \footnote{This
grid size was chosen as a practical optimization between runtime and numerical
accuracy for the smooth benchmark potentials considered here. It is kept fixed
throughout each scan so that changes in the action reflect the Fourier-mode
optimization rather than changes in the discretization.} points. Defining
\begin{equation}
    \Delta s_i
    =
    \left|
    \bm{\phi}_{i+1}-\bm{\phi}_i
    \right|,
    \qquad
    \Delta \widetilde V_{t,i}
    =
    \widetilde V_{t,i+1}-\widetilde V_{t,i},
\end{equation}
and midpoint averages
\begin{equation}
    \overline{\widetilde V}_i
    =
    \frac{1}{2}
    \left[
        \widetilde V(\bm{\phi}_i)
        +
        \widetilde V(\bm{\phi}_{i+1})
    \right],
    \qquad
    \overline{\widetilde V}_{t,i}
    =
    \frac{1}{2}
    \left[
        \widetilde V_{t,i}
        +
        \widetilde V_{t,i+1}
    \right],
\end{equation}
The midpoint averages are used because each term in the sum represents the
contribution from the interval $[t_i,t_{i+1}]$; the derivative
$\dd \widetilde V_t/\dd s$ is approximated by a finite difference on this
interval, so the numerator is evaluated at the same interval midpoint. The discretised functional used in the code is
\begin{equation}
    \widehat S
    =
    54\pi^2
    \sum_{i=0}^{N_q-2}
    \frac{
        \left[
            \overline{\widetilde V}_i
            -
            \overline{\widetilde V}_{t,i}
        \right]^2
        \Delta s_i^4
    }{
        \left[
            -\Delta \widetilde V_{t,i}
        \right]^3
    } .
    \label{eq:reduced_action_discrete}
\end{equation}
The factor $\Delta s_i^4$ arises from approximating
$\dd \widetilde V_t/\dd s$ by
$\Delta \widetilde V_{t,i}/\Delta s_i$ and from the remaining line element in
the integral.
The role of $\widehat S$ is therefore practical: it is an action-informed
objective that selects smooth Fourier deformations of the path. The resulting actions are tested through mode convergence and, where available,
by direct comparison with established numerical codes.


The Fourier construction has two complementary uses. First, it can be used as a
standalone finite-dimensional ansatz for the tunnelling path. In this
interpretation, the Fourier coefficients are optimised directly, and the
resulting path provides an action-informed estimate of the dominant curved
trajectory in field space. This use is helpful for studying mode convergence,
comparing endpoint-safe basis choices, and diagnosing how far the tunnelling
trajectory deviates from the straight-line path between the two vacua.

Second, and more conservatively, the Fourier path can be used as a
preconditioner for existing bounce solvers. Many multi-field tunnelling
algorithms begin from an initial path connecting the false and true vacua and
then deform the path, solve an effective one-dimensional tunnelling problem
along it, or use supplied path points to construct a polygonal bounce
configuration. A straight-line path is universal, but it can be a poor initial
guess when the potential contains a curved valley or when the tunnelling
trajectory avoids a high barrier in field space.
Low Fourier modes provide a simple way to improve this initial guess. Since the
sine modes vanish at the endpoints, the false and true vacua are preserved
exactly. The lowest modes then allow the path to bend smoothly over the full
interval $t\in[0,1]$, capturing the large-scale geometric correction to the
straight path without introducing highly oscillatory deformations. In the
preconditioning interpretation, the optimised Fourier path is not claimed to be
the final bounce solution. It is an inexpensive, endpoint-preserving,
action-informed path that is passed to a more complete solver.

This is the sense in which the Fourier path acts as a low-mode preconditioner:
It does not change the physical tunnelling problem or replace the final solver,
but it supplies a better starting geometry. The benchmarks in
Sec.~\ref{sec:solver_initialization} test this interpretation directly for
CosmoTransitions and FindBounce.
\section{Standalone Fourier benchmarks}
\label{sec:benchmarks}

We first assess the Fourier-deformed path ansatz as a standalone
finite-dimensional variational method. The benchmarks serve two purposes. The
two-field example provides a direct visualisation of the optimised trajectory
and tests whether the endpoint-safe Fourier basis follows the expected
low-potential valley. The higher-dimensional benchmarks quantify the agreement
of the resulting action estimates with established tunnelling codes on standard
test potentials.

The results in this section should be interpreted as tests of the Fourier path
ansatz together with the fixed-$V_t$ reduced functional introduced in
Sec.~\ref{sec:reduced_action_ansatz}. In Sec.~\ref{sec:solver_initialization},
we use the same Fourier paths in a more conservative role, as initial
conditions or path information for existing bounce solvers.
\subsection{Two-field path visualization}
\label{subsec:two_field_visualization}

\begin{figure}[t]
    \centering
    \begin{subfigure}[b]{0.48\textwidth}
        \centering
        \includegraphics[width=\linewidth]{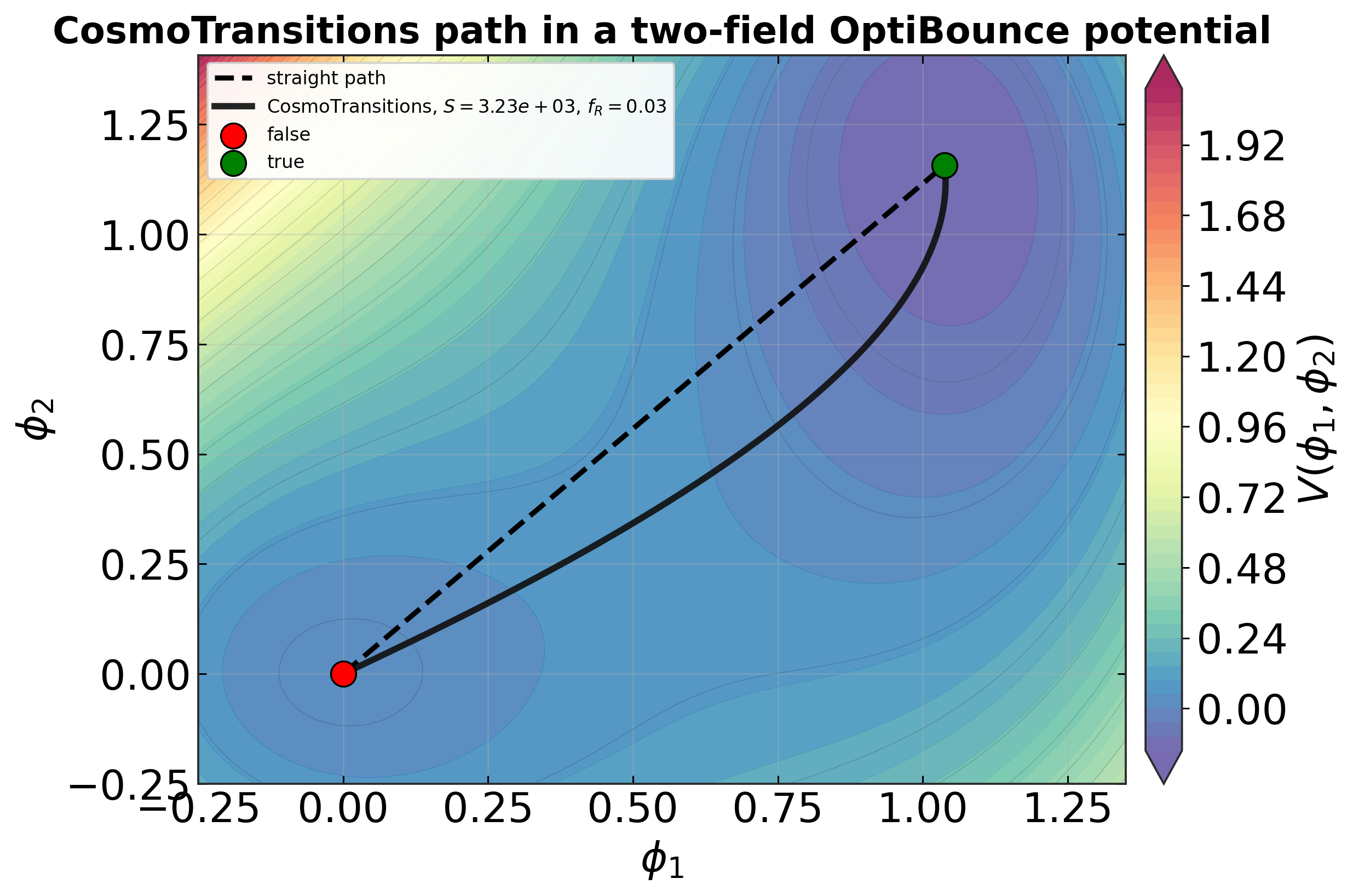}
        \caption{Fourier-deformed path}
        \label{fig:two_field_fourier_path}
    \end{subfigure}
    \hfill
    \begin{subfigure}[b]{0.48\textwidth}
        \centering
        \includegraphics[width=\linewidth]{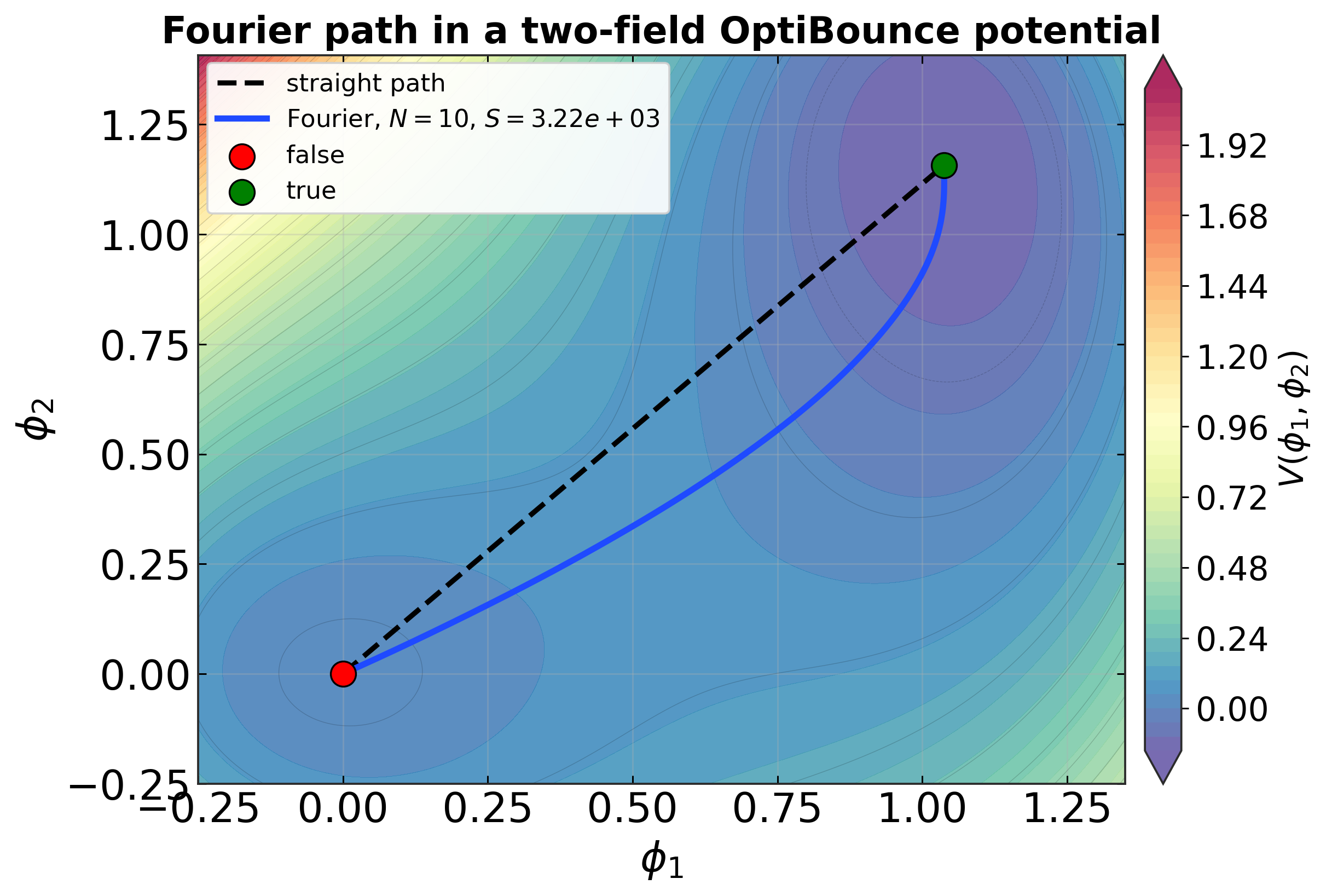}
        \caption{CosmoTransitions path}
        \label{fig:two_field_ct_path}
    \end{subfigure}
    \caption{
    Two-field visualisation for the potential in
    Eq.~\eqref{eq:two_field_optibounce_like}. The dashed black line denotes the
    straight path between the false and true vacua. The Fourier-deformed path
    and the CosmoTransitions path both bend through the same low-potential
    valley, showing that the endpoint-safe Fourier ansatz captures the relevant
    field-space trajectory.
    }
    \label{fig:two_field_fourier_vs_ct}
\end{figure}

We begin with a two-field example, where the tunnelling path can be visualised
directly. We use the two-field member of the OptiBounce-type potential family \cite{BARDSLEY},
\begin{equation}
    V(\phi_1,\phi_2)
    =
    \left[
        c_1(\phi_1-1)^2
        +
        c_2(\phi_2-1)^2
        -
        \delta
    \right]
    \left(
        \phi_1^2+\phi_2^2
    \right),
    \label{eq:two_field_optibounce_like}
\end{equation}
with
\begin{equation}
    c_1=0.684373,
    \qquad
    c_2=0.181928,
    \qquad
    \delta=0.065 .
    \label{eq:two_field_optibounce_params}
\end{equation}
The origin is an exact stationary false vacuum because of the overall factor
$\phi_1^2+\phi_2^2$. The lower-energy true vacuum is found numerically. This
example is not used as a precision benchmark for the action; rather, it is used
as a transparent two-dimensional visualisation of the field-space trajectory.
Figure~\ref{fig:two_field_fourier_vs_ct} compares the optimized Fourier path
with the path obtained using CosmoTransitions. In both panels, the dashed black
line denotes the straight-line interpolation between the false and true vacua.
The Fourier-deformed path bends away from this line and follows the
lower-potential valley. The CosmoTransitions path follows the same valley,
providing an independent geometric check that the Fourier optimisation is
selecting the physically relevant tunnelling direction.

Figure~\ref{fig:two_field_fourier_vs_ct} provides an independent geometric
cross-check of the optimised path. CosmoTransitions obtains the trajectory by
iteratively deforming a path and solving the effective one-dimensional bounce
problem along it, while the Fourier calculation optimises the fixed-$V_t$
reduced functional over a finite set of endpoint-safe basis coefficients. The
agreement of the two trajectories shows that both constructions select the
same low-potential tunnelling valley. Action-level comparisons are discussed
separately in the benchmark tables below, where the numerical procedures and
reference values can be compared directly.

\begin{figure}[t]
    \centering
    \includegraphics[width=0.7\textwidth]{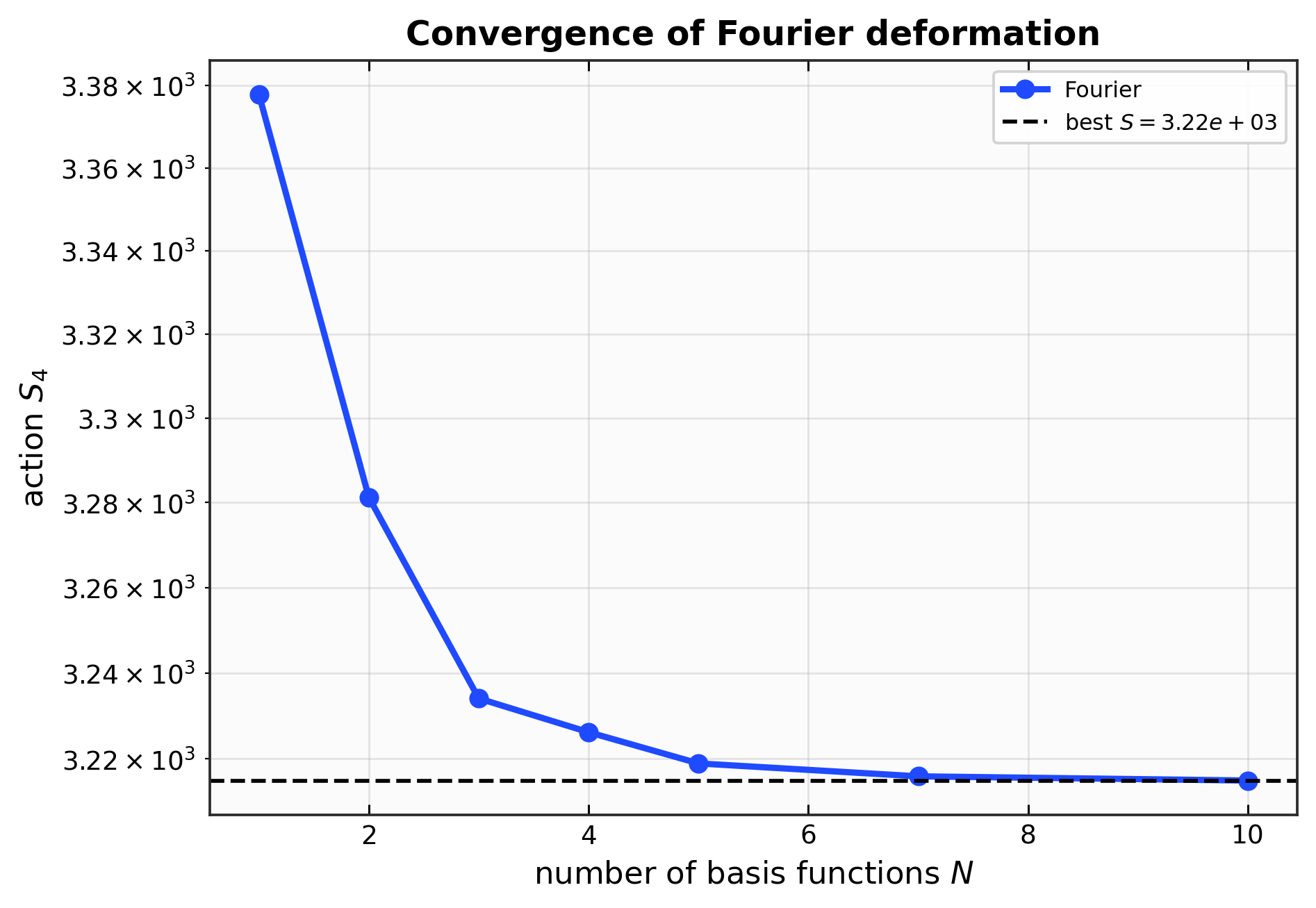}
    \caption{
    Convergence of the fixed-$V_t$ reduced action with the number of Fourier
    modes for the two-field potential in
    Eq.~\eqref{eq:two_field_optibounce_like}. The dashed horizontal line shows
    the lowest value of action value reached in the scan in all algorithms. The rapid saturation indicates that a small number of Fourier modes is sufficient to capture the dominant smooth
    bending of the path in this example.
    }
    \label{fig:fourier_mode_convergence}
\end{figure}

We also study the convergence of the reduced action as the number of Fourier
modes are increased. Figure~\ref{fig:fourier_mode_convergence} shows the
corresponding mode-convergence diagnostic for the same two-field potential.
The reduced action decreases rapidly with the first few Fourier modes and then
saturates, indicating that the dominant correction to the straight path is
captured by a small number of smooth, low-frequency deformations. The lowest value of action obtained from other algorithms is $3.22 \times 10^3$.
This two-field example gives the basic intuition behind the method. The
straight path crosses a higher-potential region, while the optimised
endpoint-safe path bends through a lower-potential valley without moving the
vacua. Quantitative action-level validation is performed in the next subsection
using the benchmark potential of Ref.~\cite{BARDSLEY}.

\subsection{Multi-field benchmark potential in d = 3}
\label{subsec:optibounce_validation}

We next validate the Fourier ansatz on a standard multi-field benchmark
potential used in the OptiBounce study~\cite{BARDSLEY}. The potential
is a nested family of $N_\phi$-field models,
\begin{equation}
    V_{N_\phi}(\bm{\phi})
    =
    \left[
        \sum_{i=1}^{N_\phi}
        c_i(\phi_i-1)^2
        -
        \delta
    \right]
    \left[
        \sum_{i=1}^{N_\phi}
        \phi_i^2
    \right],
    \label{eq:optibounce_potential}
\end{equation}
with the coefficients $c_i$ and $\delta$ chosen as in
Ref.~\cite{BARDSLEY}. Since the reference table reports the $d=3$
action, we evaluate the same $d=3$ reduced action for this comparison.
Table~\ref{tab:optibounce_validation} compares our adaptive JAX--Fourier result
with the OptiBounce, FindBounce, and CosmoTransitions values all in d=3. Apart from the
marked $N_\phi=4$ entry, the agreement with FindBounce is at the sub-percent
level over the tested range (the relative difference is below $0.1\%$ in several
cases and at most about $0.6\%$). The $N_\phi=4$ entry is
shown separately because it corresponds to a different local branch of the
finite-mode optimisation, and is not included in the quoted accuracy summary.

\begin{table}[h]
    \centering
    \caption{
Comparison on the benchmark potential of Eq.~\eqref{eq:optibounce_potential}
for the $d=3$ action. The reference columns show the OptiBounce (OB), FindBounce
(FB), and CosmoTransitions (CT) values from Ref.~\cite{BARDSLEY}; these
abbreviations are used as subscripts throughout the tables that follow.
$\Delta_{\rm FB}$ denotes the relative difference between the JAX--Fourier
result and the FindBounce value. The adaptive Fourier scan used a tolerance of
$10^{-2}$ and patience $3$. The timing is split into JAX setup/compilation time
$t_{\rm comp}$, coefficient-optimization time $t_{\rm sol}$, and total time
$t_{\rm FD}=t_{\rm comp}+t_{\rm sol}$. The CosmoTransitions time $t_{\rm CT}$
is available up to $N_\phi=10$. The dagger marks the $N_\phi=4$ row, where the
default local Fourier scan selects a higher-action branch.
}
    \label{tab:optibounce_validation}
    \resizebox{\textwidth}{!}{
    \begin{tabular}{cccccccccccc}
        \toprule
        $N_\phi$
        & $S_{\rm FD}$
        & $S_{\rm OB}$
        & $S_{\rm FB}$
        & $S_{\rm CT}$
        & $\Delta_{\rm FB}$ [\%]
        & $t_{\rm comp}$ [s]
        & $t_{\rm sol}$ [s]
        & $t_{\rm FD}$ [s]
        & $t_{\rm OB}$ [s]
        & $t_{\rm FB}$ [s]
        & $t_{\rm CT}$ [s] \\
        \midrule
        3  & 240.399 & 240.049 & 240.403 & 240.324 & 0.002 & 0.654 & 0.204 & 0.859 & 0.200 & 0.294 & 0.492 \\
        4  & $251.643^\dagger$ & 227.023 & 230.864 & 230.397 & 9.000 & 0.875 & 0.274 & 1.149 & 0.298 & 0.615 & 3.223 \\
        5  & 234.046 & 233.523 & 233.716 & 233.357 & 0.141 & 0.787 & 0.333 & 1.121 & 0.284 & 0.408 & 0.976 \\
        6  & 271.172 & 270.363 & 270.578 & 271.769 & 0.219 & 0.730 & 0.331 & 1.061 & 0.331 & 0.442 & 3.917 \\
        7  & 250.899 & 250.054 & 250.222 & 249.845 & 0.271 & 0.706 & 0.367 & 1.073 & 0.401 & 0.482 & 3.802 \\
        8  & 269.034 & 268.259 & 268.486 & 269.368 & 0.204 & 0.777 & 0.403 & 1.180 & 0.489 & 0.606 & 4.143 \\
        9  & 205.527 & 204.609 & 204.796 & 205.888 & 0.357 & 0.659 & 0.351 & 1.010 & 0.535 & 0.675 & 0.769 \\
        10 & 262.141 & 261.468 & 261.703 & 261.273 & 0.167 & 0.823 & 0.532 & 1.356 & 0.595 & 0.779 & 1.279 \\
        11 & 274.221 & 273.271 & 273.564 & --      & 0.240 & 0.604 & 0.370 & 0.974 & 0.681 & 0.861 & -- \\
        12 & 250.156 & 249.691 & 249.928 & --      & 0.091 & 1.016 & 0.621 & 1.637 & 0.776 & 0.950 & -- \\
        13 & 294.605 & 293.383 & 293.653 & --      & 0.324 & 0.565 & 0.400 & 0.965 & 0.824 & 1.012 & -- \\
        14 & 295.622 & 294.677 & 294.877 & --      & 0.252 & 0.781 & 0.520 & 1.302 & 1.259 & 1.114 & -- \\
        15 & 314.188 & 312.760 & 313.039 & --      & 0.367 & 0.706 & 0.330 & 1.036 & 1.271 & 1.222 & -- \\
        16 & 368.576 & 366.870 & 366.978 & --      & 0.436 & 0.774 & 0.511 & 1.285 & 1.352 & 1.885 & -- \\
        17 & 343.688 & 342.537 & 342.893 & --      & 0.232 & 0.634 & 0.331 & 0.965 & 1.695 & 1.395 & -- \\
        18 & 392.943 & 390.925 & 391.231 & --      & 0.438 & 0.654 & 0.360 & 1.013 & 1.832 & 1.544 & -- \\
        19 & 341.539 & 339.287 & 339.467 & --      & 0.610 & 0.620 & 0.387 & 1.006 & 2.653 & 2.436 & -- \\
        20 & 383.137 & 381.886 & 382.153 & --      & 0.257 & 0.811 & 0.477 & 1.288 & 2.487 & 1.864 & -- \\
        \bottomrule
    \end{tabular}
    }
\end{table}

\begin{figure}[t]
    \centering
    \begin{subfigure}[b]{0.48\textwidth}
        \centering
        \includegraphics[width=\linewidth]{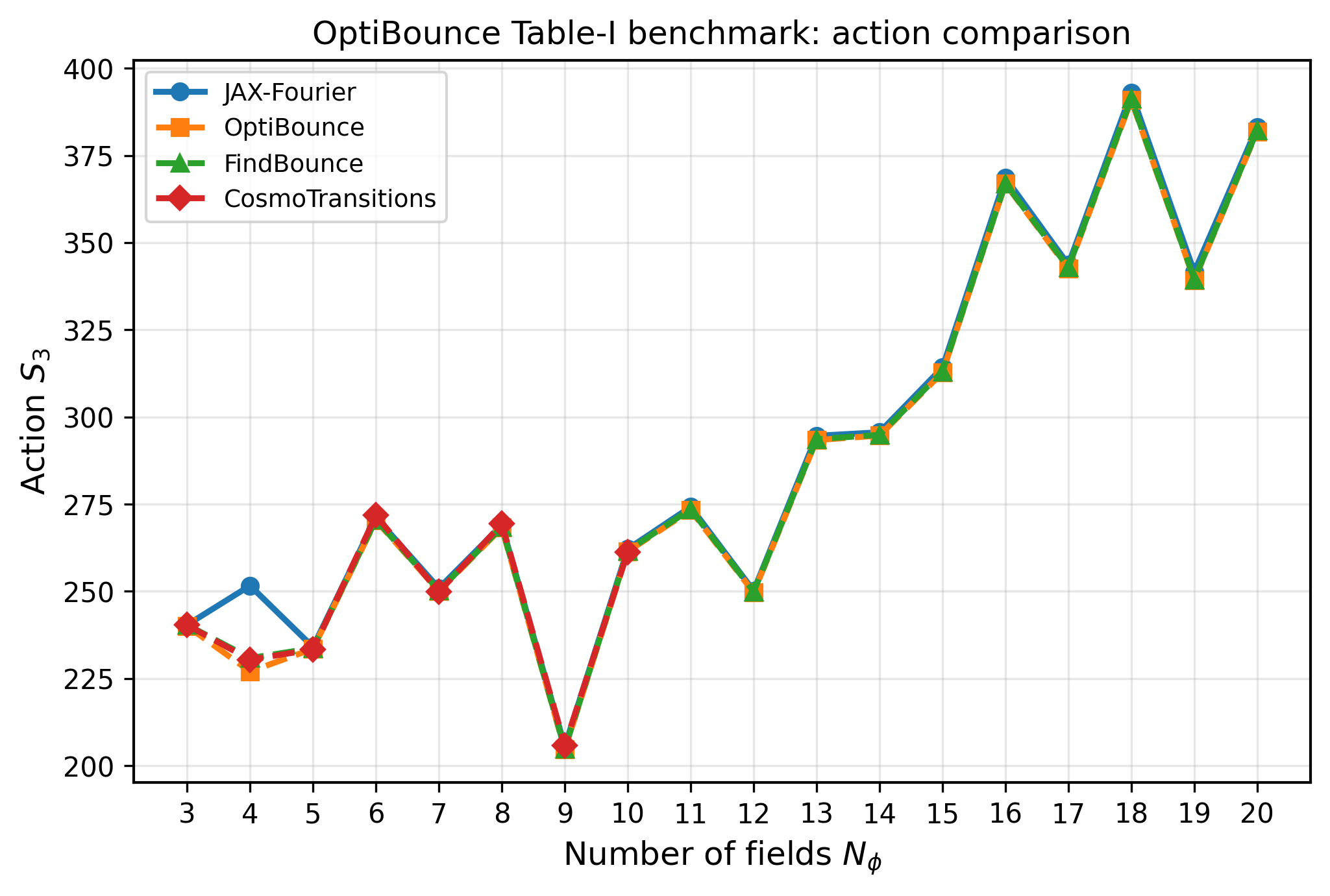}
        \caption{Action comparison}
        \label{fig:optibounce_d3_action}
    \end{subfigure}
    \hfill
    \begin{subfigure}[b]{0.48\textwidth}
        \centering
        \includegraphics[width=\linewidth]{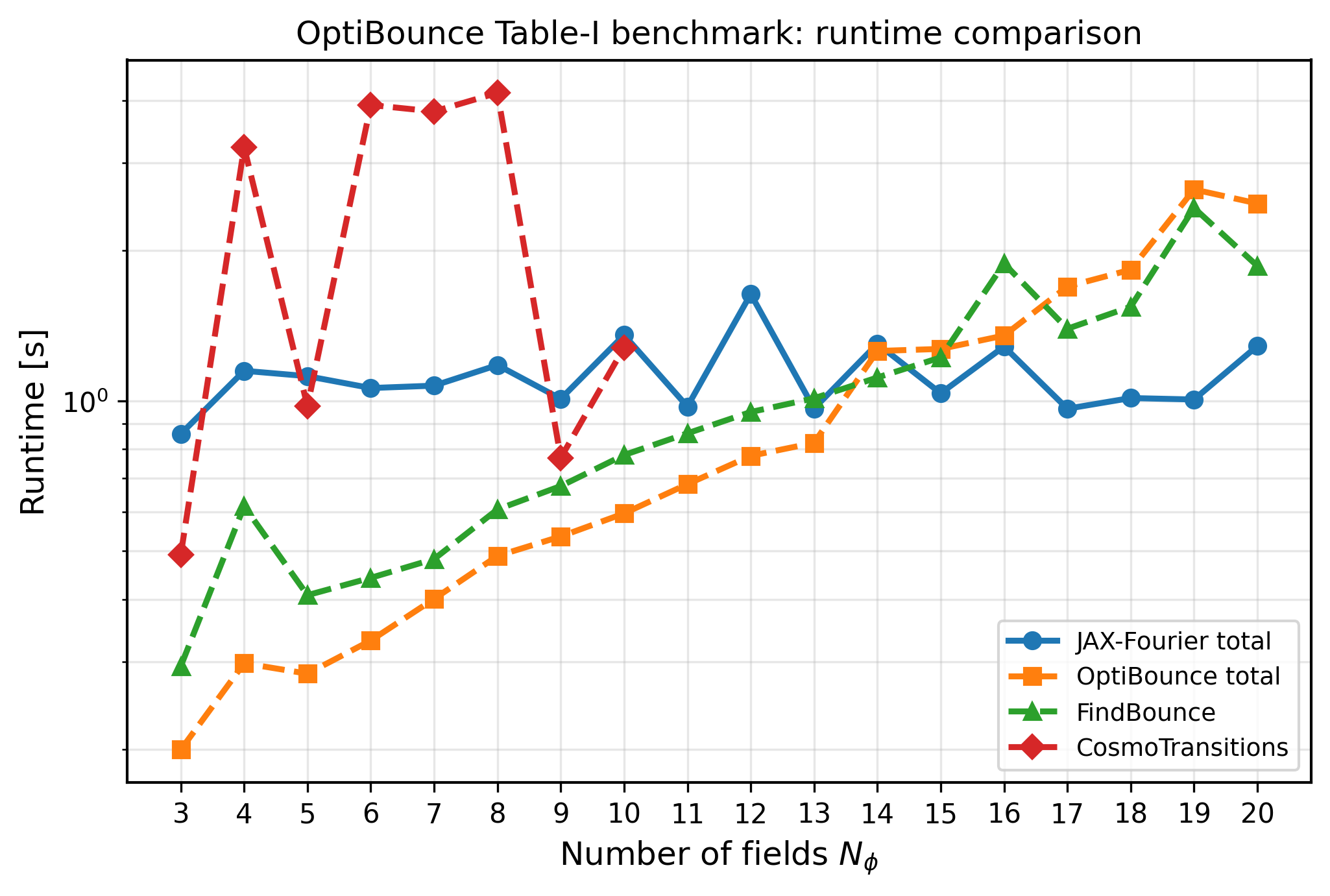}
        \caption{Runtime comparison}
        \label{fig:optibounce_d3_time}
    \end{subfigure}
    \caption{
Benchmark comparison for the $d=3$ action using the multi-field potential of
Ref.~\cite{BARDSLEY}. The JAX--Fourier scan is compared with the
OptiBounce, FindBounce, and CosmoTransitions reference actions. Apart from the
marked $N_\phi=4$ entry, the actions agree at the sub-percent level across the
tested range. The runtime panel shows the total cost of the adaptive Fourier
scan, including compilation and optimisation time.
}
    \label{fig:optibounce_d3_action_time}
\end{figure}

The Fourier scan was performed with an adaptive stopping criterion. In this run,
we used a tolerance of $10^{-2}$ and a patience value of $3$, meaning that the
scan was stopped when increasing the number of Fourier modes did not improve
the action by more than the chosen tolerance for three consecutive mode steps. Figure \ref{fig:optibounce_d3_action_time} shows the actions obtained (left) and the time taken (right) by the Fourier deformation and other algorithms for comparison.
The timing numbers in the right panel should therefore be interpreted as the
cost of the full adaptive scan, not as a fixed-mode evaluation time. Changing
the tolerance, patience, or maximum number of allowed modes can change the
runtime, while the action remains stable once sufficiently many modes have been
included.
\subsection{High-dimensional random benchmark in d = 4}
\label{subsec:high_dimensional_benchmark}

We next test the solver on a nested random-coefficient potential family with
field dimension up to $N_\phi=50$. This benchmark probes whether the
endpoint-safe Fourier ansatz remains accurate and numerically stable beyond the
range of the standard OptiBounce comparison. We use the same functional form as
Eq.~\eqref{eq:optibounce_potential},
\begin{equation}
    V_{N_\phi}(\bm{\phi})
    =
    \left[
        \sum_{i=1}^{N_\phi}
        c_i(\phi_i-1)^2
        -
        \delta_{N_\phi}
    \right]
    \left[
        \sum_{i=1}^{N_\phi}
        \phi_i^2
    \right],
    \label{eq:random_high_dim_potential}
\end{equation}
but with a new fixed set of random coefficients. The $N_\phi=N$ potential uses
the first $N$ entries of a master coefficient list, so the benchmark is nested:
increasing $N_\phi$ adds one new scalar direction without changing the
coefficients of the previous fields. For reproducibility, the values of
$\delta_{N_\phi}$ and the master list of $c_i$ coefficients are given in
Appendix~\ref{app:benchmark_coefficients}.
The false vacuum is fixed at
\begin{equation}
    \bm{\phi}_F=\bm{0},
\end{equation}
and the true vacuum is obtained numerically by minimising the potential from an
initial point near $\bm{\phi}=\bm{1}$. The comparison is performed for the
zero-temperature $d=4$ action.
Table~\ref{tab:high_dim_selected} shows selected results from the scan up to
$N_\phi=50$. The JAX--Fourier solver uses the adaptive mode-selection criterion
described in Sec.~\ref{app:jax_implementation}; it does not use the
FindBounce or CosmoTransitions values as inputs or stopping targets.
CosmoTransitions results are included where the comparison is complete, while
FindBounce provides an independent reference for the higher-dimensional cases.
\begin{table}[t]
    \centering
    \caption{
Selected high-dimensional benchmark results for the random-coefficient
potential family in Eq.~\eqref{eq:random_high_dim_potential}. The
JAX--Fourier solver uses adaptive mode selection and does not use the
FindBounce or CosmoTransitions values as inputs. The relative difference is
defined as
$\Delta_{\rm FB}=100\,|S^{\rm JAX}_4-S^{\rm FB}_4|/S^{\rm FB}_4$.
CosmoTransitions results are shown for the dimensions where the comparison was 
completed.
}
    \label{tab:high_dim_selected}
    \resizebox{\textwidth}{!}{
    \begin{tabular}{cccccccccc}
        \toprule
        $N_\phi$
        & $S^{\rm FD}_4$
        & modes
        & $S^{\rm CT}_4$
        & $t_{\rm CT}$ [s]
        & $S^{\rm FB}_4$
        & $\Delta_{\rm FB}$ [\%]
        & $t_{\rm FD}$ [s]
        & $t_{\rm FB}$ [s]
        & $t_{\rm FB}/t_{\rm FD}$ \\
        \midrule
        2  & $1.8517\times10^{1}$ & 14 & $1.7974\times10^{1}$ & 0.027 & $1.8042\times10^{1}$ & 2.63  & 3.79 & 0.168 & 0.04 \\
        3  & $3.8919\times10^{1}$ & 12 & $3.9530\times10^{1}$ & 0.042 & $3.8726\times10^{1}$ & 0.50  & 2.44 & 0.090 & 0.04 \\
        4  & $1.2844\times10^{3}$ & 6  & $1.2822\times10^{3}$ & 0.130 & $1.2845\times10^{3}$ & 0.008 & 0.58 & 0.075 & 0.13 \\
        5  & $1.6424\times10^{2}$ & 10 & $1.6399\times10^{2}$ & 0.112 & $1.6425\times10^{2}$ & 0.002 & 1.20 & 0.136 & 0.11 \\
        10 & $1.1560\times10^{3}$ & 8  & $1.1549\times10^{3}$ & 0.152 & $1.1565\times10^{3}$ & 0.04  & 0.90 & 0.189 & 0.21 \\
        20 & $8.2930\times10^{3}$ & 10 & -- & -- & $8.2850\times10^{3}$ & 0.10 & 1.28 & 0.837 & 0.65 \\
        30 & $2.6269\times10^{4}$ & 8  & -- & -- & $2.6250\times10^{4}$ & 0.08 & 1.03 & 2.144 & 2.08 \\
        40 & $7.6212\times10^{4}$ & 8  & -- & -- & $7.6146\times10^{4}$ & 0.09 & 1.10 & 4.098 & 3.71 \\
        50 & $4.1902\times10^{4}$ & 9  & -- & -- & $4.1878\times10^{4}$ & 0.06 & 1.28 & 8.990 & 7.00 \\
        \bottomrule
    \end{tabular}
    }
\end{table}

\begin{figure}[t]
    \centering
    \begin{subfigure}[b]{0.48\textwidth}
        \centering
        \includegraphics[width=\linewidth]{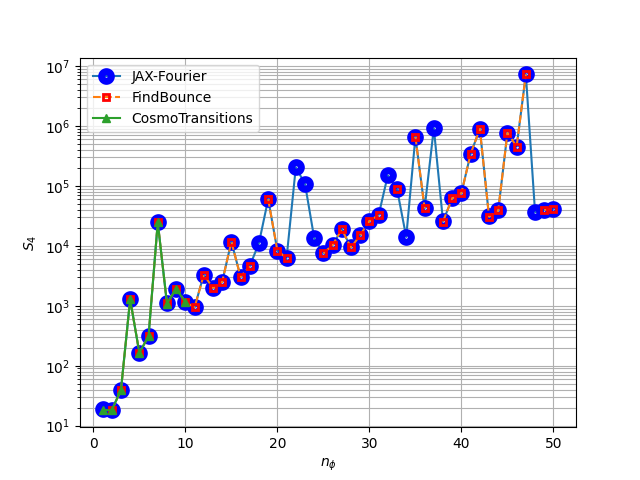}
        \caption{Action comparison}
        \label{fig:high_dim_action}
    \end{subfigure}
    \hfill
    \begin{subfigure}[b]{0.48\textwidth}
        \centering
        \includegraphics[width=\linewidth]{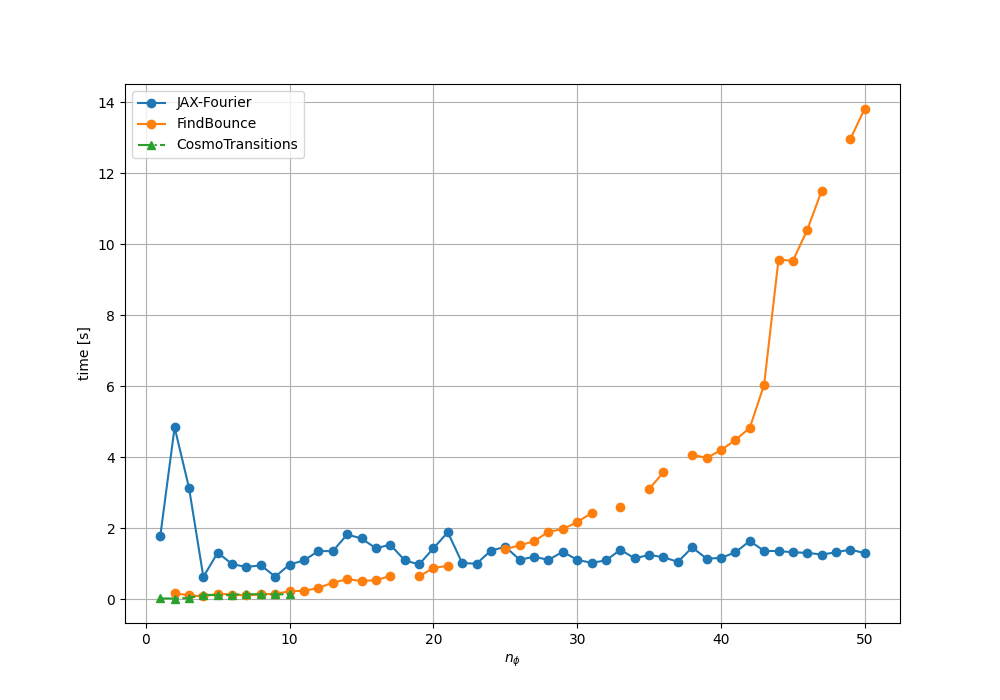}
        \caption{Runtime comparison}
        \label{fig:high_dim_time}
    \end{subfigure}
    \caption{
High-dimensional random-coefficient benchmark in $d=4$. Left: action
comparison between the JAX--Fourier result and the independent FindBounce
values where available. Right: corresponding wall-clock runtimes. The
JAX--Fourier scan remains at the second scale through $N_\phi=50$ for this
benchmark family. The timings are implementation-level wall times and should
not be interpreted as a universal scaling law.
}
    \label{fig:high_dim_action_time}
\end{figure}

The agreement with FindBounce improves rapidly beyond the lowest-dimensional
cases. The largest relative difference occurs at $N_\phi=2$, where the action
is small and a modest absolute shift appears as a larger percentage. From
$N_\phi=4$ onward, the agreement is below the percent level, and in the
representative high-dimensional cases, it is typically at the $10^{-3}$ level or
better. For example,
\begin{align}
    N_\phi=30:\qquad
    S^{\rm FD}_4 &= 2.6269\times10^4,
    &
    S^{\rm FB}_4 &= 2.6250\times10^4,
    \\
    N_\phi=40:\qquad
    S^{\rm FD}_4 &= 7.6212\times10^4,
    &
    S^{\rm FB}_4 &= 7.6146\times10^4,
    \\
    N_\phi=50:\qquad
    S^{\rm FD}_4 &= 4.1902\times10^4,
    &
    S^{\rm FB}_4 &= 4.1878\times10^4 .
\end{align}
At $N_\phi=50$ this corresponds to a relative difference of approximately
$5.7\times10^{-4}$. The Fourier scan uses only its internal no-improvement
stopping criterion, without reference to the FindBounce value.
For $N_\phi\leq10$, CosmoTransitions provides an additional check and agrees
with the JAX--Fourier result at the percent level or better, reaching
sub-percent agreement from $N_\phi=4$ onward. For a larger field dimension $>10$, we use
FindBounce as the independent reference comparison.
The selected mode count remains modest across the scan. In the representative
higher-dimensional cases, the adaptive scan chooses $N_m\simeq8$--$10$ modes.
This indicates that, for this smooth nested potential family, the dominant
tunnelling path deformation remains low-mode even when the ambient field space
is high-dimensional.
Figure~\ref{fig:high_dim_action_time} summarises the same comparison. The
action panel shows that the JAX--Fourier result tracks the successful
FindBounce results across the scan. The timing panel shows that the
JAX--Fourier runtime remains at the second scale through $N_\phi=50$, while the
independent FindBounce pipeline becomes increasingly expensive in the completed
runs.

As before, the timing comparison should be interpreted as an implementation-level
comparison for this benchmark, rather than as a universal scaling statement.
The absolute runtimes depend on the preprocessing steps, solver settings, and
details of the potential family. The main point is that the endpoint-safe
JAX--Fourier solver completes the full $N_\phi=1,\ldots,50$ scan with an
intrinsic stopping criterion and agrees with FindBounce at a sub-percent level
wherever the independent comparison is available.
\subsection{Basis comparison and mode convergence}
\label{subsec:basis_comparison}

The Fourier sine basis is a natural default for the path deformation because it
is smooth, global, and endpoint-safe. However, the Fourier basis is not unique.
The same endpoint-preserving construction can be implemented with polynomial,
local, hybrid, or wavelet-inspired basis functions. We therefore compare several
families of endpoint-safe deformations:
\begin{equation}
    \phi_i(t)
    =
    \phi_{i,F}
    +
    t(\phi_{i,T}-\phi_{i,F})
    +
    \sum_{a=1}^{N_{\rm basis}}
    c_{ia} B_a(t),
    \qquad
    B_a(0)=B_a(1)=0 .
    \label{eq:general_endpoint_safe_basis}
\end{equation}
The scan includes Fourier sine modes, Chebyshev and Legendre polynomials
expansions with endpoint envelopes, local B-splines, local Gaussian bumps,
hybrid Fourier-local bases, and shifted wavelet-inspired bases. The aim is not
to identify a universally optimal basis. Rather, the preferred representation
can depend on the geometry of the tunnelling trajectory. Smooth global
deformations are naturally captured by a small number of Fourier modes, whereas
localised turns may benefit from local or hybrid functions. The comparison
should therefore be viewed as a diagnostic of how different endpoint-safe
parametrisations perform on the benchmark potentials considered here.

Table~\ref{tab:selected_basis_comparison} gives a compact representative
comparison on the OptiBounce benchmark. For each basis, the table reports the
adaptively selected basis size, the best action reached by the scan, and the
relative difference from the FindBounce reference action. The full basis-scan
plots, including action, runtime, selected mode count, and additional wavelet
variants, are collected in Appendix~\ref{app:additional_plots}.

\begin{table}[t]
    \centering
    \caption{
Selected comparison of endpoint-safe basis families on the benchmark potential
of Ref.~\cite{BARDSLEY}. The table reports the adaptively selected
basis size, the best action, and the relative difference with respect to the
FindBounce reference. Fourier sine modes give stable sub-percent agreement with
a small number of modes. Local B-spline and hybrid Fourier-local bases are
competitive and provide useful cross-checks. The shifted wavelet-inspired basis
tested here converges quickly but gives larger actions for these smooth
benchmark paths.
}
    \label{tab:selected_basis_comparison}
   
    \begin{tabular}{clccc}
        \toprule
        $N_\phi$ & Basis & $N_{\rm basis}$ & $S_{\rm best}$ & $\Delta_{\rm FB}$ [\%] \\
        \midrule
        3  & Fourier             & 8  & 240.229 & 0.072 \\
        3  & Chebyshev           & 17 & 244.267 & 1.607 \\
        3  & B-spline local      & 6  & 240.142 & 0.109 \\
        3  & Fourier+B-spline    & 8  & 240.142 & 0.109 \\
        3  & Wavelet db2         & 3  & 290.711 & 20.927 \\
        \midrule
        10 & Fourier             & 8  & 261.807 & 0.040 \\
        10 & Chebyshev           & 11 & 269.319 & 2.910 \\
        10 & B-spline local      & 8  & 261.611 & 0.035 \\
        10 & Fourier+B-spline    & 8  & 261.648 & 0.021 \\
        10 & Wavelet db2         & 3  & 319.302 & 22.009 \\
        \midrule
        20 & Fourier             & 8  & 382.502 & 0.091 \\
        20 & Chebyshev           & 12 & 390.848 & 2.275 \\
        20 & B-spline local      & 8  & 382.107 & 0.012 \\
        20 & Fourier+Gaussian    & 10 & 382.141 & 0.003 \\
        20 & Fourier+B-spline    & 8  & 382.186 & 0.009 \\
        20 & Wavelet db2         & 3  & 448.204 & 17.284 \\
        \bottomrule
    \end{tabular}
\end{table}

The basis comparison shows that Fourier sine modes provide a reliable default
for the smooth benchmark paths considered here. In the representative cases in
Table~\ref{tab:selected_basis_comparison}, the Fourier basis reaches
sub-percent agreement with the FindBounce reference using only a small number
of modes. This supports the main working assumption of the method: the dominant
correction to the straight-line path is often a smooth, low-frequency
deformation.
Local smooth bases provide useful cross-checks. The B-spline basis gives
accuracy comparable to Fourier, and hybrid Fourier-local bases also perform
well. For example, at $N_\phi=20$ the Fourier+Gaussian and Fourier+B-spline
bases give relative differences of $0.003\%$ and $0.009\%$, respectively. This
suggests that the essential requirement is not the Fourier basis itself, but a
smooth, well-conditioned, endpoint-safe parametrisation of the field-space
trajectory.
The polynomial and wavelet-inspired bases tested here are less efficient on
these smooth examples. Chebyshev and Legendre modes, made endpoint-safe by an
overall envelope, give percent-level deviations at comparable or larger basis
sizes. The shifted wavelet-inspired ansatz converges quickly, but to larger
actions for these benchmarks. This should be interpreted as a statement about
the particular endpoint-safe wavelet construction tested here, not as a general
limitation of multiscale bases. More adaptive local or multiresolution bases
may be useful for potentials with sharp turns or multi-channel structure.

In the remainder of the paper, we focus on the Fourier basis because it gives
stable convergence, has simple endpoint behaviour, and is straightforward to use
as an initialiser for external tunnelling solvers.

\section{Fourier preconditioning of existing solvers}
\label{sec:solver_initialization}

The previous section treated the Fourier ansatz as a standalone
finite-dimensional path optimiser. We now use the same construction in a more
conservative role, as a preconditioner for established tunnelling solvers. In
this setup, the Fourier calculation is not the final bounce calculation.
Instead, it supplies a smooth endpoint-preserving path that is closer to the
tunnelling valley than the straight-line interpolation.
This section tests whether such Fourier preconditioning reduces the numerical
work required by existing solvers while leaving the final action stable.
\subsection{CosmoTransitions initialization}
\label{subsec:ct_initialization}
CosmoTransitions computes multi-field tunnelling paths using an iterative
path-deformation algorithm. Starting from an initial path between the false and
true vacua, the code solves an effective one-dimensional tunnelling problem along
the path and then deforms the path using the transverse force. The quality of
the initial path can therefore affect the number of deformation steps, the
residual force ratio, and the total runtime.
We test whether a low-mode Fourier path provides a better initial condition
than the straight-line interpolation. The Fourier-preconditioned initialisation
uses a single Fourier mode, $m=1$, to generate a smooth endpoint-preserving
curved path before calling the standard CosmoTransitions path-deformation
routine. The final bounce path and action are still obtained by
CosmoTransitions.

We use a controlled spectator-field embedding of a two-field tunnelling problem.
The purpose of this benchmark is to isolate the effect of the initial path
geometry on the subsequent CosmoTransitions deformation. The first two fields
$(h,s)$ define the curved tunnelling valley through
\begin{equation}
    V_2(h,s)
    =
    \frac{1}{4}(h^2-1)^2
    +
    \frac{1}{4}(s^2-1)^2
    +
    \lambda h^2 s^2
    -
    \epsilon h ,
    \qquad
    \lambda=0.3,\quad \epsilon=0.1 .
    \label{eq:ct_base_potential}
\end{equation}
For $N_\phi>2$, the additional fields are introduced as spectators,
\begin{equation}
    V_{N_\phi}(\bm{\phi})
    =
    V_2(h,s)
    +
    \frac{1}{2}m_{\rm spec}^2
    \sum_{k=1}^{N_\phi-2}
    \left[
        z_k - c_k B(h,s)
    \right]^2 ,
    \label{eq:ct_spectator_potential}
\end{equation}
with
\begin{equation}
    B(h,s)
    =
    (h-h_T)(h-h_F)s .
    \label{eq:ct_spectator_B}
\end{equation}
Here $(h_T,s_T)$ and $(h_F,s_F)$ are the true- and false-vacuum values of the
two-field subsystem. In the numerical scan we take $m_{\rm spec}=1$ and choose
alternating spectator coefficients scaled as $|c_k|\propto 1/\sqrt{N_\phi-2}$,
which keeps the overall spectator-sector strength approximately fixed as
$N_\phi$ is increased. By construction, increasing $N_\phi$ embeds the same
curved two-field tunnelling valley into a larger field space, so the benchmark
isolates the effect of the initial path geometry on the subsequent
CosmoTransitions deformation.
This benchmark is not intended to represent a generic random high-dimensional
potential. Rather, it provides a controlled test of whether a Fourier-curved
initial path reduces the deformation work required by CosmoTransitions when the
same curved two-field tunnelling valley is embedded in larger field spaces. The
near-constant reduction in deformation steps seen below should be interpreted in
this light.
Table~\ref{tab:ct_initialization} compares the straight-line and
Fourier-preconditioned initialisations for representative field dimensions. The
reported values are medians over five repeated runs, including the preprocessing
time required to construct the Fourier path.

\begin{table}[t]
    \centering
    \caption{
    Comparison of CosmoTransitions performance for straight-line and
    Fourier-preconditioned initial paths in the curved-valley $N_\phi$-field
    benchmark. The reported values are medians over five repeated runs. The
    Fourier-preconditioned path uses one Fourier mode, $m=1$, and the total
    time includes the Fourier preprocessing step.
    }
    \label{tab:ct_initialization}
    \begin{tabular}{cccccc}
        \toprule
        $N$
        & Initial path
        & $S_E^{\rm final}$
        & final $f_{\rm ratio}$
        & Deformation steps
        & Total time [s] \\
        \midrule
        5  & Straight line  & 16.1590 & $3.918\times10^{-2}$ & 51 & 0.1744 \\
        5  & Fourier, $m=1$ & 16.1405 & $1.962\times10^{-2}$ & 16 & 0.1357 \\
        \midrule
        10 & Straight line  & 16.1592 & $3.917\times10^{-2}$ & 51 & 0.1834 \\
        10 & Fourier, $m=1$ & 16.1405 & $1.962\times10^{-2}$ & 16 & 0.1540 \\
        \bottomrule
    \end{tabular}
\end{table}

\begin{figure}[t]
    \centering
    \begin{subfigure}[b]{0.48\textwidth}
        \centering
        \includegraphics[width=\linewidth]{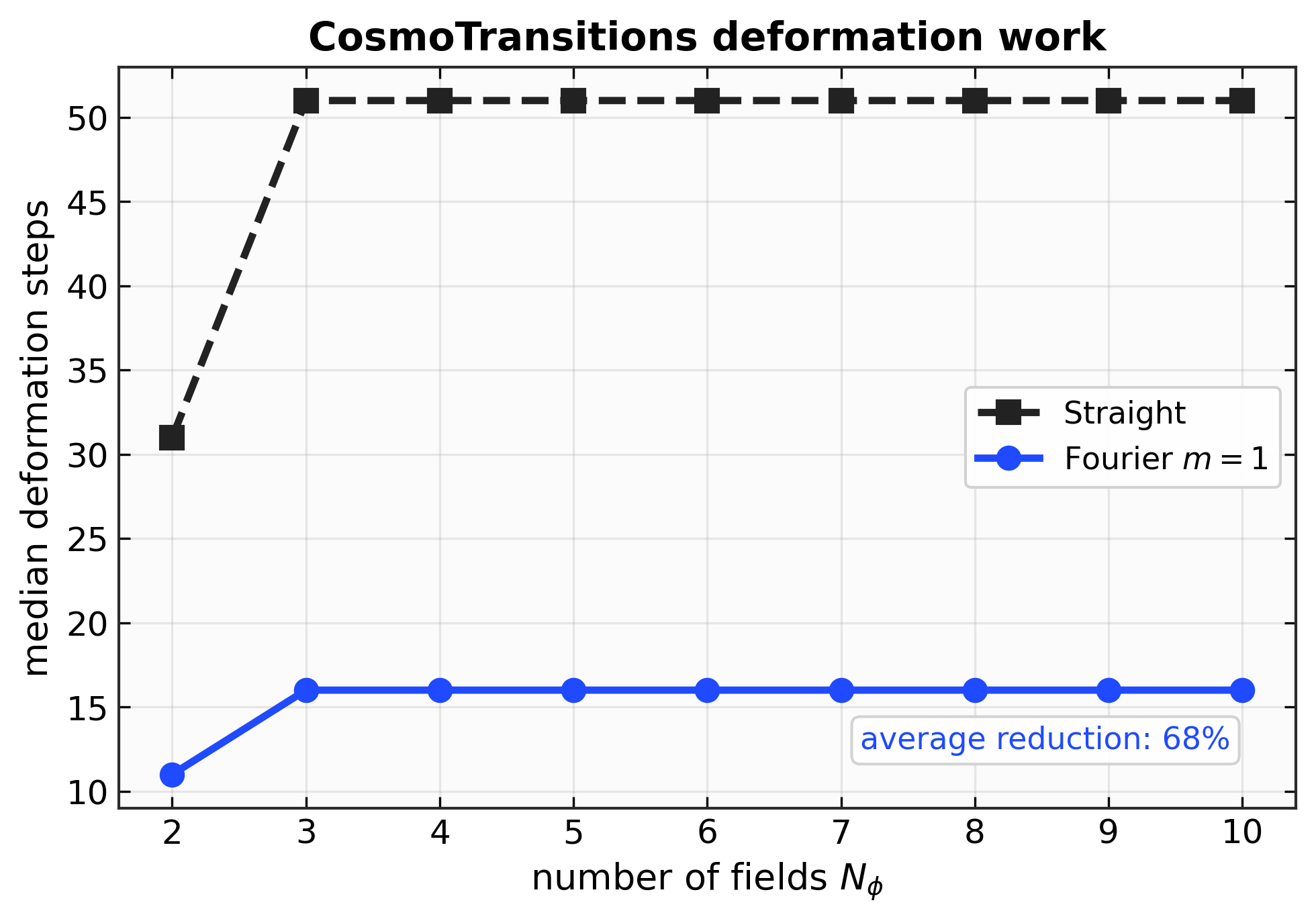}
        \caption{Deformation steps}
        \label{fig:ct_preconditioning_steps}
    \end{subfigure}
    \hfill
    \begin{subfigure}[b]{0.48\textwidth}
        \centering
        \includegraphics[width=\linewidth]{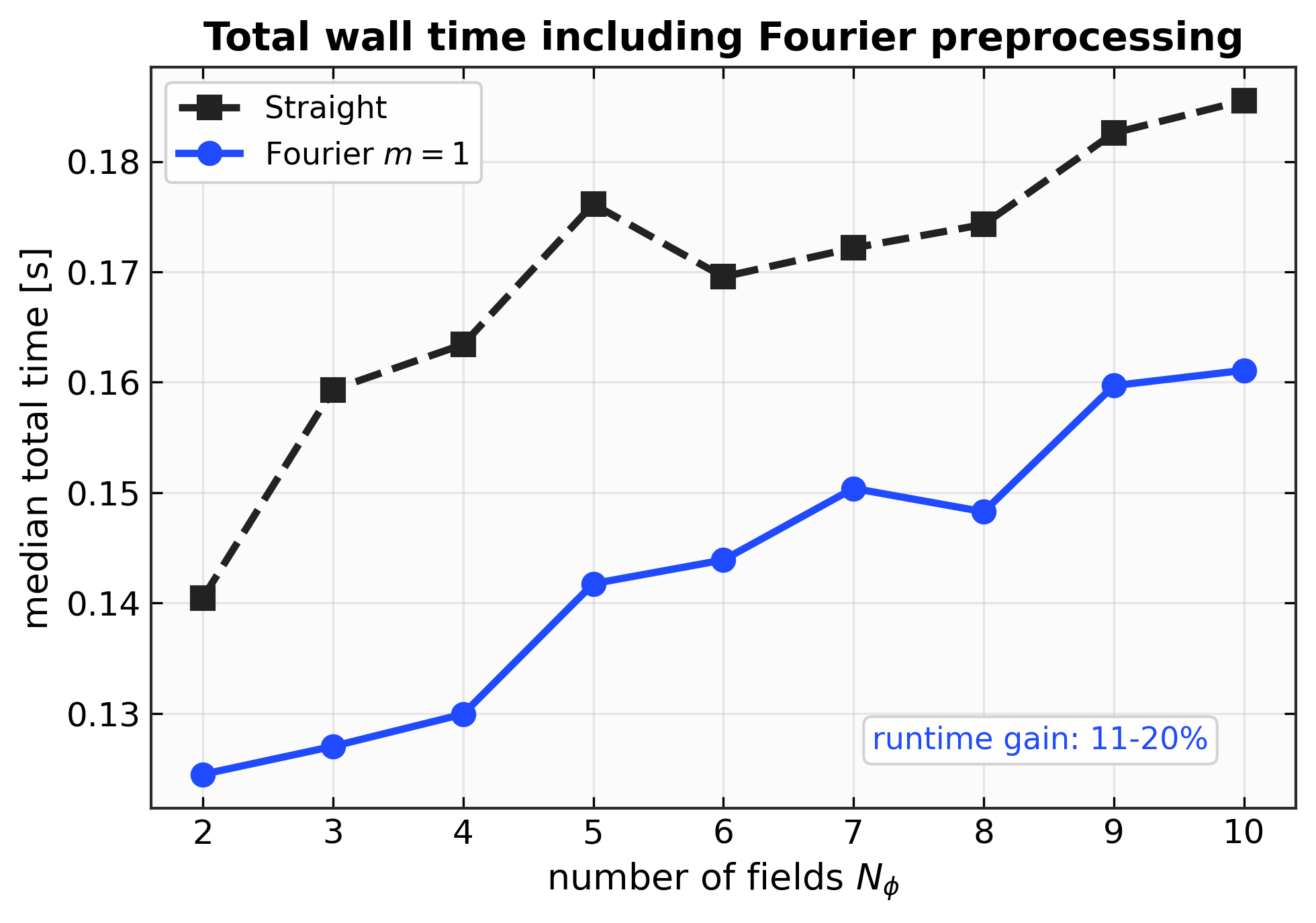}
        \caption{Total wall time}
        \label{fig:ct_preconditioning_time}
    \end{subfigure}
    \caption{
    Effect of Fourier preconditioning on the CosmoTransitions path-deformation
    run for the spectator-field benchmark in
    Eq.~\eqref{eq:ct_spectator_potential}. The straight-line initialisation is
    compared with a one-mode Fourier initialisation, $m=1$. Left: median number
    of path-deformation steps. Right: median total wall time, including the cost
    of constructing the Fourier initial path. For $N_\phi\geq3$, the Fourier
    initialisation reduces the deformation steps from $51$ to $16$. The runtime
    improvement is more modest because the Fourier preprocessing time is
    included. The near-constant step reduction reflects the controlled
    spectator-field construction: increasing $N_\phi$ embeds the same curved
    two-field tunnelling valley into a larger field space with approximately
    fixed spectator-sector strength.
    }
    \label{fig:ct_preconditioning_steps_time}
\end{figure}
The reduction in deformation work and the corresponding total wall time are
shown graphically in Fig.~\ref{fig:ct_preconditioning_steps_time}, which
emphasises that the main gain is a reduction in the amount of subsequent path
deformation required by CosmoTransitions. For the spectator-field
benchmark with $N_\phi\geq3$, the Fourier initialisation reduces the median
number of CosmoTransitions deformation steps from $51$ to $16$, corresponding
to a reduction of approximately $68.6\%$. The near-constant step reduction is a
consequence of the controlled benchmark construction: increasing $N_\phi$
embeds the same curved two-field tunnelling valley into a larger field space with
approximately fixed spectator-sector strength.
The final residual force ratio also improves. In the representative
$N_\phi=5$ and $N_\phi=10$ cases shown in Table~\ref{tab:ct_initialization},
$f_{\rm ratio}$ decreases from about $3.9\times10^{-2}$ for the straight-line
initialization to about $2.0\times10^{-2}$ for the Fourier-preconditioned
initialization.

The wall-clock improvement is smaller than the reduction in deformation steps,
because the total time includes the Fourier preprocessing. Nevertheless, the
Fourier-initialised runs remain faster over the tested range, as shown in
Fig.~\ref{fig:ct_preconditioning_steps_time}. For the representative
$N_\phi=5$ and $N_\phi=10$ entries in Table~\ref{tab:ct_initialization}, the
runtime reduction is roughly $15$--$25\%$.
The final actions obtained from the two initialisations remain close. The Fourier path is not
intended to change the tunnelling solution, but to provide CosmoTransitions with
a better initial trajectory from which to begin path deformation. In this sense,
the Fourier ansatz acts as a low-mode preconditioner for the path deformation
algorithm. In the controlled curved-valley benchmark considered here, a single
smooth Fourier mode is already sufficient to move the initial path closer to
the tunnelling valley.


\subsection{Hybrid fixed-path evaluation with CosmoTransitions}
\label{subsec:hybrid_fixed_path_ct}

As a further diagnostic, we test a hybrid fixed-path workflow. In this setup,
the Fourier optimiser is used to determine a field-space trajectory, and
CosmoTransitions is then used only to solve the effective one-dimensional
bounce problem along that fixed path. This differs from the path-deformation
test in Sec.~\ref{subsec:ct_initialization}. There, the Fourier path is used
only as an initial condition, and CosmoTransitions subsequently deforms the path.
Here, instead, the Fourier path is held fixed.
This test separates two tasks: finding a good multi-field path, and solving the
one-dimensional radial bounce problem along that path. If the fixed-path
Fourier result agrees with the fully deformed CosmoTransitions result in the
range where the latter is available, then the same hybrid workflow can be used
as a practical fixed-path estimate in higher dimensions. 

We perform this comparison on the OptiBounce-like nested potential
\begin{equation}
    V_{N_\phi}(\bm{\phi})
    =
    \left[
        \sum_{i=1}^{N_\phi}
        c_i(\phi_i-1)^2
        -
        \delta
    \right]
    \left[
        \sum_{i=1}^{N_\phi}
        \phi_i^2
    \right],
    \qquad
    \delta=0.065 .
    \label{eq:hybrid_ct_optibounce_potential}
\end{equation}
For $N_\phi=2,\ldots,10$, we compare the hybrid fixed-path result with the full
CosmoTransitions path-deformation result. The Fourier path is optimised using
the mode scan $N_m=1,2,3,5,8$, and the best proxy-action path is then passed to
the one-dimensional CosmoTransitions bounce evaluator.
Table~\ref{tab:hybrid_ct_summary} summarises the comparison. The hybrid action
agrees with the full CosmoTransitions action at the sub-percent level in all
tested cases. The largest relative difference is approximately $0.63\%$, and
the mean relative difference over $N_\phi=2,\ldots,10$ is approximately
$0.36\%$. The hybrid runtime includes both the Fourier path optimisation time
and the subsequent one-dimensional CosmoTransitions bounce action evaluation time. In this
benchmark, the hybrid workflow is also faster than the full path-deformation
run in all tested dimensions, with speedups ranging from approximately $16\%$
to $56\%$.
\begin{table}[t]
    \centering
    \caption{
    Hybrid fixed-path workflow compared with the full CosmoTransitions path
    deformation for the OptiBounce-like potential in
    Eq.~\eqref{eq:hybrid_ct_optibounce_potential}. In the hybrid workflow, the
    Fourier path is held fixed, and CosmoTransitions solves only the effective
    one-dimensional bounce problem along that path. The hybrid time is defined
    as
    $t_{\rm hybrid}=t_{\rm Fourier}+t_{\rm CT~1D}$,
    where $t_{\rm Fourier}$ is the Fourier path-optimization time and
    $t_{\rm CT~1D}$ is the one-dimensional CosmoTransitions evaluation time
    along the fixed path. The relative action difference is computed with
    respect to the full CosmoTransitions action.
    }
    \label{tab:hybrid_ct_summary}
    \resizebox{\textwidth}{!}{
    \begin{tabular}{cccccccccc}
        \toprule
        $N_\phi$
        & $N_m$
        & $S_{\rm hybrid}$
        & $S_{\rm CT,full}$
        & $\Delta S_{\rm CT}$ [\%]
        & $t_{\rm Fourier}$ [s]
        & $t_{\rm CT~1D}$ [s]
        & $t_{\rm hybrid}$ [s]
        & $t_{\rm CT,full}$ [s]
        & speedup [\%] \\
        \midrule
        2  & 8 & 110.205  & 110.887  & 0.616 & 0.015 & 0.094 & 0.109 & 0.248 & 56.0 \\
        3  & 8 & 228.248  & 227.412  & 0.368 & 0.018 & 0.073 & 0.092 & 0.170 & 46.1 \\
        4  & 8 & 383.710  & 384.025  & 0.082 & 0.021 & 0.092 & 0.112 & 0.163 & 31.0 \\
        5  & 8 & 587.037  & 589.883  & 0.483 & 0.022 & 0.088 & 0.110 & 0.153 & 27.9 \\
        6  & 8 & 1096.355 & 1094.210 & 0.196 & 0.020 & 0.095 & 0.116 & 0.167 & 30.8 \\
        7  & 8 & 1866.717 & 1855.048 & 0.629 & 0.027 & 0.087 & 0.114 & 0.250 & 54.5 \\
        8  & 8 & 2451.630 & 2437.051 & 0.598 & 0.029 & 0.081 & 0.109 & 0.166 & 34.1 \\
        9  & 8 & 3567.499 & 3564.635 & 0.080 & 0.030 & 0.101 & 0.131 & 0.156 & 16.0 \\
        10 & 8 & 5015.452 & 5024.489 & 0.180 & 0.033 & 0.106 & 0.139 & 0.171 & 18.9 \\
        \bottomrule
    \end{tabular}
    }
\end{table}
As an additional diagnostic, we also compare the residual perpendicular-force
ratio of the fixed Fourier path with that of the fully deformed
CosmoTransitions solution. For the fixed-path runs this ratio is
$f_{\rm ratio}\simeq 0.28$--$0.33$, while the fully deformed runs reach
$f_{\rm ratio}\simeq 0.03$--$0.09$. This shows that the Fourier path already
captures the dominant action-relevant geometry, while the subsequent
CosmoTransitions deformation provides an additional transverse relaxation of
the path.

For illustration, the same hybrid workflow was also run beyond the full
CosmoTransitions comparison range. With a Fourier mode scan extended to
$N_m=15$, the selected mode is $N_m=15$ for all
$N_\phi=15,\ldots,20$. The resulting fixed-path hybrid actions are
\begin{equation}
\begin{aligned}
    S_{\rm hybrid}(15) &= 1.3462\times10^4, &
    S_{\rm hybrid}(16) &= 1.6097\times10^4, &
    S_{\rm hybrid}(17) &= 1.9671\times10^4, \\
    S_{\rm hybrid}(18) &= 2.1672\times10^4, &
    S_{\rm hybrid}(19) &= 2.5581\times10^4, &
    S_{\rm hybrid}(20) &= 2.9104\times10^4 .
\end{aligned}
\end{equation}
These values demonstrate that the hybrid pipeline continues to run beyond the
range where the full CosmoTransitions comparison is performed. They should be
interpreted as Fourier-fixed-path plus one-dimensional CosmoTransitions
estimates, not as independent full CosmoTransitions multi-field bounce actions.

\subsection{FindBounce point injection}
\label{subsec:findbounce_injection}

\begin{figure}[t]
    \centering
    \includegraphics[width=0.72\textwidth]{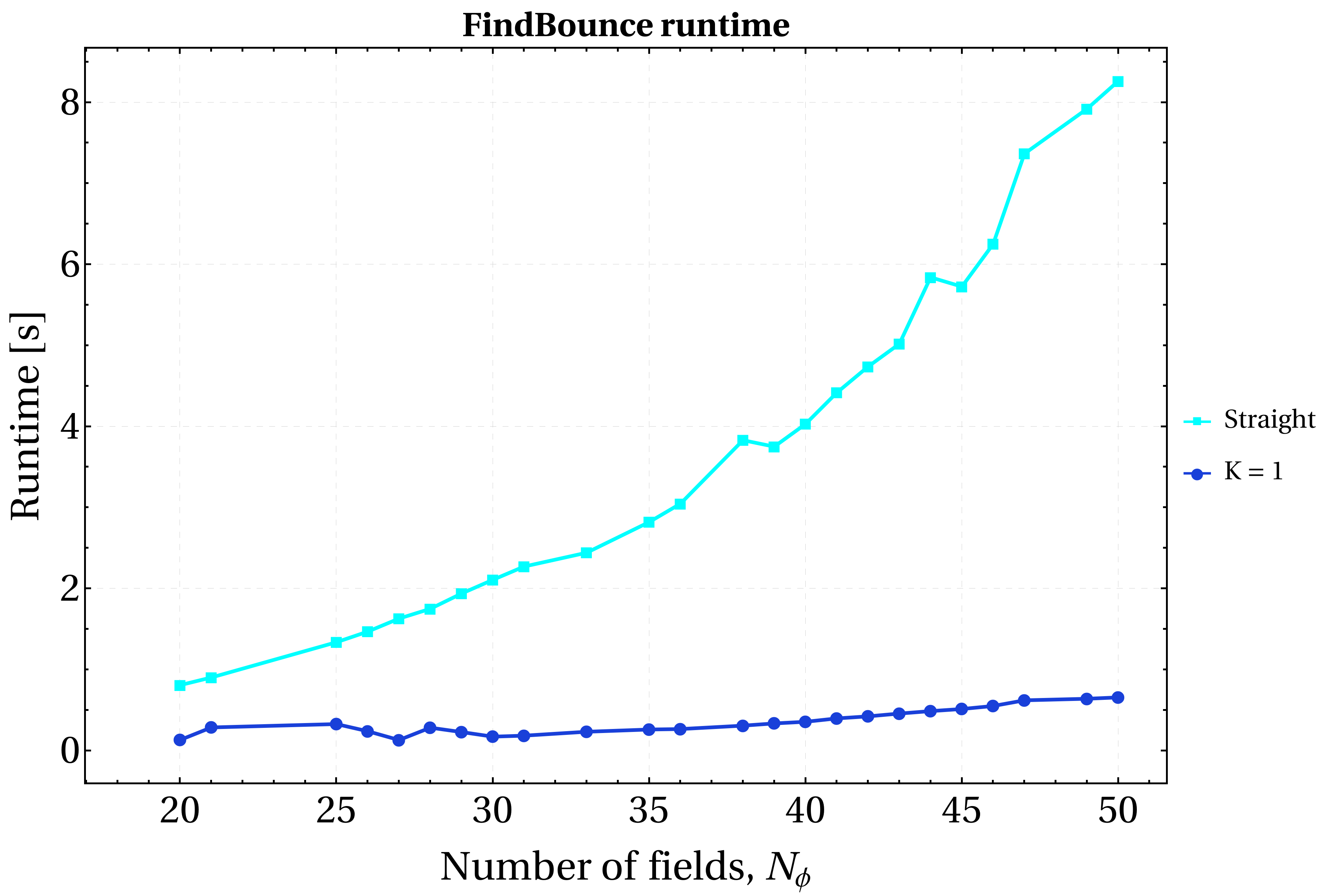}
    \caption{
    Runtime comparison between the straight-line FindBounce initialisation and
    the one-point Fourier-informed initialisation for the high-dimensional cases
    shown in Table~\ref{tab:findbounce_k1}. The label ``Straight'' denotes the
    explicit straight-line FindBounce initialisation, whose runtime is
    $t_{\rm str}$. The label ``$K=1$'' denotes the run in which one point sampled
    from the Fourier-deformed path is injected into the polygonal
    initialization, whose runtime is $t_{K=1}$. The one-point Fourier injection
    substantially reduces the runtime while the corresponding actions remain
    comparable, as shown in Table~\ref{tab:findbounce_k1}.
    }
    \label{fig:findbounce_k1_runtime}
\end{figure}

We next test whether the Fourier-deformed path can improve FindBounce
initialisation. FindBounce uses a polygonal representation of the tunnelling
path, so geometric information can be supplied through intermediate points
between the false and true vacua.
Starting from an optimised Fourier path, we sample $K$ intermediate points and
pass them to FindBounce as part of the polygonal initialisation. Thus $K=1$
denotes a one-point Fourier-informed initialisation, while the standard
straight-line initialisation contains no injected Fourier point. In the main
text, we focus on the conservative $K=1$ case, which keeps the modification of
the external solver minimal and tests whether a small amount of Fourier path
information is already sufficient to reduce the runtime.
Table~\ref{tab:findbounce_k1} compares the straight-line FindBounce run with
the one-point Fourier-informed run for selected high-dimensional cases. Here
$S_{\rm JAX}$ is the independent JAX--Fourier action estimate and is not used
as an input to FindBounce. The quantities $S_{\rm str}$ and $t_{\rm str}$
denote the action and runtime for the straight-line FindBounce initialisation,
while $S_{K=1}$ and $t_{K=1}$ denote the corresponding quantities after
injecting one point sampled from the Fourier-deformed path. The speedup is
defined as
\begin{equation}
    \text{speedup}
    =
    100\,
    \frac{t_{\rm str}-t_{K=1}}{t_{\rm str}} .
    \label{eq:findbounce_speedup}
\end{equation}
\begin{table}[t]
    \centering
    \caption{
    FindBounce benchmark using a one-point Fourier-informed initialisation.
    $S_{\rm JAX}$ is the independent JAX--Fourier result. $S_{\rm str}$ and
    $t_{\rm str}$ are the action and runtime obtained from the explicit
    straight-line FindBounce initialisation. $S_{K=1}$ and $t_{K=1}$ are the
    action and runtime obtained when one point sampled from the Fourier-deformed
    path is injected into the FindBounce polygonal initialisation. The speedup
    is defined by Eq.~\eqref{eq:findbounce_speedup}. We show the
    high-dimensional cases where the injected and straight-path actions remain
    comparable.
    }
    \label{tab:findbounce_k1}
    \begin{tabular}{ccccccc}
        \toprule
        $N_\phi$
        & $S_{\rm FD}$
        & $S_{\rm str}$
        & $S_{K=1}$
        & $t_{\rm str}$ [s]
        & $t_{K=1}$ [s]
        & speedup [\%] \\
        \midrule
        20 & $8.293\times10^{3}$ & $8.285\times10^{3}$ & $8.198\times10^{3}$ & 0.866 & 0.148 & 82.9 \\
        30 & $2.627\times10^{4}$ & $2.625\times10^{4}$ & $2.605\times10^{4}$ & 2.242 & 0.191 & 91.5 \\
        40 & $7.621\times10^{4}$ & $7.615\times10^{4}$ & $7.651\times10^{4}$ & 4.259 & 0.400 & 90.6 \\
        50 & $4.190\times10^{4}$ & $4.188\times10^{4}$ & $4.192\times10^{4}$ & 8.566 & 0.742 & 91.3 \\
        \bottomrule
    \end{tabular}
\end{table}
The one-point Fourier-informed initialisation substantially reduces the
FindBounce runtime in all cases shown in Table~\ref{tab:findbounce_k1}. The
reduction is about $83\%$ at $N_\phi=20$ and about $91\%$ for
$N_\phi=30,40,50$. The corresponding actions remain close to both the
straight-path FindBounce results and the independent JAX--Fourier estimates,
indicating that the runtime gain is obtained without a significant shift in the
computed action.
Figure~\ref{fig:findbounce_k1_runtime} shows the same comparison graphically.
It compares $t_{\rm str}$, the runtime of the straight-line FindBounce
initialization, with $t_{K=1}$, the runtime obtained after injecting one
Fourier-informed point. The reduction in runtime shows that even one point sampled from the optimized
Fourier path can guide the initial polygon closer to the relevant tunnelling
trajectory.

The $K=1$ comparison is intentionally conservative. It does not optimise over
all possible polygonal initialisations, nor does it modify the FindBounce
algorithm. It only tests whether a minimal amount of Fourier-generated path
information can improve the solver. In the benchmark cases shown here, one
Fourier-informed point substantially reduces the runtime while preserving
agreement with the straight-path FindBounce and JAX--Fourier action estimates.
The full scan over $K=1,\ldots,8$ is reported in
Appendix~\ref{app:findbounce_full_scan}. These extended results illustrate the
sensitivity of the polygonal initialisation to the amount of supplied path
information and point toward future adaptive strategies for choosing the number
and placement of injected points.
\section{Conclusions and outlook}
\label{sec:conclusions}

We have presented an endpoint-safe Fourier path method for multi-field vacuum
tunnelling. The tunnelling trajectory is written as a straight-line interpolation
between the false and true vacua, plus sine-mode deformations that vanish at the
endpoints. The boundary conditions are therefore satisfied exactly for any
choice of Fourier coefficients, and the path optimisation becomes a
finite-dimensional variational problem. In the implementation used here, the
coefficients are optimised using a reduced action functional motivated by the
tunnelling-potential formulation.

The method has two complementary uses. First, it can be used as a standalone
ansatz for curved tunnelling paths. Second, it can be used as a preconditioner
for existing tunnelling solvers, where the Fourier path supplies an improved
initial trajectory or a small number of intermediate path points. In the second
interpretation, the Fourier calculation is not the final bounce calculation;
rather, it provides a smooth, endpoint-preserving, action-informed starting
geometry for a more complete solver.

As a standalone path ansatz, the method reproduces the expected curved
trajectory in the two-field visualisation example and agrees with established
codes on standard benchmarks. On the multi-field benchmark potential of
Ref.~\cite{BARDSLEY}, the JAX--Fourier implementation agrees with
OptiBounce, FindBounce, and CosmoTransitions at the sub-percent level in the
regular cases. On the nested random-coefficient benchmark, the scan completes
up to $N_\phi=50$ using an internal mode-convergence criterion. Wherever the
independent FindBounce comparison is available, the actions agree at the
sub-percent level, with representative high-dimensional differences typically
below $0.1\%$. The selected number of modes remains modest, usually
$N_m\simeq 8$--$10$ for the larger field dimensions.

The basis comparison shows that Fourier sine modes are a robust default for
the smooth benchmark paths considered here. Local B-spline and hybrid
Fourier-local bases provide useful cross-checks and can be competitive, while
the simple polynomial and shifted-wavelet constructions tested here are less
efficient on these examples. The practical lesson is not that Fourier modes are
universally optimal, but that smooth endpoint-safe parametrisations can capture
the dominant path deformation with relatively few degrees of freedom.

The preconditioning tests show that Fourier paths can reduce the work required
by existing solvers. For CosmoTransitions, a one-mode Fourier initialisation
reduces the number of path-deformation steps from $51$ to $16$ in the
controlled curved-valley tests, while leaving the final action stable. For
FindBounce, a conservative one-point Fourier-informed initialisation reduces
the runtime by about $83\%$--$91\%$ in the displayed high-dimensional cases,
with actions remaining comparable to the straight-path initialisation. These
results support the interpretation of Fourier-deformed paths as a useful
preconditioning layer rather than a replacement for established bounce
algorithms.

There are several natural extensions. The present implementation uses a fixed
smooth tunnelling-potential profile, rather than solving the full variational
problem for $V_t$. Combining the endpoint-safe Fourier path ansatz with a full
optimisation of $V_t$ would give a more complete Fourier-based
tunnelling-potential solver. Finite-temperature $O(3)$ tunnelling is another
direct extension, since the field-space path construction is independent of the
spacetime symmetry. More adaptive local or multiresolution bases may also be
useful for potentials with sharp turns, nearly flat directions, or multiple
competing tunnelling channels.

Overall, Fourier-deformed paths provide a simple, endpoint-safe, and
computationally efficient representation of curved multi-field tunnelling
trajectories. They can be used directly as a reduced variational ansatz and can
also improve the initialisation of existing bounce solvers. The method is
therefore best viewed as a complementary path representation that can make
multi-field tunnelling calculations faster, more stable, and easier to diagnose.


\appendix
%

\section{JAX implementation and numerical optimisation}
\label{app:jax_implementation}

The numerical implementation used for the high-dimensional benchmark is based on a hybrid
JAX--SciPy workflow. JAX is used to evaluate the reduced action and its gradient with respect
to the Fourier coefficients, while the finite-dimensional minimisation itself is performed using
the L-BFGS-B algorithm implemented in \texttt{scipy.optimize.minimize}. Thus, the method is
not a fixed-step gradient-descent algorithm. Instead, it is a gradient-based quasi-Newton
optimisation, where the gradient is supplied exactly by automatic differentiation.
For an $N_\phi$-field problem with $N_m$ Fourier modes, the path is written as
\begin{equation}
    \phi_i(t)
    =
    (\phi_F)_i
    +
    t\left[(\phi_T)_i-(\phi_F)_i\right]
    +
    \sum_{k=1}^{N_m}
    a_{ik}\sin(k\pi t),
    \label{eq:jax_fourier_path}
\end{equation}
where $i=1,\ldots,N_\phi$. In the code, the coefficients are stored as a two-dimensional array
\begin{equation}
    a_{ik},
    \qquad
    i=1,\ldots,N_\phi,
    \qquad
    k=1,\ldots,N_m,
\end{equation}
with shape $(N_\phi,N_m)$. For the interface with \texttt{scipy.optimize.minimize}, this array is
flattened into a one-dimensional vector of length $N_\phi N_m$. At every action evaluation, the
flat vector is reshaped back into the coefficient matrix.
The path parameter is discretised on a uniform grid
\begin{equation}
    0=t_0<t_1<\cdots<t_{N_q-1}=1,
\end{equation}
with $N_q=260$ in the benchmark runs. For each value of $N_m$, the sine basis is precomputed as
\begin{equation}
    B_{kq}
    =
    \sin(k\pi t_q),
    \qquad
    k=1,\ldots,N_m,
    \qquad
    q=0,\ldots,N_q-1 .
\end{equation}
The full path on the grid is then reconstructed by the vectorised matrix operation
\begin{equation}
    \Phi_{qi}
    =
    (\Phi_{\rm str})_{qi}
    +
    \sum_{k=1}^{N_m}B_{kq}a_{ik},
    \label{eq:jax_path_matrix_form}
\end{equation}
where
\begin{equation}
    (\Phi_{\rm str})_{qi}
    =
    (\phi_F)_i
    +
    t_q\left[(\phi_T)_i-(\phi_F)_i\right].
\end{equation}
In the implementation, Eq.~\eqref{eq:jax_path_matrix_form} is evaluated on the
entire grid at once using vectorised array operations over the grid index $q$,
the field index $i$, and the Fourier-mode index $k$. Thus, for fixed
$(N_\phi,N_m,N_q)$, the path reconstruction is reduced to standard matrix
algebra rather than a Python-level loop over fields or grid points. This
array-based structure is well suited to JAX automatic differentiation and
just-in-time compilation~\cite{jax2018github,jaxdocs}.
For the random-coefficient benchmark, the potential considered is
\begin{equation}
    V(\bm\phi)
    =
    \left[
        \sum_{i=1}^{N_\phi}
        c_i(\phi_i-1)^2
        -
        \delta_{N_\phi}
    \right]
    \sum_{i=1}^{N_\phi}\phi_i^2 .
    \label{eq:jax_random_potential}
\end{equation}
The false vacuum is fixed at
\begin{equation}
    \bm\phi_F=\bm 0.
\end{equation}
For each value of $N_\phi$, the true vacuum is found numerically before the Fourier path
minimization. This preliminary minimization is performed with \texttt{scipy.optimize.minimize}
using the BFGS method, starting from $\bm\phi=\bm 1$. The analytic gradient of the potential is
supplied to the optimiser:
\begin{equation}
    \frac{\partial V}{\partial \phi_i}
    =
    2c_i(\phi_i-1)
    \sum_j\phi_j^2
    +
    2\phi_i
    \left[
        \sum_j c_j(\phi_j-1)^2-\delta
    \right].
    \label{eq:random_potential_gradient}
\end{equation}
The BFGS minimisation of the true vacuum uses a gradient tolerance $10^{-12}$ and a maximum
of $5000$ iterations. After this step, the vacuum energies $V_F$, $V_T$, and the energy
difference
\begin{equation}
    \Delta V=V_F-V_T
\end{equation}
are recorded. Cases with $\Delta V\leq 0$ are rejected.
The Fourier minimisation uses a fixed tunnelling-potential interpolation along the path,
\begin{equation}
    V_t(t)
    =
    V_F
    +
    t^2(3-2t)(V_T-V_F).
    \label{eq:smooth_vt_interpolation}
\end{equation}
This interpolation satisfies
\begin{equation}
    V_t(0)=V_F,
    \qquad
    V_t(1)=V_T,
    \qquad
    \frac{\dd V_t}{\dd t}\bigg|_{t=0,1}=0,
\end{equation}
and provides a smooth endpoint-safe profile for the reduced action. The implementation does
not solve the full variational tunnelling-potential problem. Instead, it minimises the reduced
action obtained from this fixed smooth $V_t(t)$ ansatz.
On the discretised path, the potential is evaluated as
\begin{equation}
    V_q
    =
    V(\bm\phi(t_q)).
\end{equation}
The line element between neighbouring grid points is
\begin{equation}
    \Delta s_q
    =
    \left|
        \bm\phi(t_{q+1})-\bm\phi(t_q)
    \right|,
    \qquad
    q=0,\ldots,N_q-2.
\end{equation}
The finite-difference change in the tunnelling potential is
\begin{equation}
    \Delta V_{t,q}
    =
    V_t(t_{q+1})-V_t(t_q).
\end{equation}
The discrete reduced action minimised in the code is
\begin{equation}
    S_4
    =
    \frac{27\pi^2}{2}
    \sum_{q=0}^{N_q-2}
    \frac{
    \left[
        \left(V_q-V_t(t_q)\right)
        +
        \left(V_{q+1}-V_t(t_{q+1})\right)
    \right]^2
    (\Delta s_q)^4
    }{
    -\left(\Delta V_{t,q}\right)^3
    }.
    \label{eq:discrete_jax_action}
\end{equation}
This is the scalar objective function passed to the optimiser. The optimisation variables are
only the Fourier coefficients $a_{ik}$. Equation~\eqref{eq:discrete_jax_action} is the same
reduced action as Eq.~\eqref{eq:reduced_action_discrete}: the prefactor differs only because the
numerator here uses the \emph{summed} midpoint values
$(V_q-V_{t,q})+(V_{q+1}-V_{t,q+1})$ rather than their average, and
$\tfrac{27\pi^2}{2}=\tfrac{1}{4}\cdot 54\pi^2$ compensates for the resulting factor of four.
The value and gradient of Eq.~\eqref{eq:discrete_jax_action} are evaluated
using JAX automatic differentiation. In the implementation, the action function
is wrapped as
\begin{equation}
    \bm x
    \mapsto
    \left(
        S(\bm x),
        \nabla_{\bm x}S(\bm x)
    \right),
\end{equation}
using \texttt{jax.value\_and\_grad}, where $\bm x$ denotes the flattened vector
of Fourier coefficients. This value-and-gradient function is then just-in-time
compiled using \texttt{jax.jit}. Double precision is enabled through
\texttt{jax.config.update("jax\_enable\_x64", True)}, since the reduced action
can be sensitive to small differences in the potential along the path.
For each fixed pair $(N_\phi,N_m)$, the JAX action-and-gradient function is
compiled once. The compilation is triggered by a first call at the
zero-coefficient point,
\begin{equation}
    a_{ik}=0.
\end{equation}
After this first compilation step, all subsequent evaluations during the
L-BFGS-B minimisation reuses the compiled function. This is useful because the optimiser evaluates the same action-and-gradient map many times while changing
only the coefficient values.
The Fourier coefficients are optimised using the L-BFGS-B method. The objective
and gradient are passed together to \texttt{scipy.optimize.minimize} with
\texttt{jac=True}. The optimiser settings used in the benchmark runs are
\begin{equation}
    N_{\rm iter}^{\rm max}=800,
    \qquad
    {\tt ftol}=10^{-10},
    \qquad
    {\tt gtol}=10^{-7},
    \qquad
    {\tt maxls}=50.
\end{equation}
The use of L-BFGS-B also allows simple box bounds to be imposed on every
Fourier coefficient. We write these bounds as
\begin{equation}
    -A_{\rm max}
    \leq
    a_{ik}
    \leq
    A_{\rm max},
    \qquad
    A_{\rm max}
    =
    \kappa\,
    |\bm\phi_T-\bm\phi_F| .
    \label{eq:coefficient_bound}
\end{equation}
Here $\kappa$ is a benchmark-dependent coefficient-bound factor. The role of
the bound is not to impose a physical constraint, but to prevent the optimiser
from exploring very large Fourier deformations that move the trial path far
away from the region connecting the two vacua.
The scan over Fourier modes is adaptive. For each value of $N_\phi$, the code
runs the Fourier minimisation for an ordered set of mode numbers
\begin{equation}
    N_m \in \mathcal{M},
\end{equation}
where the set $\mathcal{M}$ is specified for each benchmark. At each value of
$N_m$, the JAX action-and-gradient function is compiled once, and the L-BFGS-B
minimization is run from the selected initial conditions. The global best action
is updated after each mode. The scan is stopped when the relative improvement
in the global best action remains below a prescribed tolerance
$\epsilon_{\rm mode}$ for $p$ consecutive mode steps. Thus, the stopping
criterion is internal to the Fourier scan and does not require an external
reference action.
The initial conditions for the Fourier coefficients are also specified
explicitly. The zero start,
\begin{equation}
    a_{ik}=0,
\end{equation}
is always included and corresponds to the undeformed straight-line path. In
addition, the implementation can use warm starts from previously optimised
solutions. If the optimised coefficients at $N_m$ modes are known, then an
initial point for a larger mode number is obtained by copying the lower-mode
coefficients and setting the newly added higher-mode coefficients to zero. Some
benchmark runs also include deterministic random starts, with fixed seeds, as a
check against local minima. The precise set of starts used in a given benchmark
is reported with the numerical results.

The acceleration in this implementation comes from the combination of three
ingredients. First, the Fourier ansatz converts the path optimisation into a
finite-dimensional problem over $N_\phi N_m$ coefficients. Second, JAX evaluates
the full reduced action and its gradient by automatic differentiation, avoiding
finite-difference gradients. Third, just-in-time compilation fuses the batched
operations over grid points, fields, and Fourier modes into compiled array
operations. The scaling of the calculation is still controlled by the number of
fields, modes, and grid points, but the compiled implementation substantially
reduces the Python-level overhead that would appear in a direct loop-based
implementation. This makes repeated scans over field number, mode number, and
benchmark potentials computationally practical.

\section{Endpoint-safe basis functions}
\label{app:bases}

In this appendix, we list the explicit endpoint-safe basis functions used in
Sec.~\ref{subsec:basis_comparison}. For all basis families, the path is written
as
\begin{equation}
    \phi_i(t)
    =
    \phi_{i,F}
    +
    t(\phi_{i,T}-\phi_{i,F})
    +
    \sum_{a=1}^{N_{\rm basis}}
    c_{ia} B_a(t),
    \qquad
    t\in[0,1],
    \label{eq:app_general_basis_path}
\end{equation}
with
\begin{equation}
    B_a(0)=B_a(1)=0 .
\end{equation}
Thus, all variations preserve the false and true vacua exactly. In the numerical
implementation, each basis function is sampled on the same grid used for the
action integral and normalised by its maximum absolute value,
\begin{equation}
    B_a(t)
    \longrightarrow
    \frac{B_a(t)}
    {
        \max_{t\in[0,1]} |B_a(t)|
    },
    \label{eq:basis_normalization}
\end{equation}
whenever the denominator is nonzero. This normalisation makes the coefficient
scales comparable across different basis families. Throughout this appendix we
label the basis families by the superscript on $B_a$: F (Fourier sine),
G (Gaussian), BS (B-spline), and W (wavelet); hybrid families that concatenate a
Fourier block with a local block are written F$+$G (Fourier-Gaussian) and
F$+$BS (Fourier-B-spline).

\paragraph{Fourier sine basis.}
The default basis used in the main text is
\begin{equation}
    B_a^{\rm F}(t)
    =
    \sin(a\pi t),
    \qquad
    a=1,\ldots,N_{\rm basis}.
    \label{eq:app_fourier_basis}
\end{equation}
Each mode vanishes at both endpoints and is smooth on the full interval.

\paragraph{Chebyshev basis.}
Let $T_a(x)$ denote the Chebyshev polynomial of the first kind, with
$x=2t-1$. The endpoint-safe Chebyshev basis is
\begin{equation}
    B_a^{\rm Ch}(t)
    =
    t(1-t)\,T_a(2t-1),
    \qquad
    a=1,\ldots,N_{\rm basis}.
    \label{eq:app_chebyshev_basis}
\end{equation}
The envelope $t(1-t)$ enforces the endpoint conditions.

\paragraph{Legendre basis.}
Let $P_a(x)$ denote the Legendre polynomial. The endpoint-safe Legendre basis is
\begin{equation}
    B_a^{\rm L}(t)
    =
    t(1-t)\,P_a(2t-1),
    \qquad
    a=1,\ldots,N_{\rm basis}.
    \label{eq:app_legendre_basis}
\end{equation}

\paragraph{Gaussian local basis.}
For the local Gaussian basis, the centres $\mu_a$ are uniformly spaced in
$[0.1,0.9]$, and we use
\begin{equation}
    \sigma_G
    =
    \frac{0.75}{\max(N_{\rm basis},2)} .
\end{equation}
The endpoint-safe Gaussian basis functions are
\begin{equation}
    B_a^{\rm G}(t)
    =
    t(1-t)
    \exp\left[
        -\frac{(t-\mu_a)^2}{2\sigma_G^2}
    \right].
    \label{eq:app_gaussian_basis}
\end{equation}

\paragraph{B-spline local basis.}
Let $S_a^{(p)}(t)$ denote the $a$th open-uniform B-spline basis function of
degree
\begin{equation}
    p=\min(3,N_{\rm basis}-1).
\end{equation}
The endpoint-safe B-spline basis is
\begin{equation}
    B_a^{\rm BS}(t)
    =
    t(1-t)\,S_a^{(p)}(t),
    \qquad
    a=1,\ldots,N_{\rm basis}.
    \label{eq:app_bspline_basis}
\end{equation}

\paragraph{Hybrid Fourier-local bases.}
For a total of $N_{\rm basis}$ functions, we take
\begin{equation}
    N_{\rm F}
    =
    \lceil 0.6\,N_{\rm basis}\rceil,
    \qquad
    N_{\rm loc}
    =
    N_{\rm basis}-N_{\rm F}.
\end{equation}
The hybrid Fourier-Gaussian basis is
\begin{equation}
    \mathcal{B}^{\rm F+G}
    =
    \left\{
    B_1^{\rm F},\ldots,B_{N_{\rm F}}^{\rm F},
    B_1^{\rm G},\ldots,B_{N_{\rm loc}}^{\rm G}
    \right\},
    \label{eq:app_hybrid_fourier_gaussian}
\end{equation}
and the hybrid Fourier-B-spline basis is
\begin{equation}
    \mathcal{B}^{\rm F+BS}
    =
    \left\{
    B_1^{\rm F},\ldots,B_{N_{\rm F}}^{\rm F},
    B_1^{\rm BS},\ldots,B_{N_{\rm loc}}^{\rm BS}
    \right\}.
    \label{eq:app_hybrid_fourier_bspline}
\end{equation}

\paragraph{Wavelet-inspired bases.}
Let $\psi(u)$ denote the mother wavelet. In the scan, we used Daubechies
db2, db4, db6, Symlet sym4, and Coiflet coif1 wavelets. The centers $\mu_a$
are uniformly spaced in $[0.1,0.9]$, and the width is
\begin{equation}
    \sigma_W
    =
    \frac{0.8}{\max(N_{\rm basis},2)} .
\end{equation}
We define the rescaled support coordinate
\begin{equation}
    u_a(t)
    =
    \frac{t-\mu_a}{\sigma_W}
    +
    \frac{1}{2}.
\end{equation}
The endpoint-safe wavelet-inspired basis functions are
\begin{equation}
    B_a^{\rm W}(t)
    =
    t(1-t)\,\psi(u_a(t)),
    \qquad
    a=1,\ldots,N_{\rm basis}.
    \label{eq:app_wavelet_basis}
\end{equation}
Values outside the support of the mother wavelet are set to zero by
interpolation. This is a wavelet-inspired variational basis, not a full
orthonormal discrete wavelet transform. Other wavelet parametrisations or
adaptive wavelet trees may behave differently.
\section{Additional endpoint-safe basis-comparison plots}
\label{app:additional_plots}

This appendix collects the diagnostic plots used in the endpoint-safe basis
comparison discussed in Sec.~\ref{subsec:basis_comparison}. The main text keeps
only the compact numerical table, while the figures here show the corresponding
action, runtime, selected basis size, and representative path-approximation
behavior. These plots are intended as diagnostics of the basis families tested
in this work, not as a universal ranking of endpoint-safe bases.

\subsection{Basis scan on the OptiBounce \texorpdfstring{$d=3$}{d=3} benchmark}
\label{app:basis_optibounce_d3}

\begin{figure}[h]
    \centering
    \begin{subfigure}[b]{0.32\textwidth}
        \centering
        \includegraphics[width=\linewidth]{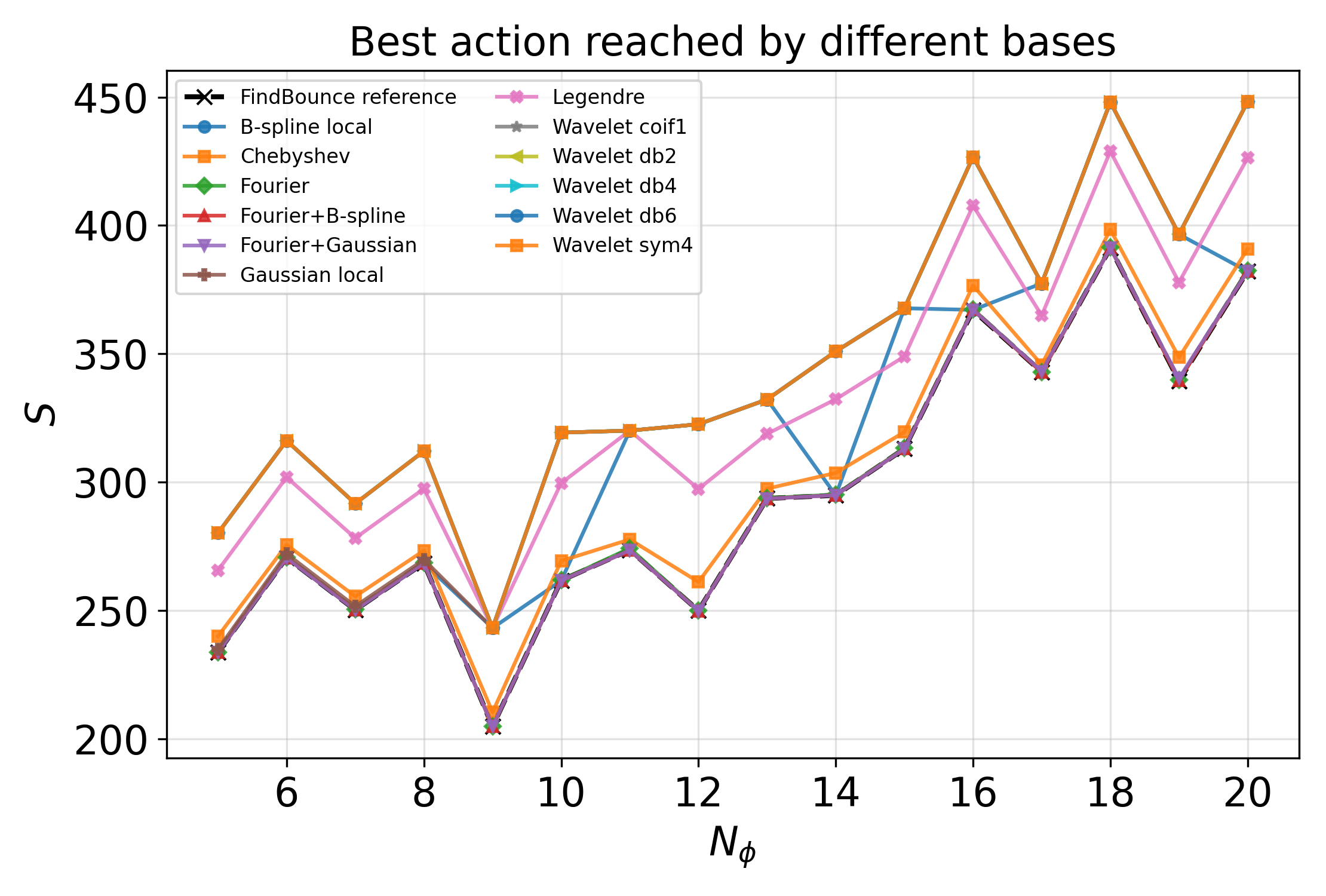}
        \caption{Best action}
        \label{fig:app_basis_optibounce_d3_action}
    \end{subfigure}
    \hfill
    \begin{subfigure}[b]{0.32\textwidth}
        \centering
        \includegraphics[width=\linewidth]{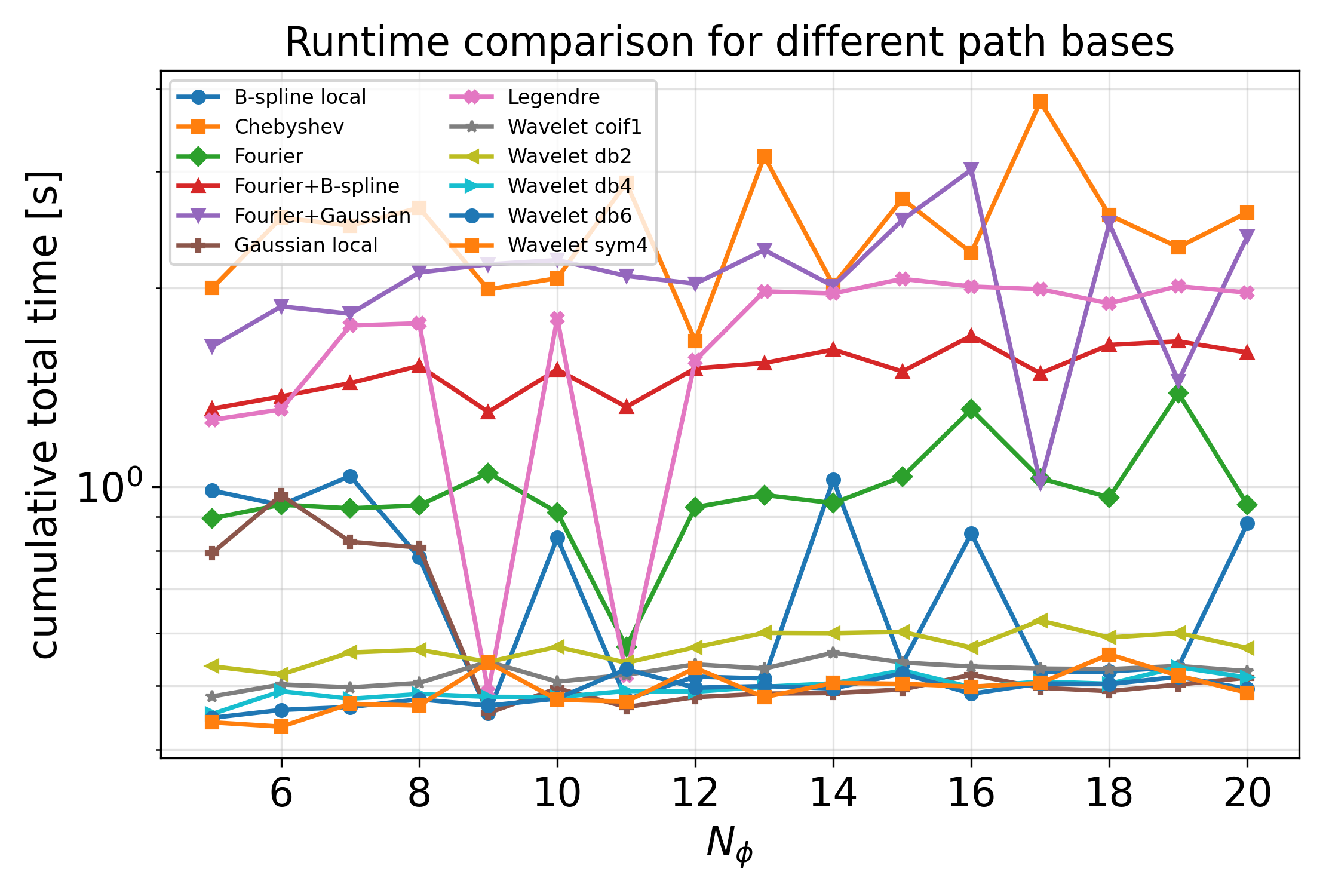}
        \caption{Runtime}
        \label{fig:app_basis_optibounce_d3_time}
    \end{subfigure}
    \hfill
    \begin{subfigure}[b]{0.32\textwidth}
        \centering
        \includegraphics[width=\linewidth]{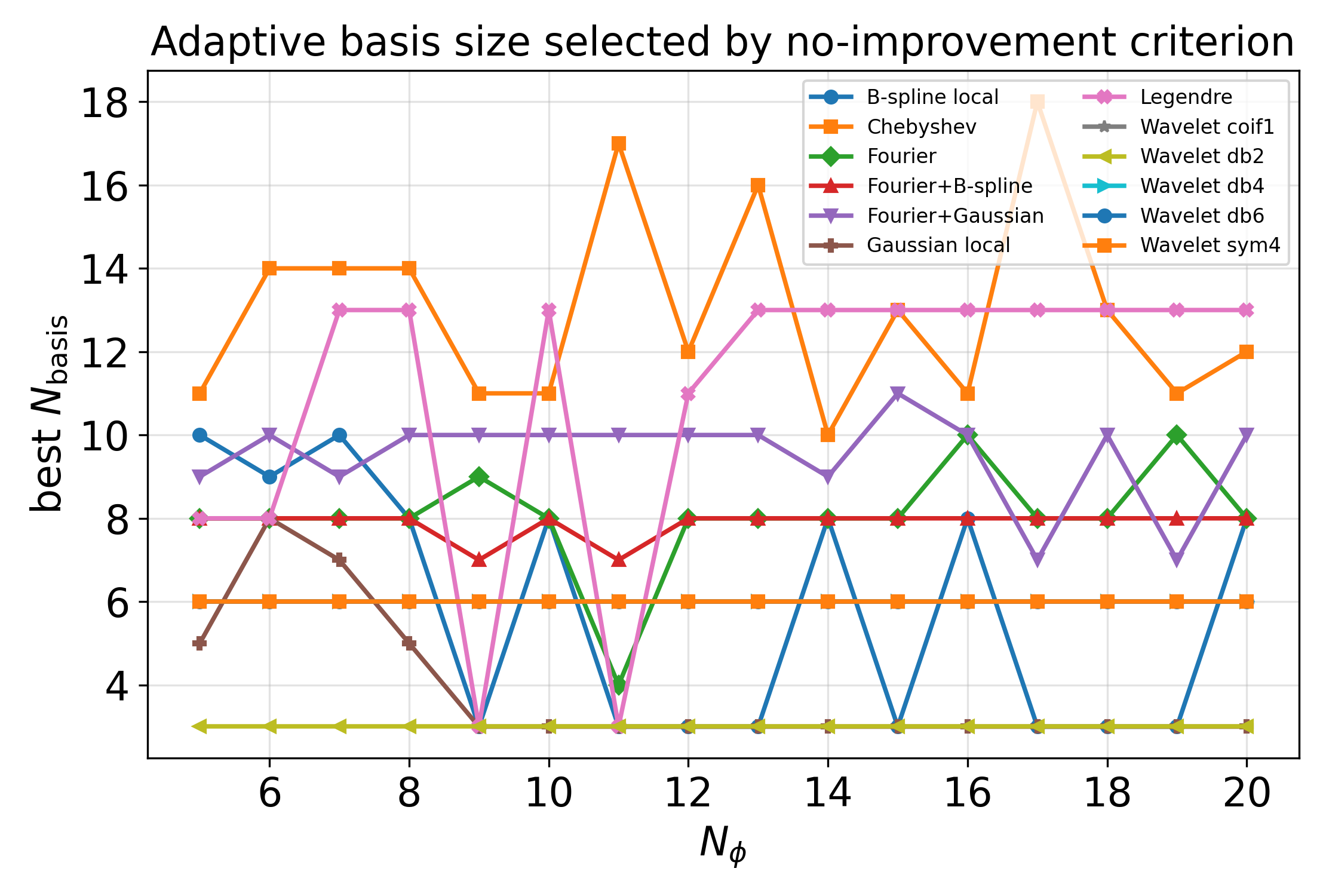}
        \caption{Selected basis size}
        \label{fig:app_basis_optibounce_d3_modes}
    \end{subfigure}
    \caption{
    Endpoint-safe basis comparison on the OptiBounce $d=3$ benchmark. The
    panels show the best action, total runtime, and the adaptively selected number
    of basis functions for the tested basis families.
    }
    \label{fig:app_basis_optibounce_d3}
\end{figure}

Figure~\ref{fig:app_basis_optibounce_d3} shows the full basis comparison on the
OptiBounce $d=3$ benchmark. The panels show the best action reached by the
adaptive scan, the total runtime, and the number of basis functions selected by
the no-improvement stopping criterion.

\subsection{Basis scan on the random \texorpdfstring{$d=4$}{d=4} benchmark}
\label{app:basis_random_d4}

\begin{figure}[h]
    \centering
    \begin{subfigure}[b]{0.32\textwidth}
        \centering
        \includegraphics[width=\linewidth]{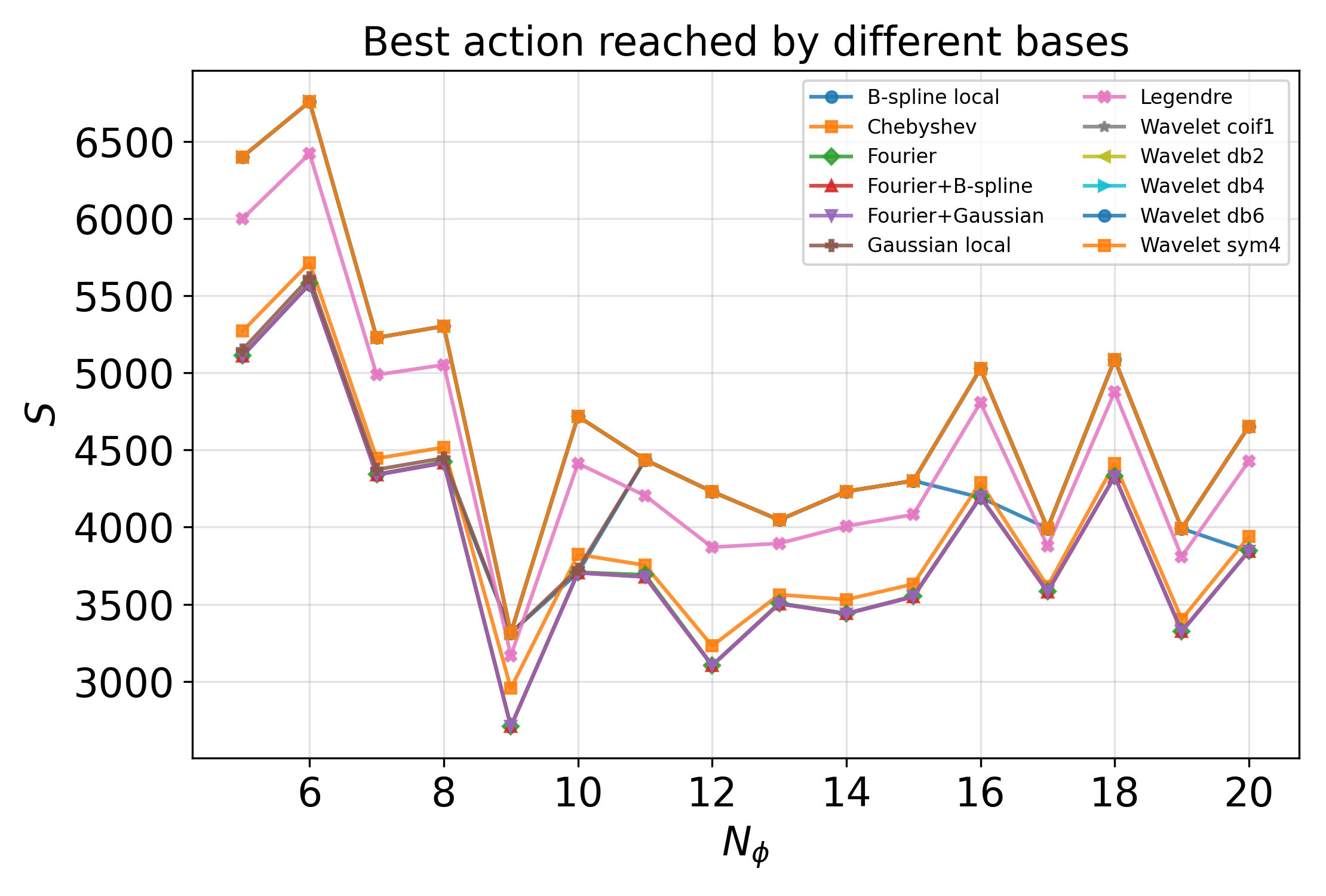}
        \caption{Best action}
        \label{fig:app_basis_random_d4_action}
    \end{subfigure}
    \hfill
    \begin{subfigure}[b]{0.32\textwidth}
        \centering
        \includegraphics[width=\linewidth]{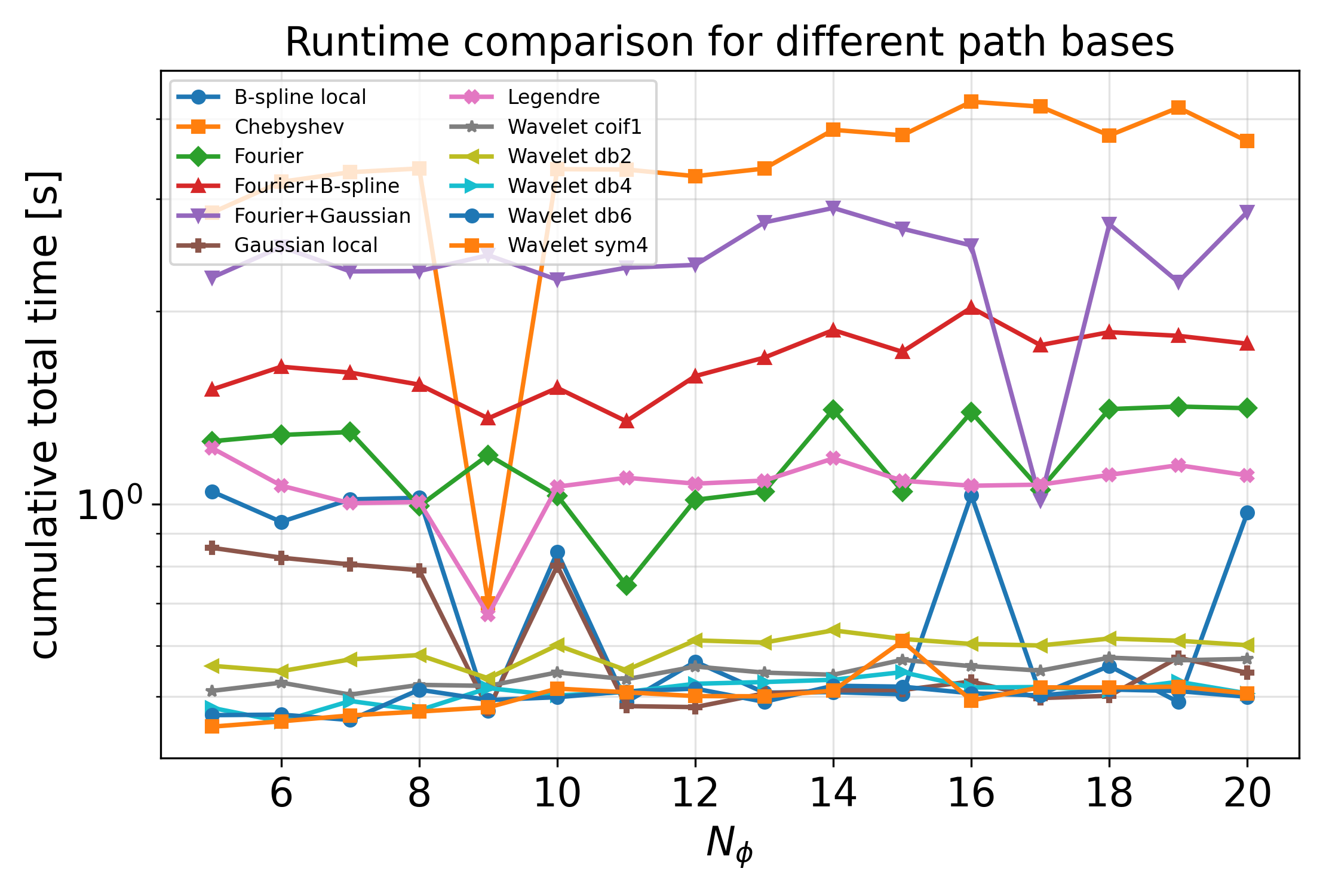}
        \caption{Runtime}
        \label{fig:app_basis_random_d4_time}
    \end{subfigure}
    \hfill
    \begin{subfigure}[b]{0.32\textwidth}
        \centering
        \includegraphics[width=\linewidth]{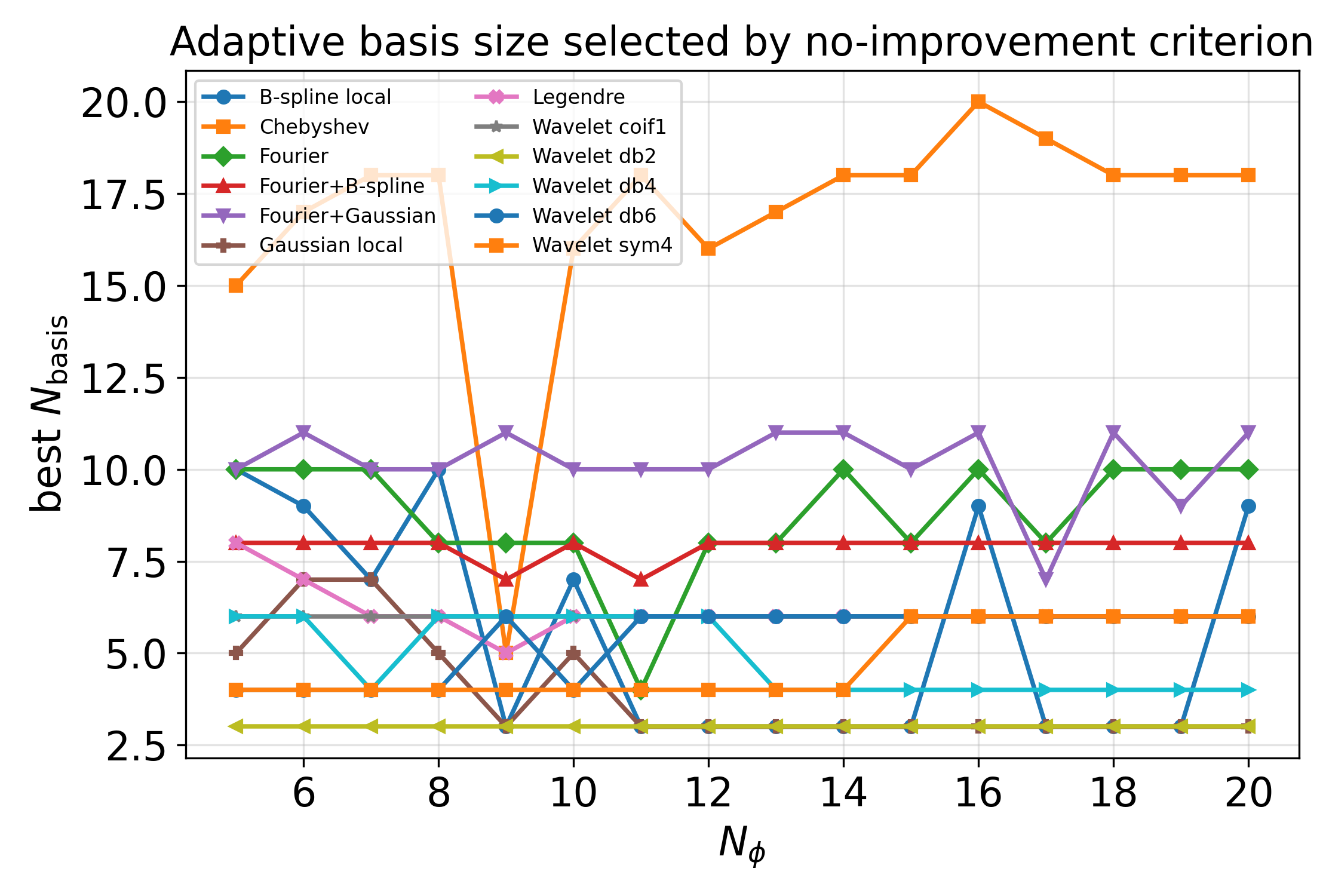}
        \caption{Selected basis size}
        \label{fig:app_basis_random_d4_modes}
    \end{subfigure}
    \caption{
    Endpoint-safe basis comparison on the random $d=4$ benchmark. The panels
    show the best action, total runtime, and adaptively selected number of
    basis functions for the tested basis families.
    }
    \label{fig:app_basis_random_d4}
\end{figure}

Figure~\ref{fig:app_basis_random_d4} shows the corresponding basis comparison
for the random $d=4$ benchmark. As in the main text, the purpose of this scan
is to compare the behaviour of different endpoint-safe basis families on the
same fixed set of smooth benchmark potentials.

\subsection{Two-field path comparison across basis families}
\label{app:basis_two_field_paths}

\begin{figure}[h]
    \centering

    \begin{subfigure}[b]{0.32\textwidth}
        \centering
        \includegraphics[width=\linewidth]{ct_optibounce2d_path_contour.png}
        \caption{CosmoTransitions}
        \label{fig:app_two_field_ct_path}
    \end{subfigure}
    \hfill
    \begin{subfigure}[b]{0.32\textwidth}
        \centering
        \includegraphics[width=\linewidth]{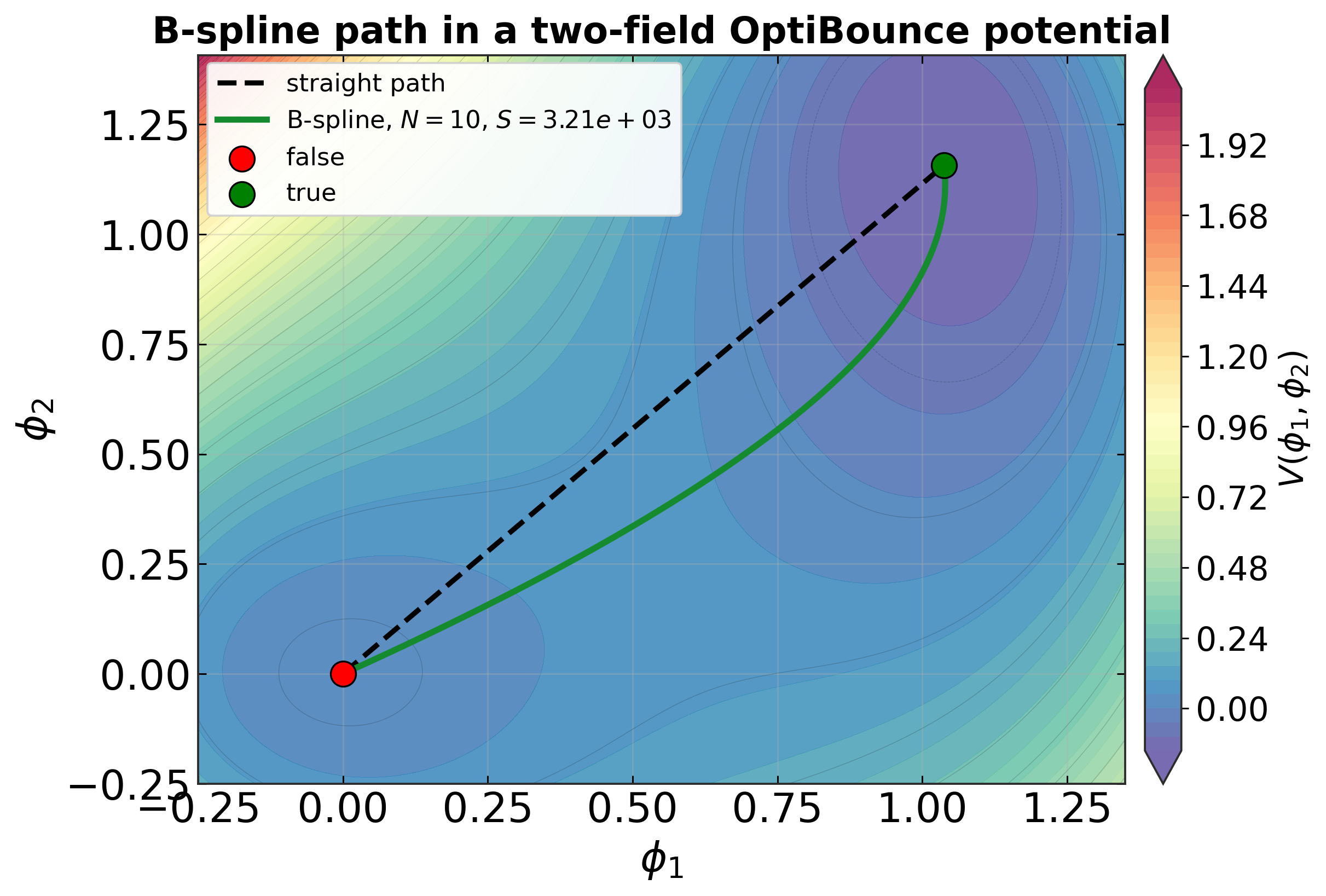}
        \caption{B-spline}
        \label{fig:app_two_field_bspline_path}
    \end{subfigure}
    \hfill
    \begin{subfigure}[b]{0.32\textwidth}
        \centering
        \includegraphics[width=\linewidth]{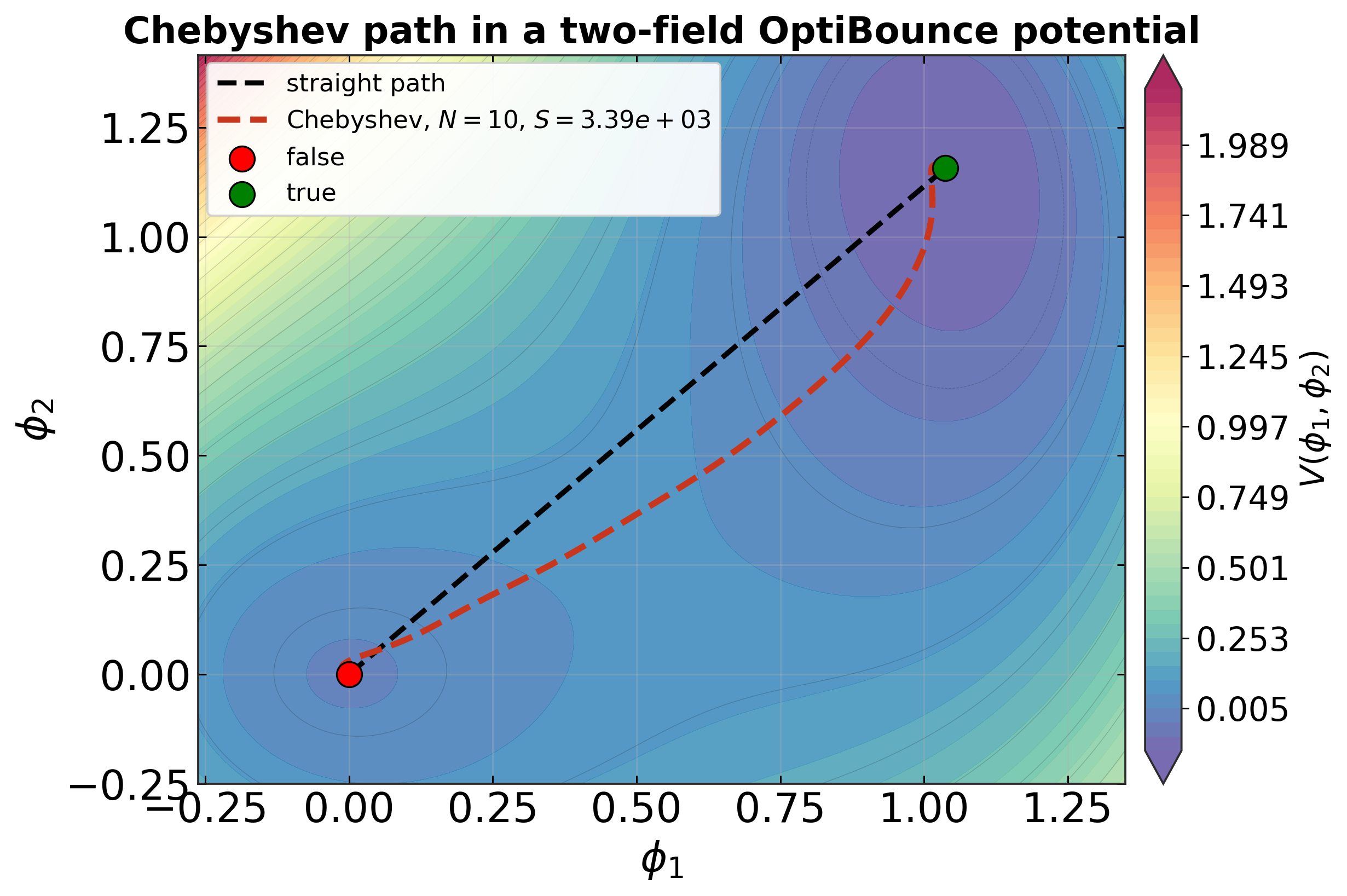}
        \caption{Chebyshev}
        \label{fig:app_two_field_chebyshev_path}
    \end{subfigure}

    \vspace{0.5em}

    \begin{subfigure}[b]{0.32\textwidth}
        \centering
        \includegraphics[width=\linewidth]{optibounce2d_fourier_path_contour.png}
        \caption{Fourier}
        \label{fig:app_two_field_fourier_path}
    \end{subfigure}
    \hfill
    \begin{subfigure}[b]{0.32\textwidth}
        \centering
        \includegraphics[width=\linewidth]{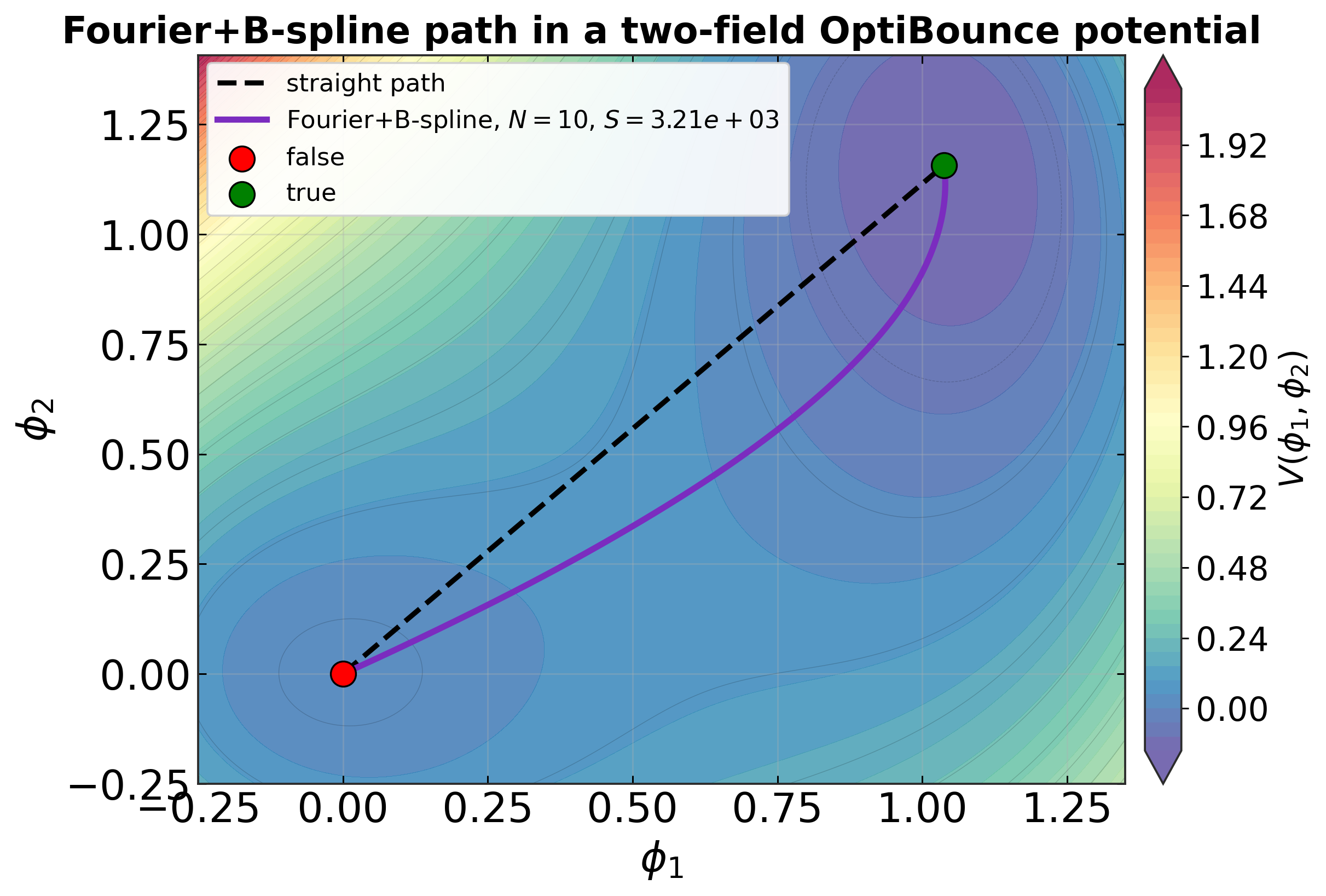}
        \caption{Fourier+B-spline}
        \label{fig:app_two_field_hybrid_path}
    \end{subfigure}
    \hfill
    \begin{subfigure}[b]{0.32\textwidth}
        \centering
        \includegraphics[width=\linewidth]{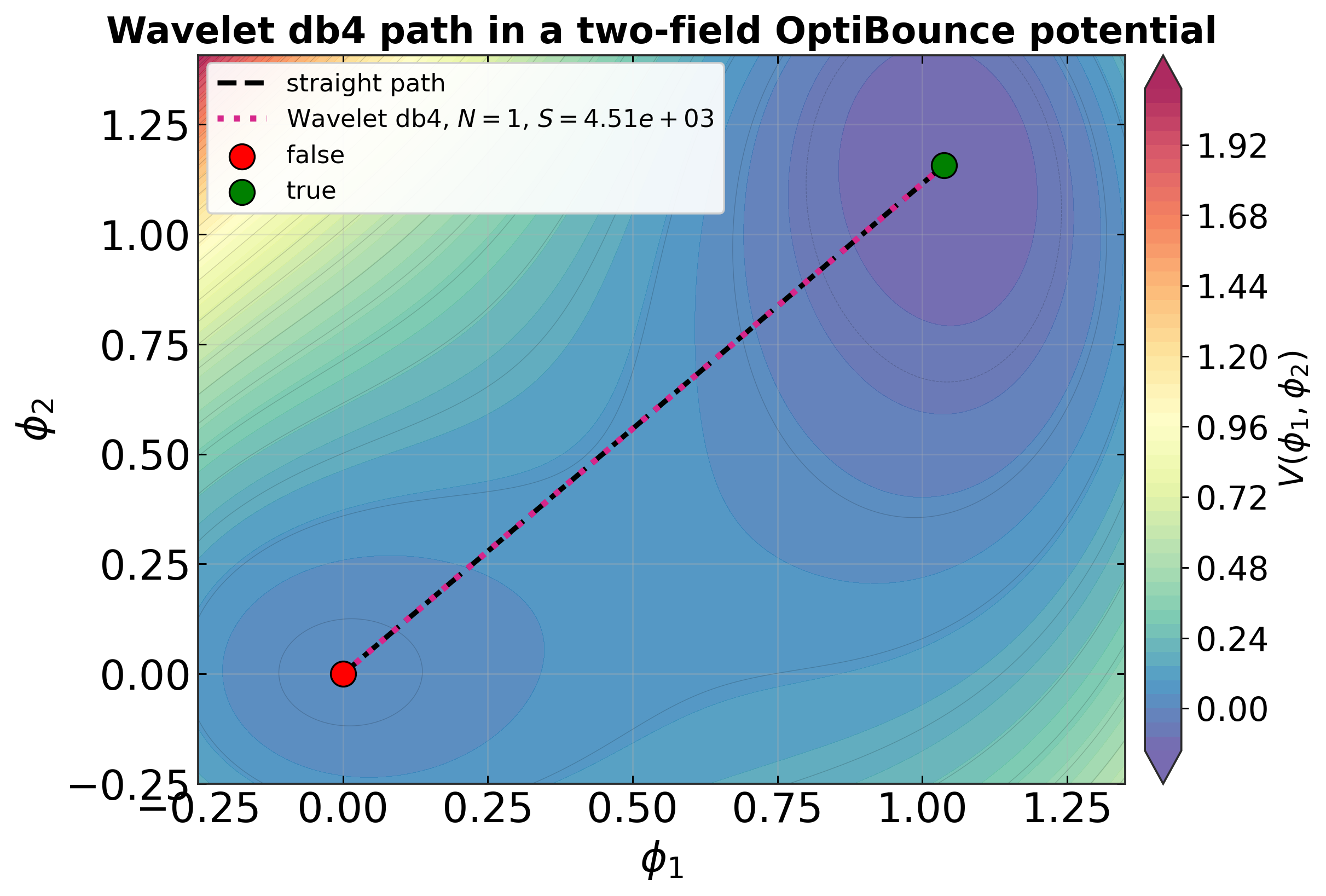}
        \caption{Wavelet db4}
        \label{fig:app_two_field_wavelet_path}
    \end{subfigure}

    \caption{
    Comparison of two-field paths obtained with different endpoint-safe basis
    families. The CosmoTransitions path is included as an external geometric
    reference. The smooth endpoint-safe bases identify the same low-potential
    valley, while the shifted wavelet-inspired basis used here is less well
    matched to this smooth global deformation.
    }
    \label{fig:app_two_field_basis_paths}
\end{figure}

Figure~\ref{fig:app_two_field_basis_paths} compares the paths obtained from
different endpoint-safe bases in the two-field visualisation potential. The
smooth bases recover similar curved paths through the low-potential valley,
while the shifted wavelet-inspired basis used in this scan gives a visibly less
effective deformation for this smooth benchmark.

\subsection{Endpoint-safe basis approximation diagnostics}
\label{app:basis_approximation_diagnostics}

\begin{figure}[h]
    \centering
    \begin{subfigure}[b]{0.32\textwidth}
        \centering
        \includegraphics[width=\linewidth]{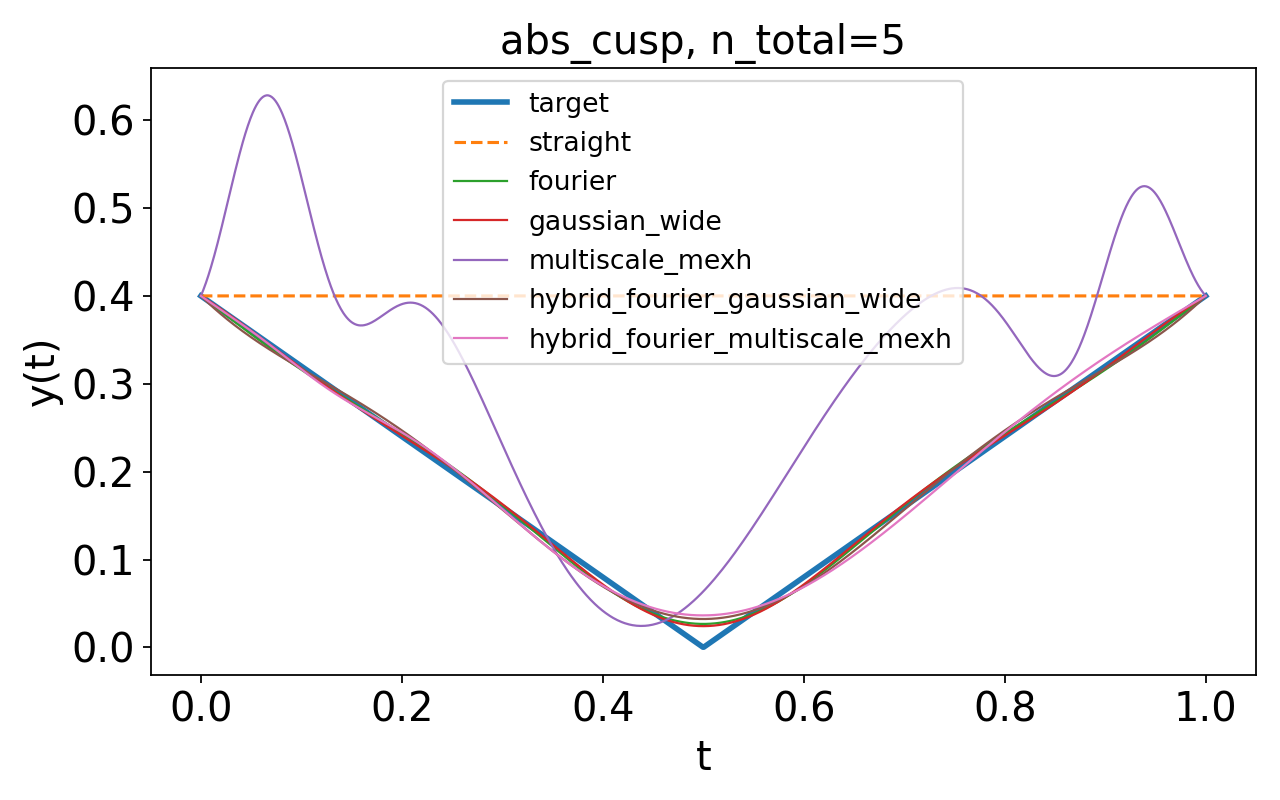}
        \caption{$N_{\rm basis}=5$}
    \end{subfigure}
    \hfill
    \begin{subfigure}[b]{0.32\textwidth}
        \centering
        \includegraphics[width=\linewidth]{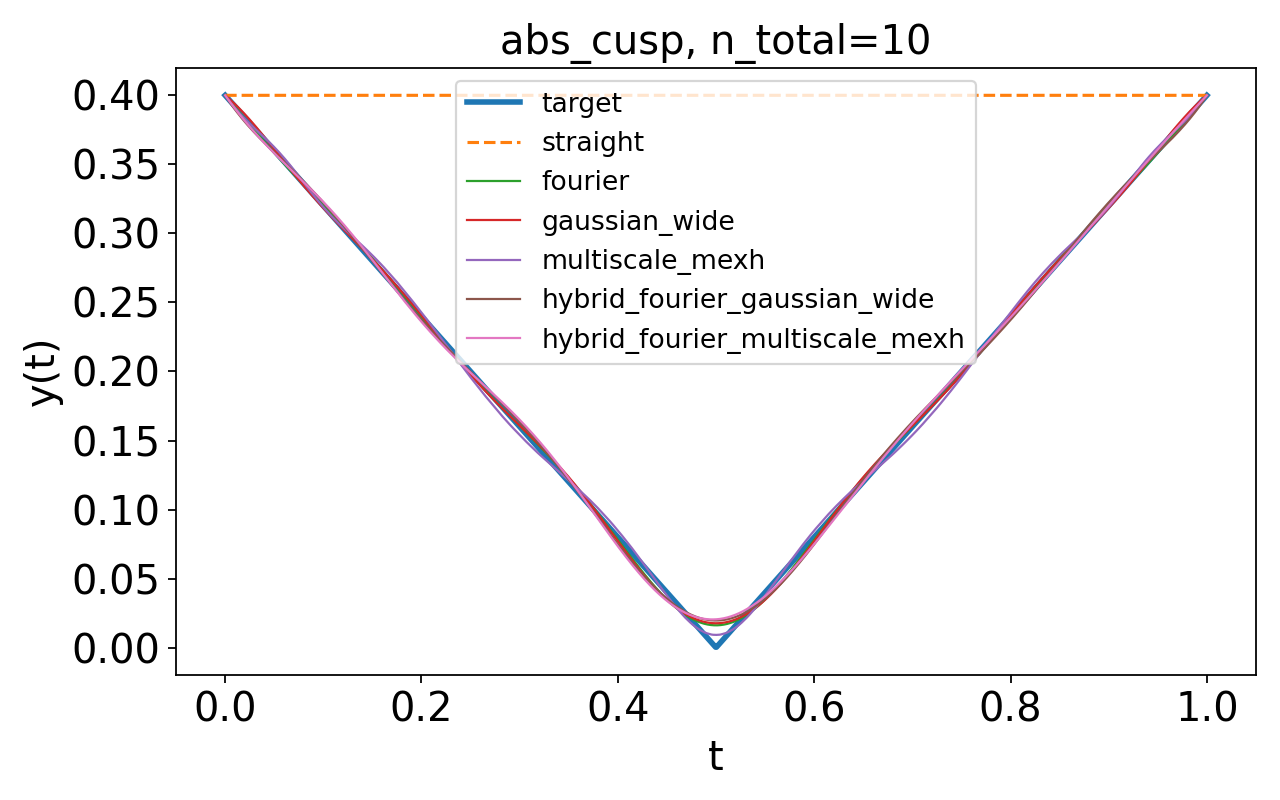}
        \caption{$N_{\rm basis}=10$}
    \end{subfigure}
    \hfill
    \begin{subfigure}[b]{0.32\textwidth}
        \centering
        \includegraphics[width=\linewidth]{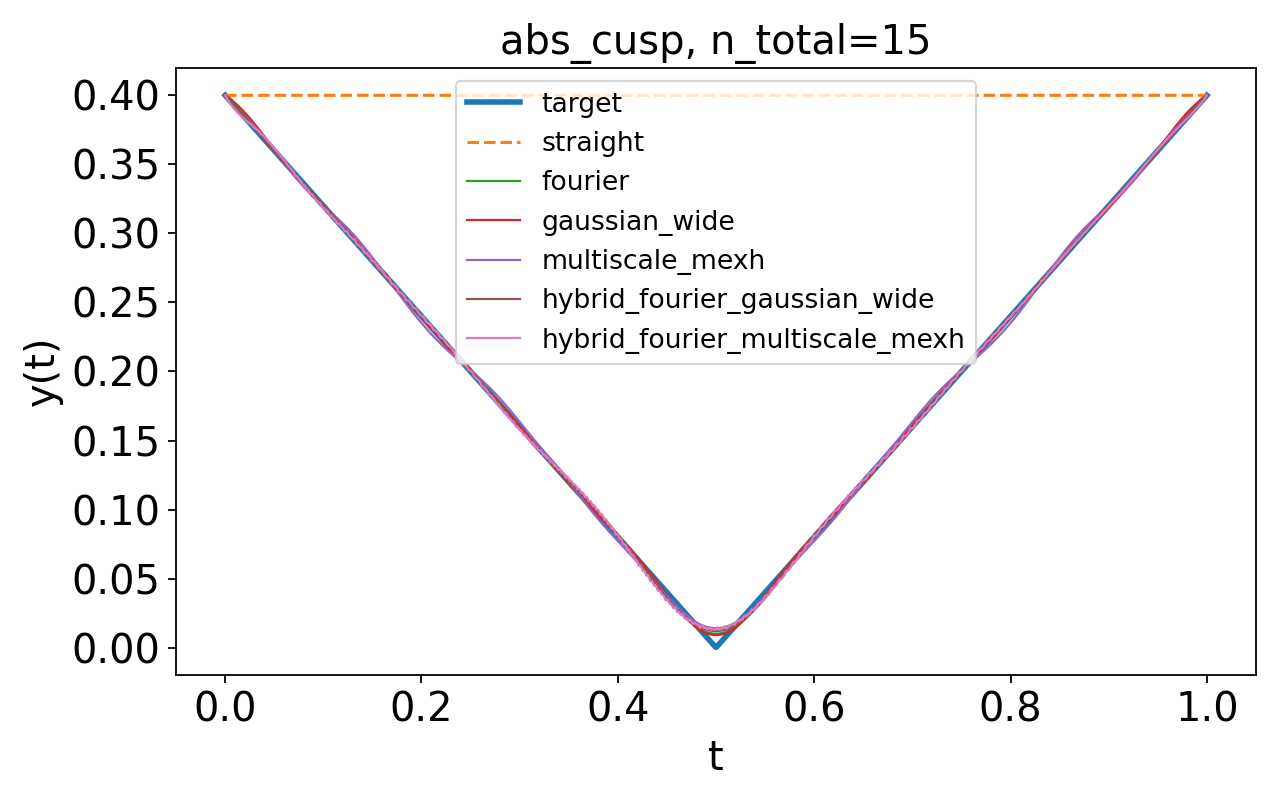}
        \caption{$N_{\rm basis}=15$}
    \end{subfigure}
    \caption{
    Endpoint-safe basis approximation diagnostic for an absolute-value cusp.
    Increasing the number of basis functions improves the representation of
    the non-smooth feature.
    }
    \label{fig:app_basis_abs_cusp}
\end{figure}

\begin{figure}[h]
    \centering
    \begin{subfigure}[b]{0.32\textwidth}
        \centering
        \includegraphics[width=\linewidth]{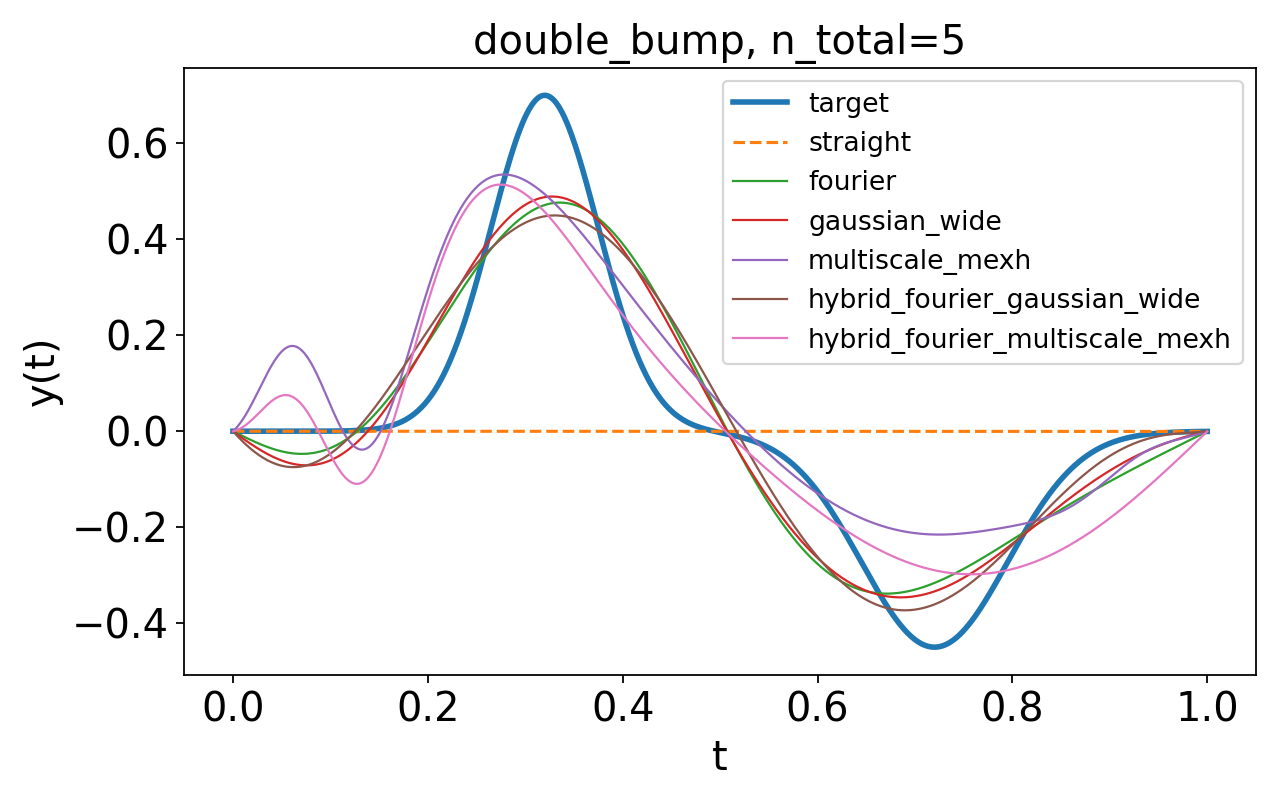}
        \caption{$N_{\rm basis}=5$}
    \end{subfigure}
    \hfill
    \begin{subfigure}[b]{0.32\textwidth}
        \centering
        \includegraphics[width=\linewidth]{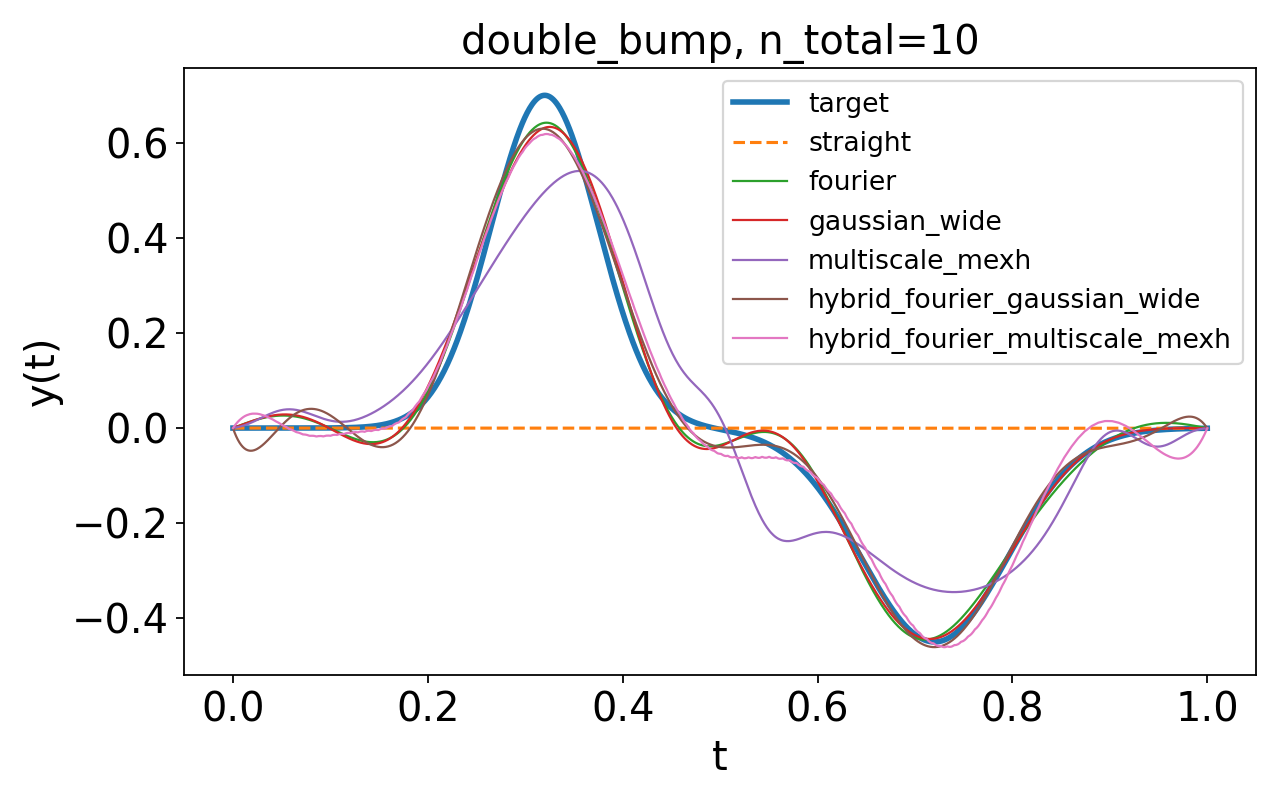}
        \caption{$N_{\rm basis}=10$}
    \end{subfigure}
    \hfill
    \begin{subfigure}[b]{0.32\textwidth}
        \centering
        \includegraphics[width=\linewidth]{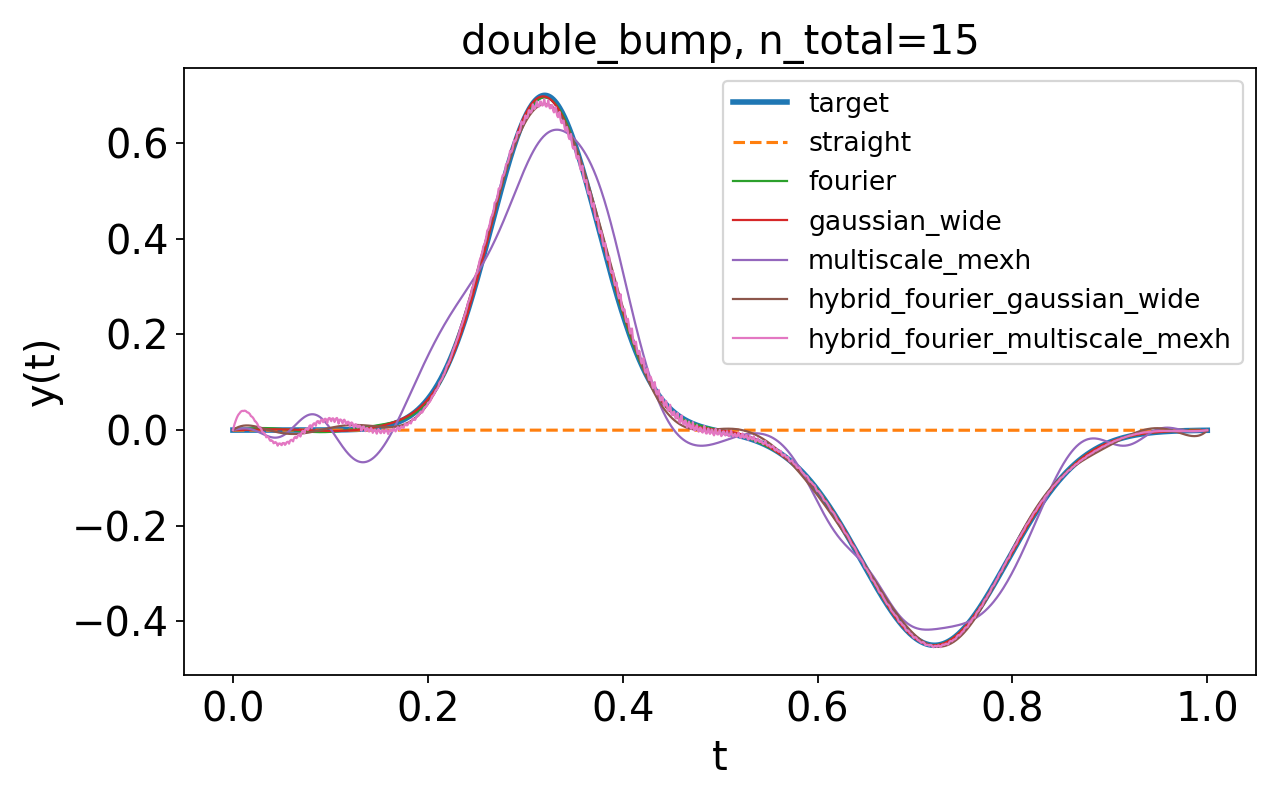}
        \caption{$N_{\rm basis}=15$}
    \end{subfigure}
    \caption{
    Endpoint-safe basis approximation diagnostic for a double-bump profile.
    This tests the ability of the basis to represent more than one localised
    region of curvature.
    }
    \label{fig:app_basis_double_bump}
\end{figure}

\begin{figure}[h]
    \centering
    \begin{subfigure}[b]{0.32\textwidth}
        \centering
        \includegraphics[width=\linewidth]{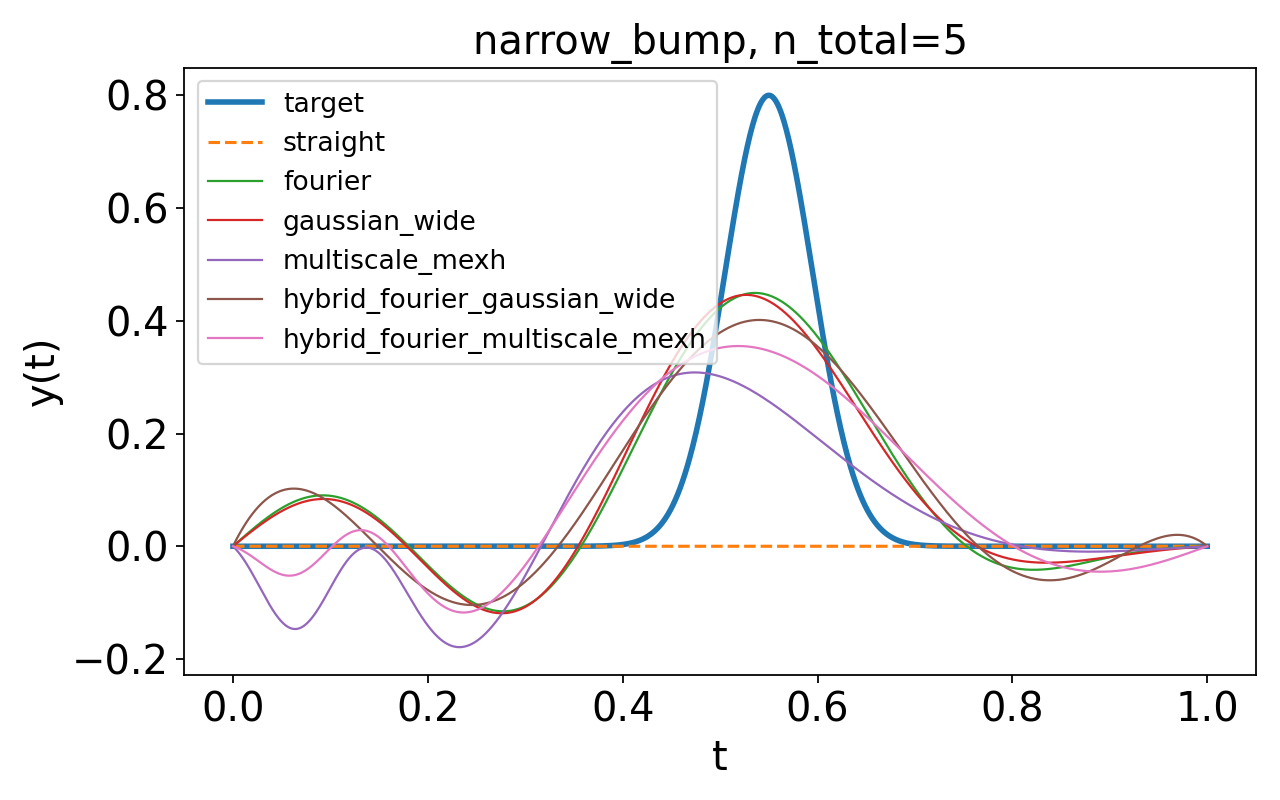}
        \caption{$N_{\rm basis}=5$}
    \end{subfigure}
    \hfill
    \begin{subfigure}[b]{0.32\textwidth}
        \centering
        \includegraphics[width=\linewidth]{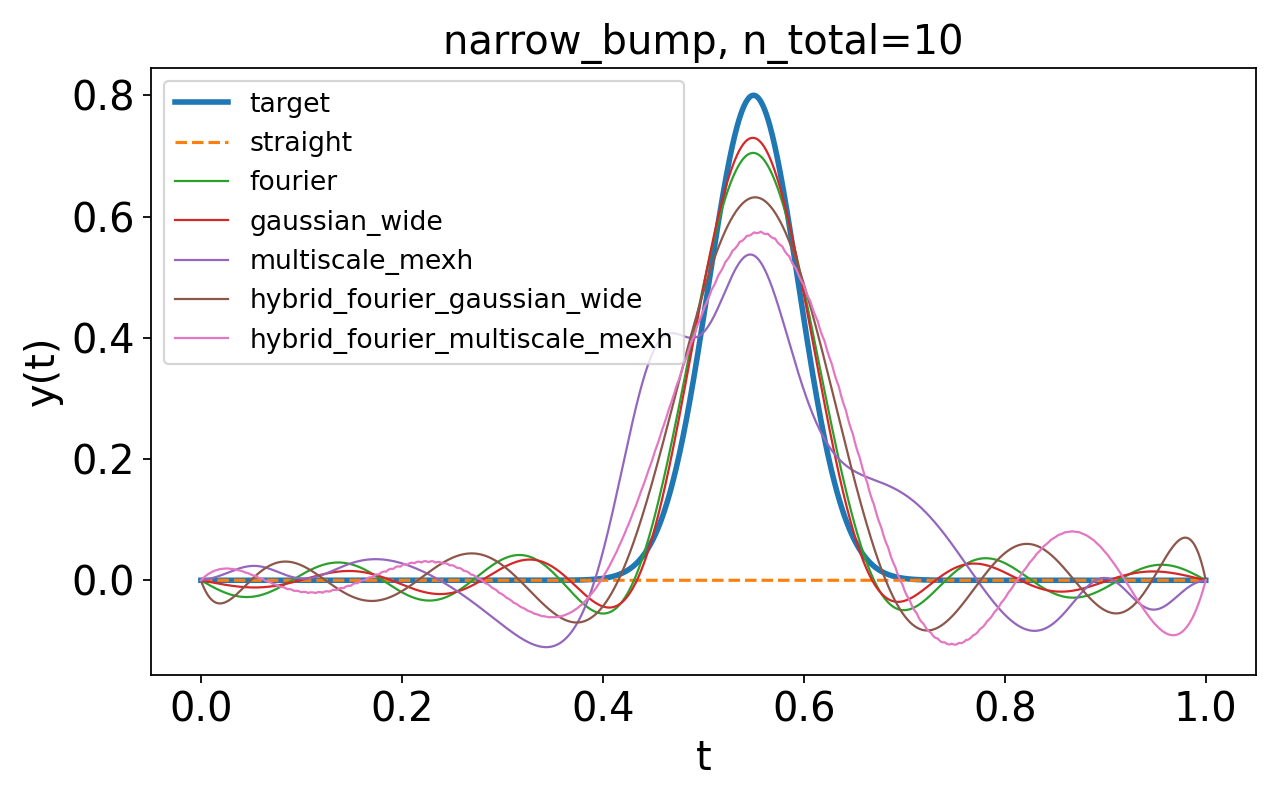}
        \caption{$N_{\rm basis}=10$}
    \end{subfigure}
    \hfill
    \begin{subfigure}[b]{0.32\textwidth}
        \centering
        \includegraphics[width=\linewidth]{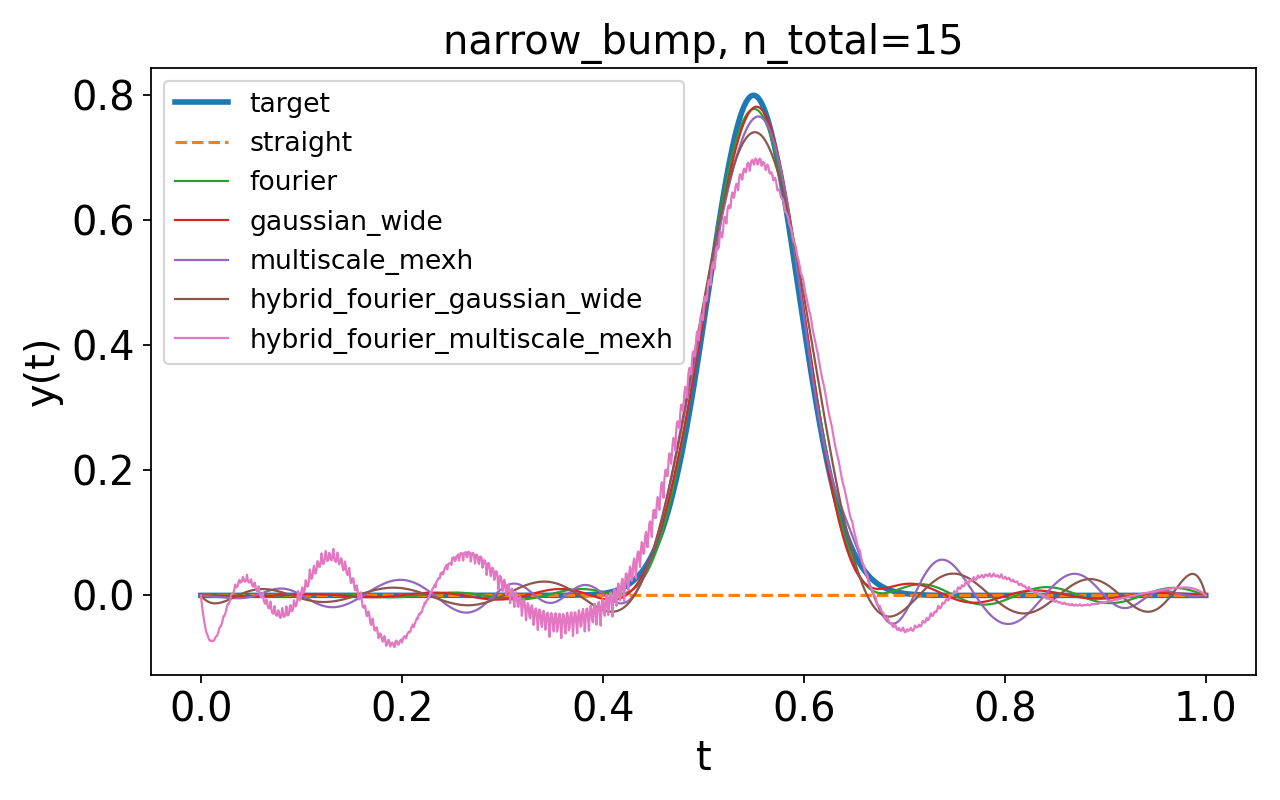}
        \caption{$N_{\rm basis}=15$}
    \end{subfigure}
    \caption{
    Endpoint-safe basis approximation diagnostic for a narrow bump. Local
    basis functions are expected to become more useful for such localised
    structures.
    }
    \label{fig:app_basis_narrow_bump}
\end{figure}

\begin{figure}[h]
    \centering
    \begin{subfigure}[b]{0.32\textwidth}
        \centering
        \includegraphics[width=\linewidth]{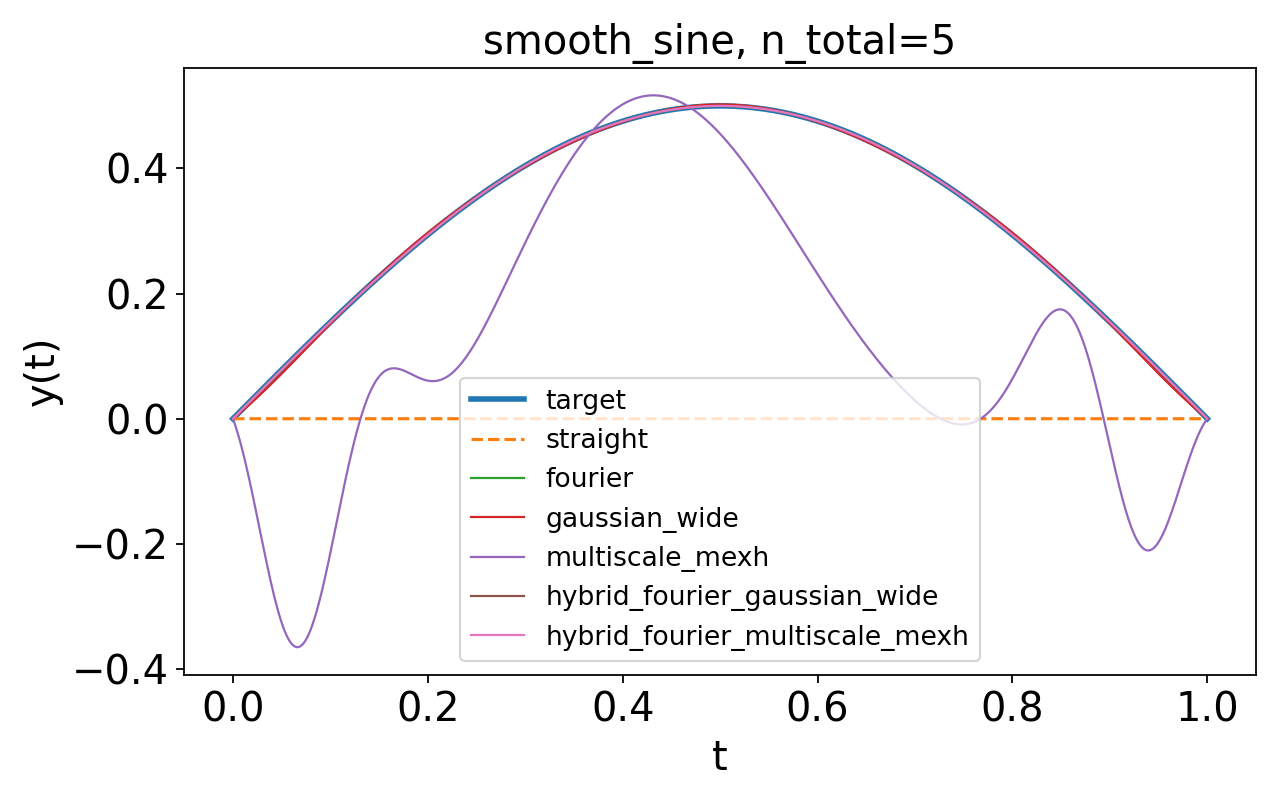}
        \caption{$N_{\rm basis}=5$}
    \end{subfigure}
    \hfill
    \begin{subfigure}[b]{0.32\textwidth}
        \centering
        \includegraphics[width=\linewidth]{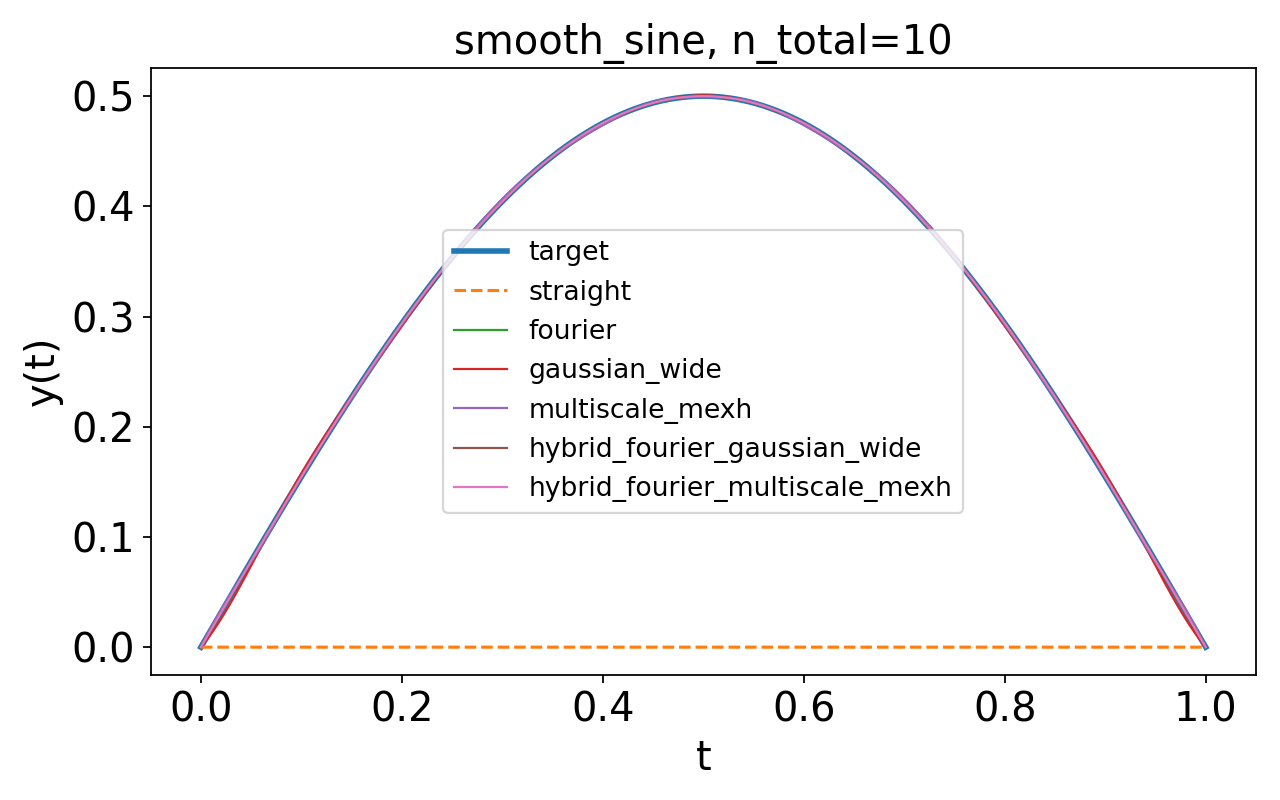}
        \caption{$N_{\rm basis}=10$}
    \end{subfigure}
    \hfill
    \begin{subfigure}[b]{0.32\textwidth}
        \centering
        \includegraphics[width=\linewidth]{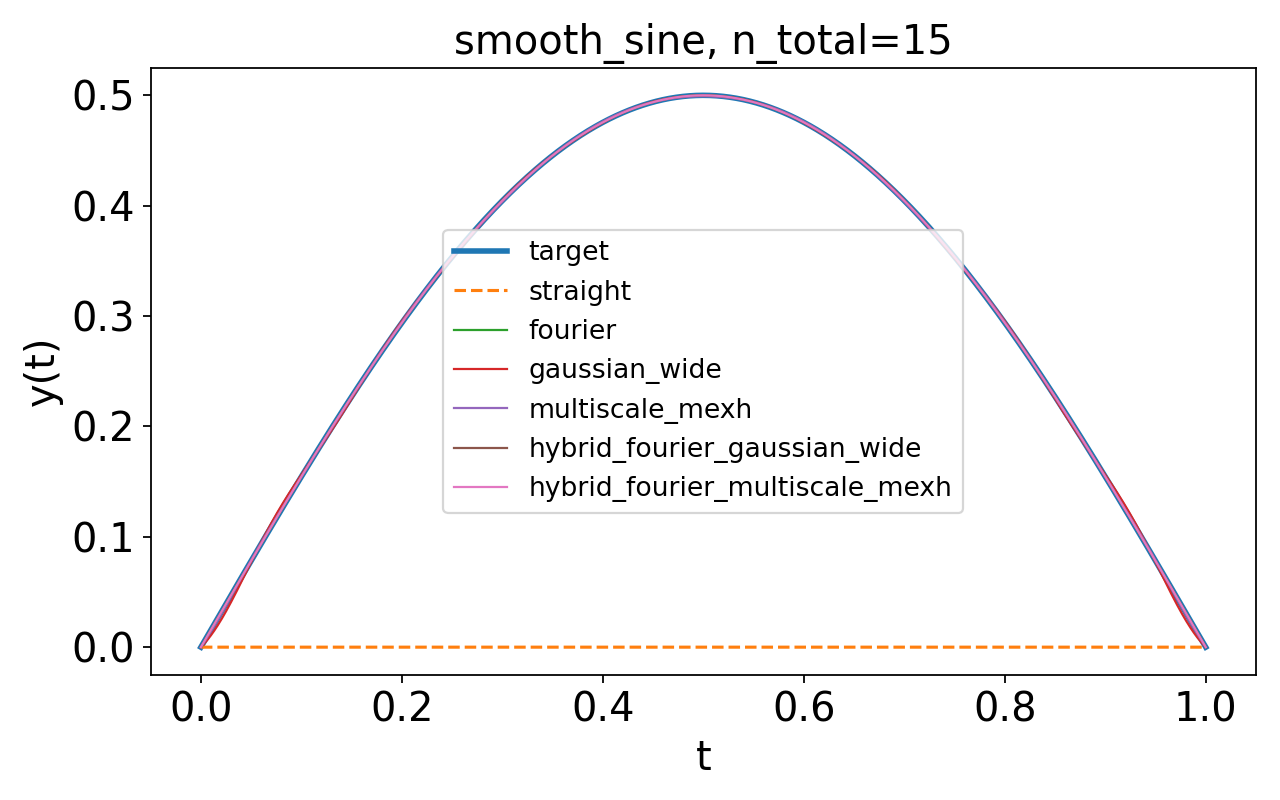}
        \caption{$N_{\rm basis}=15$}
    \end{subfigure}
    \caption{
    Endpoint-safe basis approximation diagnostic for a smooth sinusoidal
    profile. This is the type of global smooth deformation for which Fourier
    sine modes are naturally well-suited.
    }
    \label{fig:app_basis_smooth_sine}
\end{figure}

\begin{figure}[h]
    \centering
    \begin{subfigure}[b]{0.32\textwidth}
        \centering
        \includegraphics[width=\linewidth]{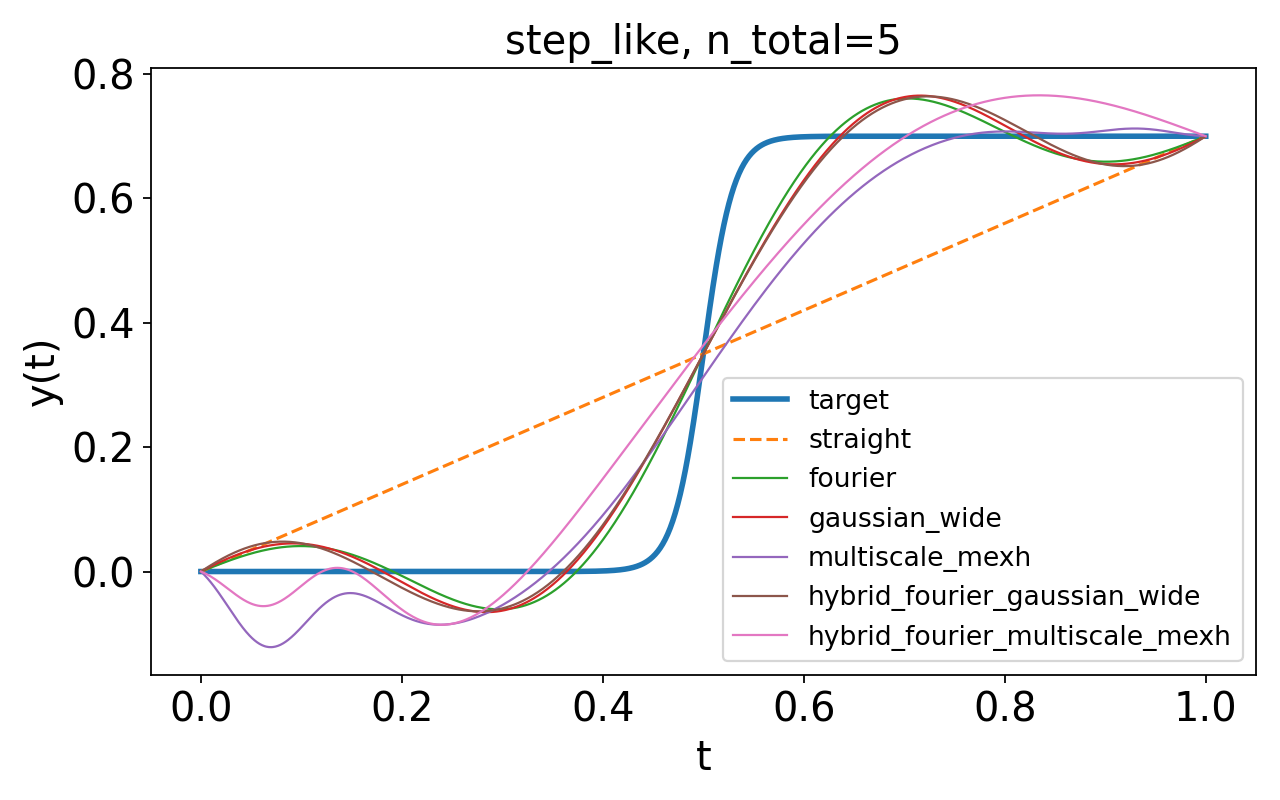}
        \caption{$N_{\rm basis}=5$}
    \end{subfigure}
    \hfill
    \begin{subfigure}[b]{0.32\textwidth}
        \centering
        \includegraphics[width=\linewidth]{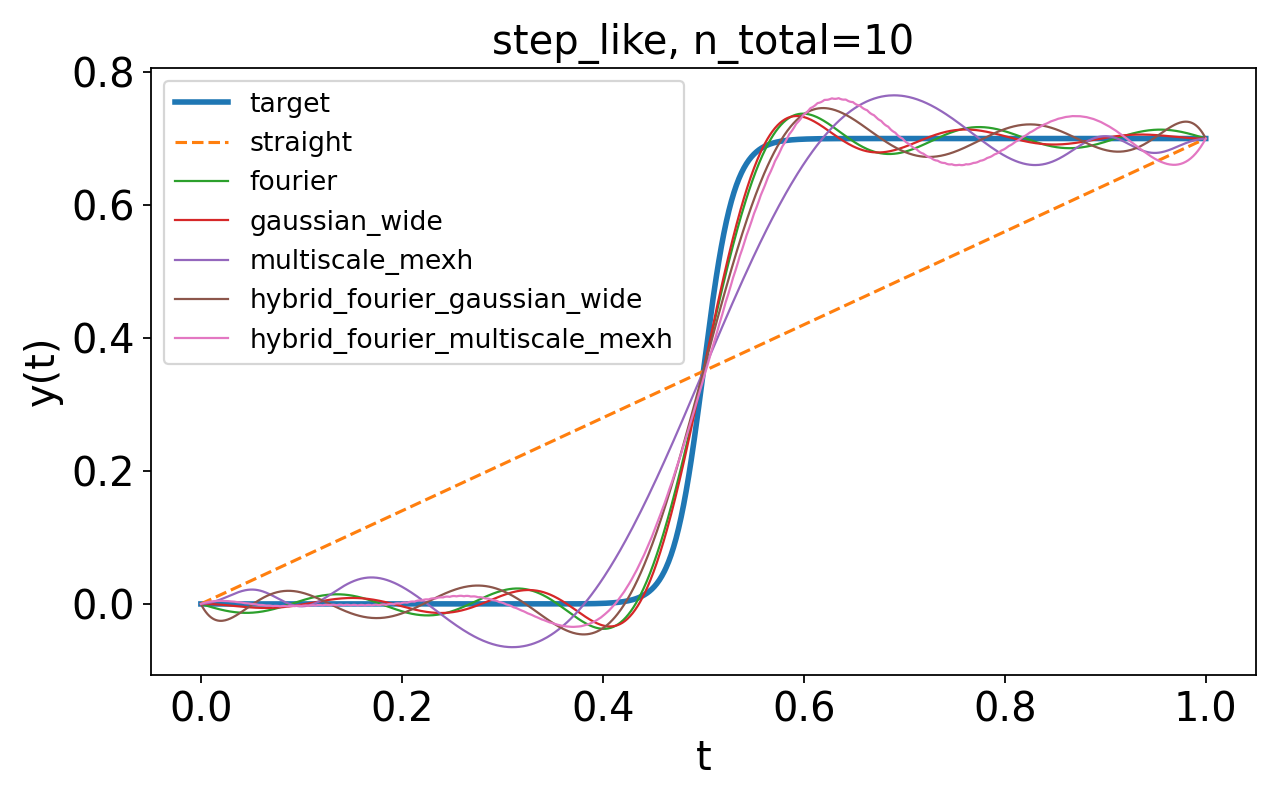}
        \caption{$N_{\rm basis}=10$}
    \end{subfigure}
    \hfill
    \begin{subfigure}[b]{0.32\textwidth}
        \centering
        \includegraphics[width=\linewidth]{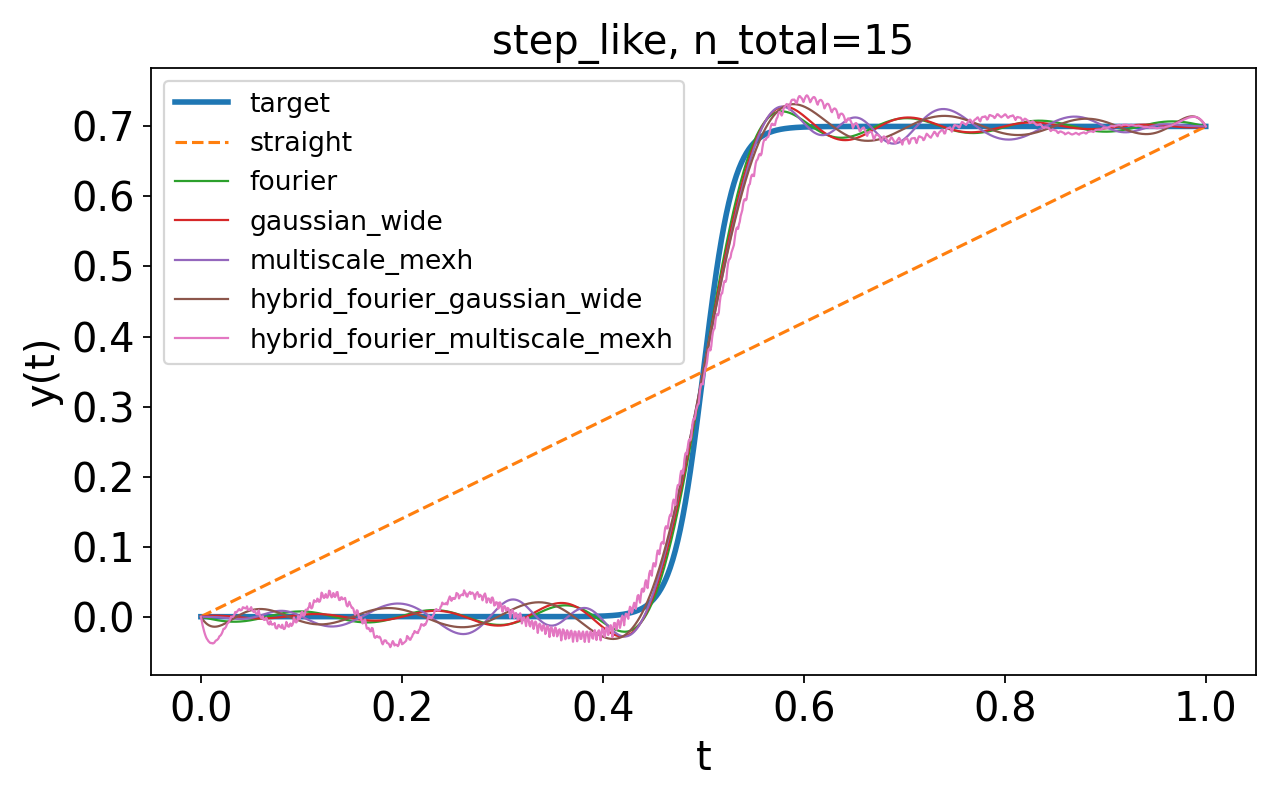}
        \caption{$N_{\rm basis}=15$}
    \end{subfigure}
    \caption{
    Endpoint-safe basis approximation diagnostic for a step-like profile.
    Such profiles are deliberately challenging for smooth global bases and are
    included to illustrate possible hard cases.
    }
    \label{fig:app_basis_step_like}
\end{figure}

\begin{figure}[h]
    \centering
    \begin{subfigure}[b]{0.32\textwidth}
        \centering
        \includegraphics[width=\linewidth]{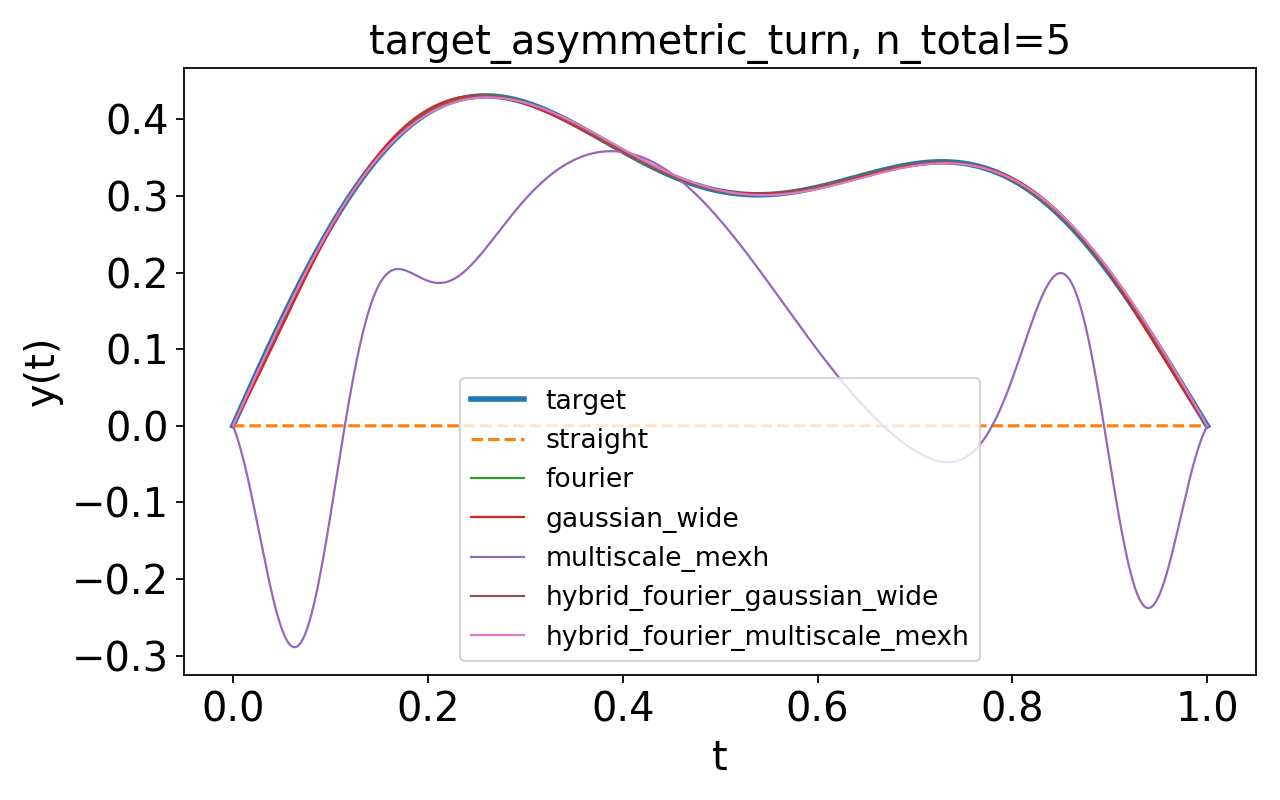}
        \caption{$N_{\rm basis}=5$}
    \end{subfigure}
    \hfill
    \begin{subfigure}[b]{0.32\textwidth}
        \centering
        \includegraphics[width=\linewidth]{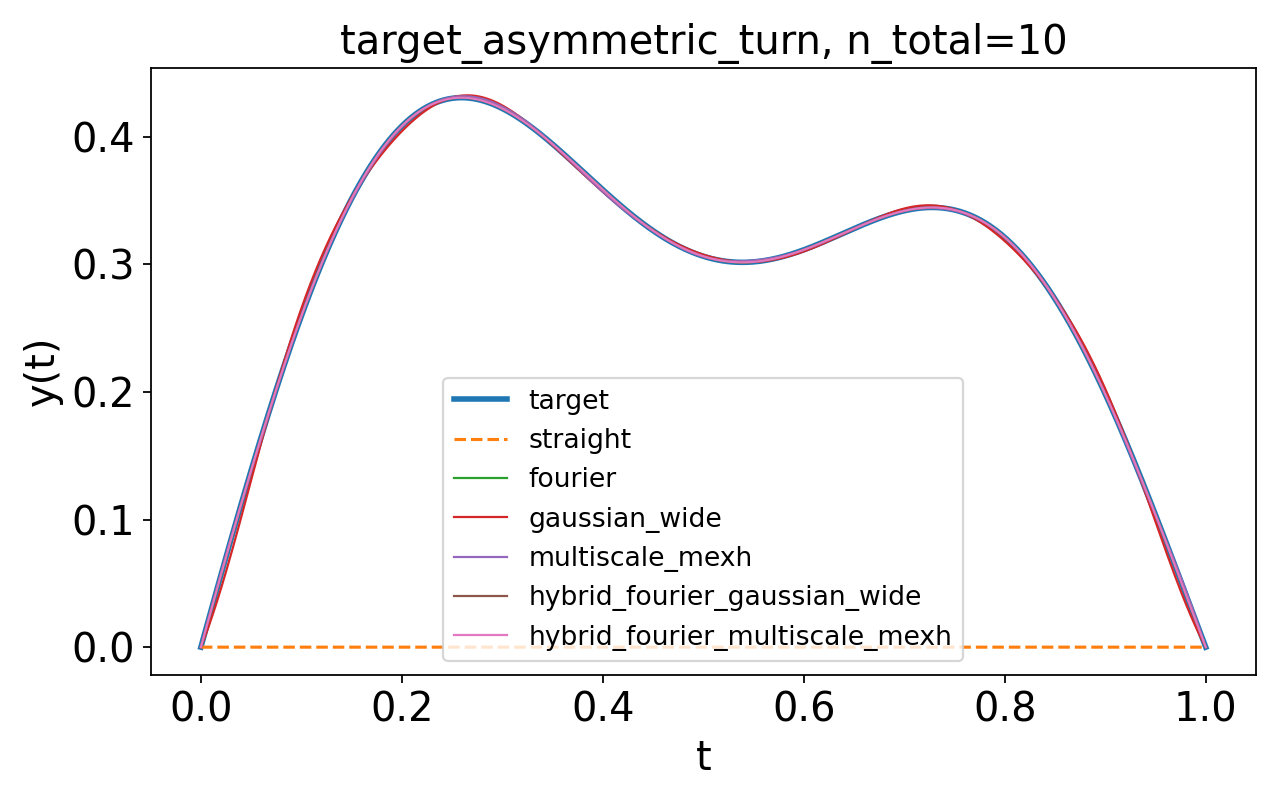}
        \caption{$N_{\rm basis}=10$}
    \end{subfigure}
    \hfill
    \begin{subfigure}[b]{0.32\textwidth}
        \centering
        \includegraphics[width=\linewidth]{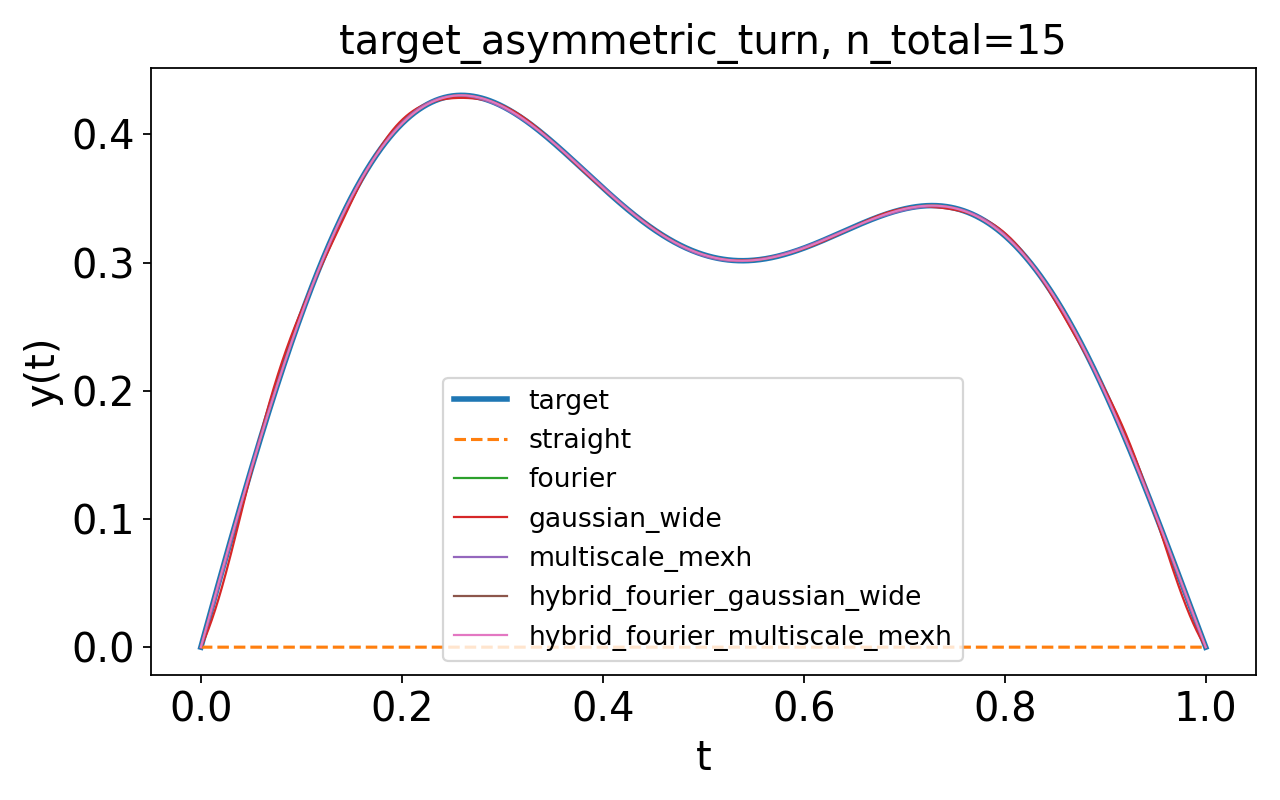}
        \caption{$N_{\rm basis}=15$}
    \end{subfigure}
    \caption{
Endpoint-safe basis approximation diagnostic for a smooth asymmetric-turn
profile, representing a nontrivial but smooth curved deformation.
} \label{fig:app_basis_tanh_wall_sharp}
\end{figure}

Figures~\ref{fig:app_basis_abs_cusp},\ref{fig:app_basis_double_bump},\ref{fig:app_basis_narrow_bump},\ref{fig:app_basis_smooth_sine},\ref{fig:app_basis_step_like},\ref{fig:app_basis_tanh_wall_sharp}
show how the tested endpoint-safe basis families approximate representative
one-dimensional path-deformation profiles as the number of basis functions is
increased. These profiles are not bounce solutions and are not obtained from a
potential. They are controlled diagnostic shapes chosen to mimic different
features that may appear in field-space tunnelling paths, such as smooth global
curvature, localized bends, multiple turns, cusp-like behavior, and sharp
step-like transitions. The purpose of the comparison is to illustrate which
basis families are naturally suited to smooth global deformations and which are
more effective for localized or non-smooth structures.

\section{Full FindBounce point-injection scan}
\label{app:findbounce_full_scan}

\begin{figure}[h]
    \centering
    \begin{subfigure}[b]{0.51\textwidth}
        \centering
        \includegraphics[width=\linewidth]{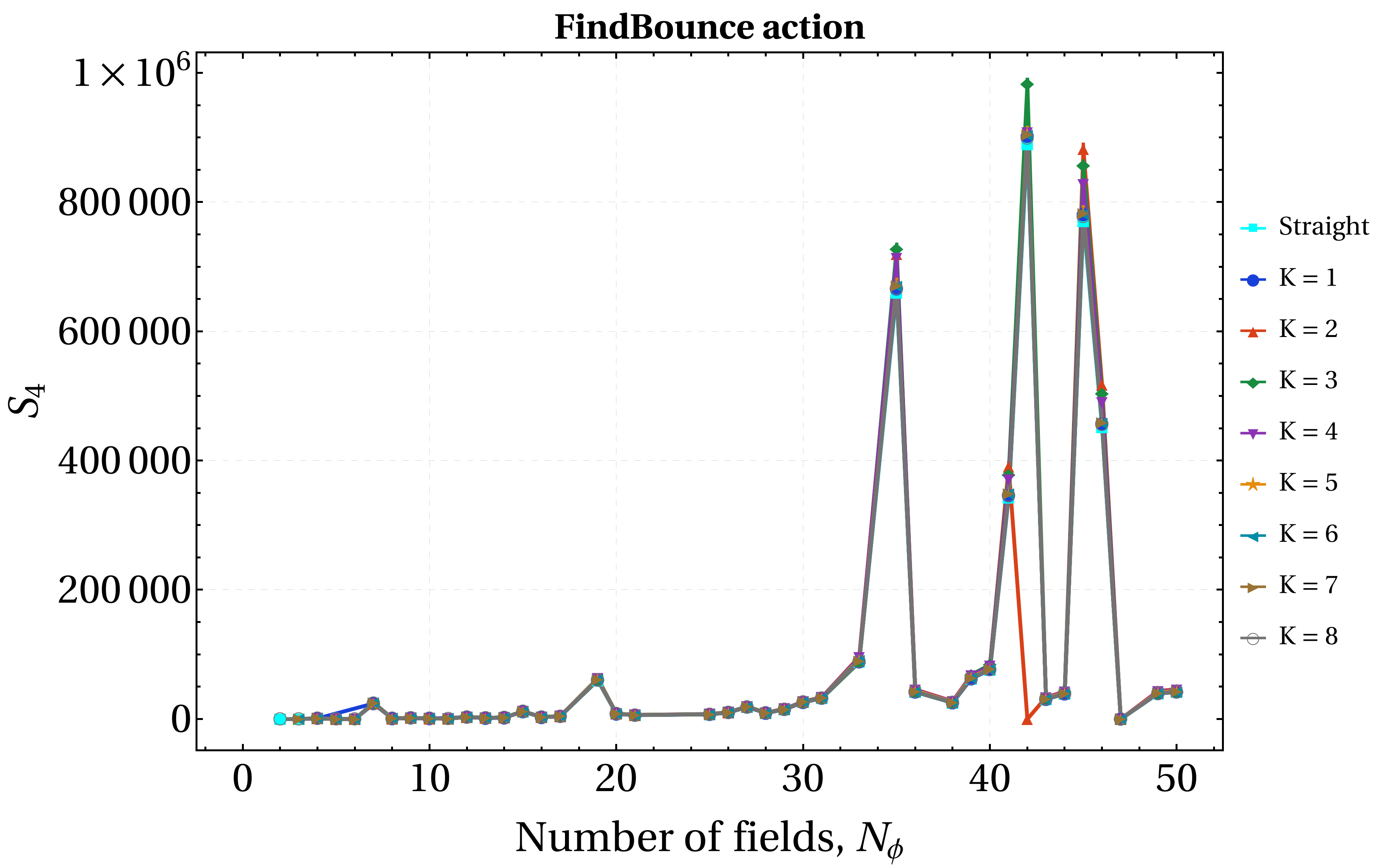}
        \caption{Action comparison}
        \label{fig:app_findbounce_k_action}
    \end{subfigure}
    \hfill
    \begin{subfigure}[b]{0.48\textwidth}
        \centering
        \includegraphics[width=\linewidth]{findbounce_k1_runtime_highdim.png}
        \caption{K=1 runtime (high N\_phi)}
        \label{fig:app_findbounce_k1_time}
    \end{subfigure}
    \\[1ex]  
    \begin{subfigure}[b]{0.98\textwidth}
        \centering
        \includegraphics[width=0.7\linewidth]{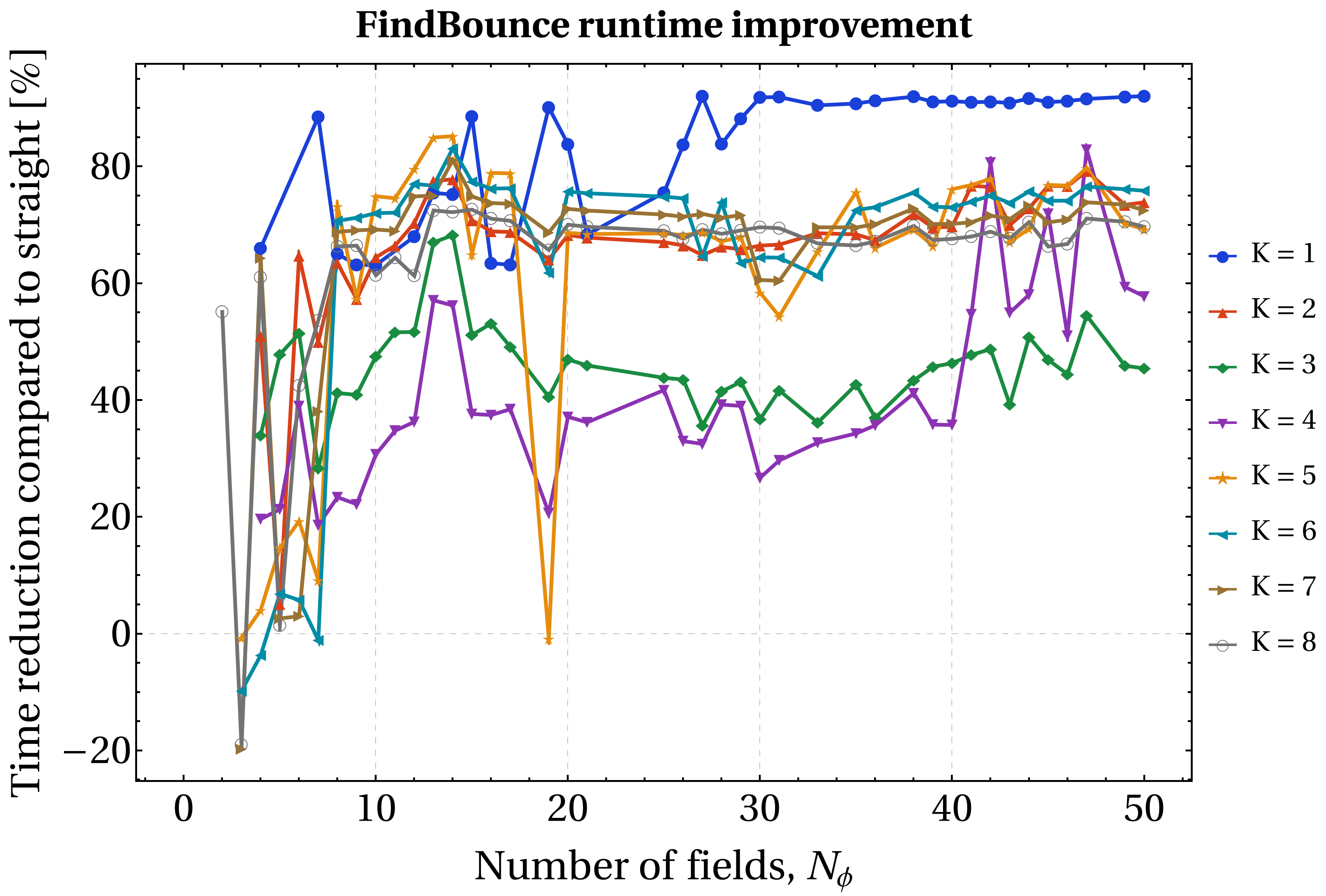}
        \caption{Runtime improvement (\% vs straight)}
        \label{fig:app_findbounce_k_improvement}
    \end{subfigure}
    \caption{
    Full FindBounce scan using Fourier-informed point injections. The label
    ``Straight'' denotes the standard straight-line initialisation, while
    $K=1,\ldots,8$ denotes an initialization in which $K$ intermediate points
    sampled from the optimised Fourier path are supplied to the FindBounce
    polygonal initialisation. The runtime is often reduced when the Fourier path
    information is included. The bottom panel quantifies the runtime reduction
    as a percentage relative to the straight path; positive values mean faster
    computation.
    }
    \label{fig:app_findbounce_full_k_scan}
\end{figure}

This appendix contains the extended FindBounce point-injection scan discussed
in Sec.~\ref{subsec:findbounce_injection}. The main text focuses on the
conservative one-point injection, $K=1$, which tests whether a minimal amount
of Fourier path information can improve the external solver. Here we show the
larger scan over $K=1,\ldots,8$.
For a given optimised Fourier path, we sample $K$ intermediate points along the
path and pass them to FindBounce as part of the polygonal initialisation. The
label ``Straight'' denotes the standard straight-line initialisation with no
injected Fourier point, while the labels $K=1,\ldots,8$ denote runs in which
$K$ intermediate points sampled from the Fourier-deformed path are supplied to
FindBounce.
Figure~\ref{fig:app_findbounce_full_k_scan} shows the action and runtime
comparison as a function of field dimension. The runtime panel shows that
Fourier-informed initialisation often reduces the FindBounce runtime. The
action panel shows that the effect of increasing $K$ is more solver-dependent:
in many high-dimensional cases, the actions remain close to the straight path
result, while some low-dimensional entries move to different action branches.
The extended scan, therefore, illustrates how the polygonal initialisation
responds to the amount of supplied Fourier path information.

The extended scan should be interpreted as a diagnostic of solver
initialisation rather than as a separate action benchmark. Increasing $K$
supplies more detailed geometric information to the polygonal initialisation,
and the response of the solver can depend on the field dimension and on the
local structure of the potential. The results, therefore, provide useful
diagnostics of this sensitivity and point toward future adaptive strategies for
choosing both the number and placement of injected points.

\section{Random benchmark coefficients}
\label{app:benchmark_coefficients}

For reproducibility, we list the numerical coefficients used in the
random-coefficient benchmark of Sec.~\ref{subsec:high_dimensional_benchmark}.
The potential is
\begin{equation}
    V_{N_\phi}(\bm{\phi})
    =
    \left[
        \sum_{i=1}^{N_\phi}
        c_i(\phi_i-1)^2
        -
        \delta_{N_\phi}
    \right]
    \left[
        \sum_{i=1}^{N_\phi}
        \phi_i^2
    \right].
    \label{eq:app_random_coeff_potential}
\end{equation}
The coefficients $c_i$ are drawn once and stored as a master list. The
$N_\phi=N$ potential uses the first $N$ entries of this list. The parameter
$\delta_{N_\phi}$ is generated separately for each field dimension and is
listed in Table~\ref{tab:random_delta_values}. The master list of $c_i$
coefficients is given in Table~\ref{tab:random_c_values}.

\begin{table}[p]
    \centering
    \caption{
    Values of $\delta_{N_\phi}$ used in the random-coefficient benchmark.
    }
    \label{tab:random_delta_values}
    \begin{tabular}{cc|cc|cc|cc|cc}
        \toprule
        $N_\phi$ & $\delta_{N_\phi}$ &
        $N_\phi$ & $\delta_{N_\phi}$ &
        $N_\phi$ & $\delta_{N_\phi}$ &
        $N_\phi$ & $\delta_{N_\phi}$ &
        $N_\phi$ & $\delta_{N_\phi}$ \\
        \midrule
        1 & 0.221847 & 11 & 0.588000 & 21 & 0.492959 & 31 & 0.380900 & 41 & 0.229588 \\
        2 & 0.453316 & 12 & 0.367058 & 22 & 0.144844 & 32 & 0.227052 & 42 & 0.166538 \\
        3 & 0.580579 & 13 & 0.489945 & 23 & 0.189226 & 33 & 0.281576 & 43 & 0.558962 \\
        4 & 0.187605 & 14 & 0.488430 & 24 & 0.419425 & 34 & 0.574295 & 44 & 0.522187 \\
        5 & 0.555685 & 15 & 0.291235 & 25 & 0.535464 & 35 & 0.149725 & 45 & 0.183746 \\
        6 & 0.521535 & 16 & 0.514730 & 26 & 0.487283 & 36 & 0.411234 & 46 & 0.227793 \\
        7 & 0.110666 & 17 & 0.469733 & 27 & 0.409699 & 37 & 0.143006 & 47 & 0.088864 \\
        8 & 0.417092 & 18 & 0.352350 & 28 & 0.552505 & 38 & 0.526793 & 48 & 0.572195 \\
        9 & 0.371763 & 19 & 0.195008 & 29 & 0.470960 & 39 & 0.392492 & 49 & 0.570118 \\
        10 & 0.502025 & 20 & 0.431052 & 30 & 0.396339 & 40 & 0.377964 & 50 & 0.569079 \\
        \bottomrule
    \end{tabular}
\end{table}

\begin{table}[p]
    \centering
    \caption{
    Master list of $c_i$ coefficients used in the random-coefficient
    benchmark. The $N_\phi=N$ potential uses the first $N$ entries.
    }
    \label{tab:random_c_values}
    \begin{tabular}{cc|cc|cc|cc|cc}
        \toprule
        $i$ & $c_i$ &
        $i$ & $c_i$ &
        $i$ & $c_i$ &
        $i$ & $c_i$ &
        $i$ & $c_i$ \\
        \midrule
        1 & 0.316925 & 11 & 0.243948 & 21 & 0.155138 & 31 & 0.424820 & 41 & 0.568111 \\
        2 & 0.330670 & 12 & 0.114556 & 22 & 0.385291 & 32 & 0.351709 & 42 & 0.113898 \\
        3 & 0.443894 & 13 & 0.391308 & 23 & 0.327356 & 33 & 0.274176 & 43 & 0.313693 \\
        4 & 0.460470 & 14 & 0.516287 & 24 & 0.288142 & 34 & 0.526207 & 44 & 0.244186 \\
        5 & 0.353939 & 15 & 0.492572 & 25 & 0.134237 & 35 & 0.443424 & 45 & 0.115652 \\
        6 & 0.502579 & 16 & 0.234313 & 26 & 0.210105 & 36 & 0.529122 & 46 & 0.571282 \\
        7 & 0.559264 & 17 & 0.473028 & 27 & 0.546836 & 37 & 0.409519 & 47 & 0.124231 \\
        8 & 0.359073 & 18 & 0.335657 & 28 & 0.357503 & 38 & 0.499048 & 48 & 0.219889 \\
        9 & 0.391590 & 19 & 0.146378 & 29 & 0.135883 & 39 & 0.496336 & 49 & 0.568844 \\
        10 & 0.337356 & 20 & 0.463183 & 30 & 0.337061 & 40 & 0.565668 & 50 & 0.294674 \\
        \bottomrule
    \end{tabular}
\end{table}


\section*{Acknowledgements}

A.S.\ acknowledges financial support from the IISc and the Council of Scientific and
Industrial Research (CSIR), Government of India, under Senior Research
Fellowship No.\ 09/0079(15487)/2022-EMR-I. SKV is supported by IISc REDA grants. SRK was supported by the Kishore Vaigyanik Protsahan Yojana (KVPY) fellowship and by Perimeter Institute for Theoretical Physics.

The numerical work in this paper made use of open-source scientific software,
including \texttt{Python}, \texttt{NumPy}, \texttt{SciPy}, \texttt{JAX},
\texttt{Matplotlib}, and \texttt{Pandas}. We also acknowledge the developers of
the public vacuum-decay and phase-transition tools used for comparison and
cross-checks, including \texttt{CosmoTransitions}, \texttt{FindBounce},
\texttt{BubbleProfiler}, \texttt{SimpleBounce}, and \texttt{OptiBounce}. All
third-party software was used under its respective open-source license.
Large Language Models were used only for language polishing, code organisation,
and assistance with routine scripting. All scientific content, numerical
checks, interpretations, and conclusions are the authors' own.
The code and scripts associated with this work are publicly available at
\href{https://github.com/AadarshSingh0/fourier-path-bounce}
{\faGithub\ \texttt{fourier-path-bounce}}. The original code written for this project is released under the MIT License.



\bibliographystyle{JHEP}
\bibliography{reference}

\end{document}